\documentclass[aps,superscriptaddress,prb,twocolumn]{revtex4-1}
\usepackage{amssymb,amsmath,mathptmx}
\usepackage{graphicx}
\usepackage{dcolumn}
\usepackage{bm}
\usepackage{hyperref}
\usepackage{color}
\usepackage[utf8x]{inputenc}
\usepackage{array}
\usepackage{ulem}
\usepackage{soul}

\newcolumntype{C}{>{\centering\arraybackslash}p{1em}}

\newcommand{\be}{\begin{equation}}
\newcommand{\ee}{\end{equation}}
\newcommand{\bea}{\begin{eqnarray}}
\newcommand{\eea}{\end{eqnarray}}

\begin{document}

\title{Ballistic Andreev interferometers}

\author{R\'egis M\'elin}
\email{regis.melin@neel.cnrs.fr}

\affiliation{Université Grenoble-Alpes, CNRS, Grenoble INP, Institut
  NEEL, Grenoble, France}
\author{Asmaul Smitha Rashid}

\affiliation{Department of Electrical Engineering, The Pennsylvania
  State University, University Park, Pennsylvania 16802, USA}

\author{Morteza Kayyalha}

\affiliation{Department of Electrical Engineering, The Pennsylvania
  State University, University Park, Pennsylvania 16802, USA}

\begin{abstract}
{A Josephson junction, formed between two phase-biased
  superconductors and a normal metal, hosts a discrete spectrum of
  Andreev bound states (ABS). In this paper, we develop a theory for
  long ballistic Andreev interferometers in two-dimensional metals. We
  consider three frameworks in our theoretical analysis: (i)
  perturbation theory in the tunneling amplitudes; (ii)
  non-perturbative transport theory; and (iii) physically motivated
  approximations to visualize the conductance maps in the (flux,
  voltage) plane. We find a non-standard phase-sensitive Andreev
  reflection process in ballistic interferometers that couples the
  supercurrent to the non-equilibrium populations of the ABS in the
  normal region. Furthermore, our model shows that conductance
  spectroscopy follows the spectrum of the ABS in long junctions.}
{We also discuss our results in terms of the
  semiclassical theory, the classical orbits being the one-dimensional
  Andreev tubes.}  {Our theoretical analysis captures
  the results of recent experiments by the Penn State and Harvard
  groups.}
\end{abstract}

\maketitle

\section{Introduction}

{Building on Anderson's theory of gauge invariance in
  superconductors \cite{Anderson1,Anderson2}, Josephson
  \cite{Josephson} demonstrated that a dissipationless DC-supercurrent
  ($I$) flows through a two-terminal weak link such that
  $I(\varphi_{ab})=I_c \sin{(\varphi_{ab})}$, where $\varphi_{ab}$ is
  the phase difference between the superconductors $S_a$ and $S_b$.
  Saint James and de Gennes \cite{deGennes,SaintJames} later
  discovered that in superconductor-normal-superconductor ($S_aNS_b$)
  junctions, microscopic Andreev reflections and the resulting Andreev
  bound states (ABS) \cite{Andreev} give rise to supercurrent.  In
  these junctions, the ABS spectrum depends on whether the junction is
  in the short- or long-junction limit, which is determined by the
  length of the $N$ region relative to the superconducting coherence
  length. Kulik, Ishii and Bagwell \cite{Kulik,Ishii,Bagwell}
  specifically investigated ballistic long junctions and found that
  they host ABS with alternating $0$- and $\pi$-shifted current-phase
  relations.}

\begin{figure*}[htb]
  \includegraphics[width=\textwidth]{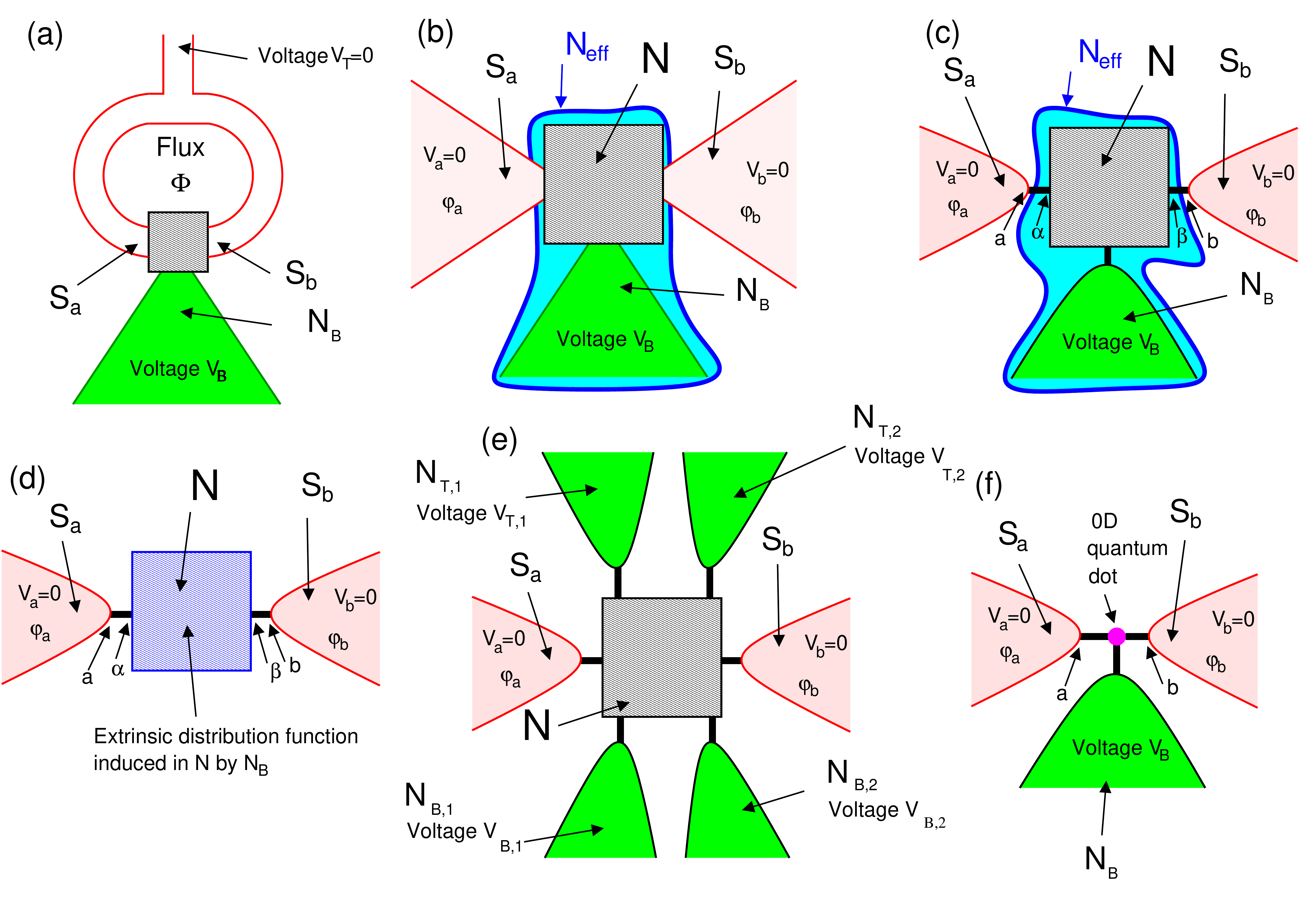}
  \caption{{{{The ballistic normal-superconducting Andreev
          interferometer.}  Panel a shows} the experimental
      configuration of the ballistic normal-superconducting Andreev
      interferometer, including the {\it external} loop connecting
      $S_a$ and $S_b$, pierced by the magnetic flux $\Phi$, and the
      control parameter of the voltage bias $V_B$ on the normal lead
      $N_B$.  Three of the considered theoretical modelings discussed
      in the paper are shown on panel b, c and d, including
      multichannel interfaces (b), single-channel contacts (c) and an
      effective model for the populations in the sequential tunneling
      channel (d). A ballistic normal-superconducting Andreev interferometer
      with four normal leads is shown {on panel e. The} theoretical
      device geometries on panels b-e are physically introduced in
      section~\ref{sec:geometry}.  A device with a quantum dot is also
      shown (f), corresponding to the simple 0D model solved in
      section~\ref{sec:example-0D} and in Appendix~\ref{sec:0D}.}
\label{fig:thedevice}
}
\end{figure*}

{Here, we extend this model to ballistic
  three-terminal Andreev interferometers consisting of a central
  normal region $N$ connected to two phase-biased superconducting
  contacts ($S_a$ and $S_b$) in a loop (Fig.~\ref{fig:thedevice}a).
  The central-$N$ is additionally connected to a normal lead $N_B$,
  biased at voltage $V_B$ with current $I_B$. We consider that this
  $\left(S_aNS_b\right)N_B$ Josephson device is in the long junction
  limit, where the dimension of the central $N$ region exceeds the
  ballistic coherence length. We find that the differential
  conductance $d I_B/d V_B$ can directly probe the ABS spectrum
  \cite{Kulik,Ishii,Bagwell}. Additionally, we identify nonstandard
  Andreev reflection processes that convert phase-sensitive
  supercurrent in the superconducting loop into normal current in
  $N_B$. This {\it phase-sensitive Andreev reflection} (phase-AR)
  process couples the pair amplitudes at $S_a$ and $S_b$ to the
  nonequilibrium distribution function in $N$. Our theoretical
  framework is three-fold:}

{(i) We consider that the central-$N$ has a continuous spectrum and we
  evaluate the Andreev interferometer conductance at the lowest order
  of perturbation theory in the coupling between the central-$N$ and
  the superconducting leads $S_a$ and $S_b$. We then find that the
  conductance $d I_B / d V_B$ features a smooth harmonic oscillatory
  checkerboard pattern in the flux and voltage plane.}

{(ii) We evaluate the general nonperturbative
  expression of the currents, valid at all orders in the tunneling
  amplitudes and for all dimensions of the central-$N$, with respect
  to the Fermi wavelength and the superconducting coherence length. We
  demonstrate that the conductance $d I_B/d V_B$ exhibits peaks
  corresponding to the ABS energy levels in the $S_aNS_b$ junction.
  We also discuss the contributions of sequential
  tunneling and elastic cotunneling to the conductance.}

{(iii) We 
consider physically-motivated approximations for the conductance
  $d I_B/d V_B$, which combines the preceding perturbative and
  nonperturbative expansions of (i) and (ii). We
  obtain colorplots of the conductance as a function of magnetic
  flux and bias voltage. We consider all the values of the normal lead
  line-width broadening, both small and large, compared to the
  energy level spacing of the corresponding $SNS$ device. The
  colorplots feature a cross-over between a regime of {\it conductance peaks}
  at small line-width broadening
  and {\it conductance oscillations} at larger line-width broadening.}

{{We also} highlight the experimental consequences of
  our theory:}

{(i) Our theory provides a complete explanation for the recent Penn
  State group experiment \cite{PSE} on three-terminal
  $(S_a,S_b,N_B)$ and $(S_a,S_b,S_B)$ Andreev interferometers.}

{(ii) Our theory can also explain the bias voltage and magnetic flux
  dependence of the quartet anomaly in the recent Harvard group
  experiment \cite{Huang2022} on four-terminal Josephson junctions.}

{Finally, we discuss the physical picture of the
  classical trajectories sustained by the one-dimensional (1D) Andreev
  tubes, in connection with Ref.~\onlinecite{Kulik}, and with the
  experiments mentioned above \cite{PSE,Huang2022}. We put forward the
  periodic energy-compression of the ABS spectrum that emerges from
  the semiclassical picture, and can be probed in the conductance
  spectra.}

{The paper is organized as follows. The Hamiltonians,
  the geometry and the methods are introduced in
  section~\ref{sec:H-geom-method}. The perturbative and
  nonperturbative calculations are presented {in
    sections~\ref{sec:lowest-order} and~\ref{sec:intrinsic1}
    respectively. The physically-motivated approximations and the
    figures for the conductance spectra are presented in
    Section~\ref{sec:num-dot}. A discussion and final remarks are
    provided in section~\ref{sec:discussion-conclusion}.}}

\begin{figure}[htb]
  \includegraphics[width=.7\columnwidth]{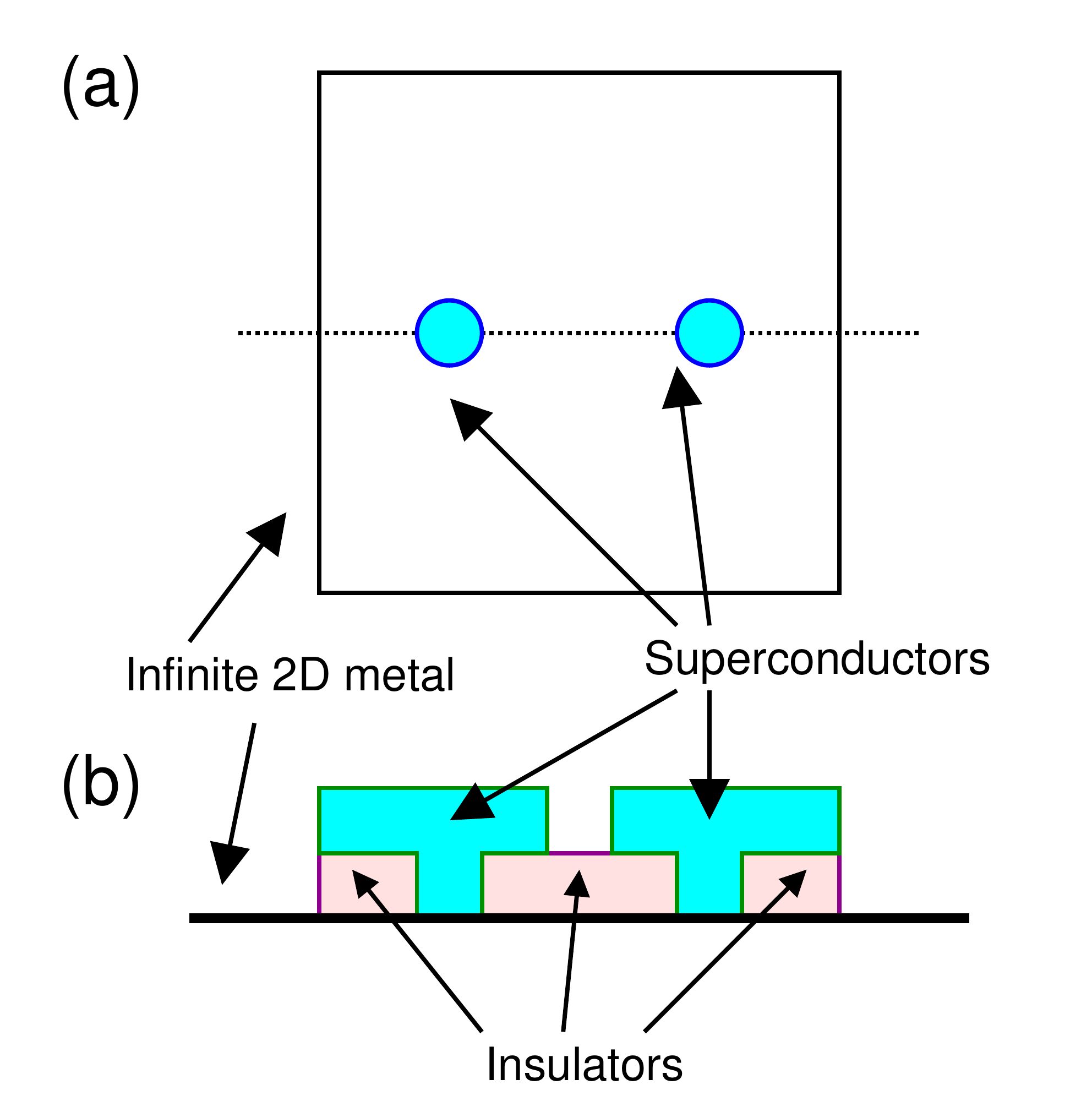}
  \caption{{{The infinite planar geometry.} The device
      consists of a ballistic infinite 2D normal metal connected to the two
      superconductors $S_a$ and $S_b$ by small-area interfaces. Those
      normal-superconducting contacts could be realized by evaporating
      the superconducting lead on top of an insulator containing two
      holes. We show a top view of the device on panel a and a cut on
      panel b. The ballistic infinite 2D normal metal is biased at the voltage
      $V_B$. The tunneling density of states of this infinite planar
      configuration does not show resonances} {as a
      function of the energy} {in the absence of
      coupling to the superconducting leads. The corresponding
      calculations are presented in section~\ref{subsec:infinite}.}
\label{fig:infinite-planar}
}
\end{figure}

\section{Hamiltonians, geometry and method}
\label{sec:H-geom-method}

In this section, we present the technical background of the paper: the
Hamiltonians (see subsection~\ref{sec:H}) and the geometry (see
subsection~\ref{sec:geometry}). More details of the methods are
presented in Appendix~\ref{app:methods}, which is summarized in
{section~\ref{sec:methods}.}

\subsection{Hamiltonians}
\label{sec:H}

In this subsection, we present the Hamiltonians of the superconducting
and normal leads, {and the coupling between them. The
  superconducting} leads are described by the standard BCS Hamiltonian
with the gap $\Delta$, the bulk hopping amplitude $W$ and the
superconducting phase variable $\varphi_k$:
\begin{eqnarray}
  \label{eq:H-BCS1}
  \hat{\cal H}_{BCS}&=&-W \sum_{\langle i,j \rangle}
  \sum_{\sigma_z=\uparrow,\downarrow} \left(c_{i,\sigma_z}^+
  c_{j,\sigma_z}+ c_{j,\sigma_z}^+ c_{i,\sigma_z}\right)\\&-& \Delta
  \sum_k \left(\exp\left(i\varphi_k\right) c_{k,\uparrow}^+
  c_{k,\downarrow}^+ + \exp\left(-i\varphi_k\right)
  c_{k,\downarrow} c_{k,\uparrow}\right) ,
  \label{eq:H-BCS2}
\end{eqnarray}
where spin-$\sigma_z$ electrons have the opportunity to be transferred
with amplitude $W$ between the neighboring tight-binding
{sites} $\langle i,j \rangle$ on a lattice. In addition
to the kinetic energy {in Eq.~(\ref{eq:H-BCS1})}, the pairing term
with amplitude $\Delta$ {in Eq.~(\ref{eq:H-BCS2})} binds opposite-spin
electrons with the phase $\varphi_k$, taken as uniform within each
superconducting lead, i.e. $\varphi_k=\varphi_a$ or
$\varphi_k=\varphi_b$ in the superconducting lead $S_a$ or $S_b$
respectively.

The normal-metallic parts of the {Andreev interferometers} are
captured by a square-lattice tight-binding model with the bulk hopping
amplitude $W$:
\begin{eqnarray}
  \label{eq:H-N}
  \hat{\cal H}_{N}=-W \sum_{\langle i,j \rangle}
  \sum_{\sigma_z=\uparrow,\downarrow} \left(c_{i,\sigma_z}^+
  c_{j,\sigma_z}+ c_{j,\sigma_z}^+ c_{i,\sigma_z}\right),
\end{eqnarray}
where we use for simplicity in Eq.~(\ref{eq:H-N}) the same hopping
amplitude $W$ as in Eq.~(\ref{eq:H-BCS1}) for the kinetic term of the
BCS Hamiltonian. The summation in Eq.~(\ref{eq:H-N}) runs over
{all pairs} of {the} neighboring tight-binding sites
in the finite central normal conductor or in the infinite normal lead
$N_B$, depending on which part of the circuit is described.

{The experiments on two-dimensional (2D) metals that are discussed
  here \cite{PSE,Huang2022} are realized with graphene gated away from
  the Dirac point. The theoretical description is legitimately based
  on generic models of 2D metal Fermi surfaces, already that of a
  square lattice gated away from nesting at half-filling.  The effect
  of changing the gate voltage $V_g$ is via the following term in the
  Hamiltonian:
  \begin{equation}
    \label{eq:Wg}
  {\cal H}_g=-W_g \sum_{i,\sigma_z} c^+_{i,\sigma_z} c_{i,\sigma_z}
  ,
\end{equation}
where the corresponding energy $W_g$ is proportional to the gate
voltage $V_g$.}

The tunneling term couples the leads $L_1$ and $L_2$
according to
\begin{eqnarray}
  \label{eq:H-t}
  \hat{\cal H}_{L_1,L_2}=-t \sum_{\langle i,j \rangle}
  \sum_{\sigma_z=\uparrow,\downarrow} \left(c_{i,\sigma_z}^+
  c_{j,\sigma_z}+ c_{j,\sigma_z}^+ c_{i,\sigma_z}\right)
  ,
\end{eqnarray}
where $i$ and $j$ belong to the $L_1$ and $L_2$ sides of the
interfaces respectively.

\subsection{Andreev interferometer geometries}
\label{sec:geometry}

In this subsection, we present the different geometries that are
investigated {in the paper. The
  normal-superconducting} Andreev interferometer shown in
Fig.~\ref{fig:thedevice}a consists of a large superconducting loop
{that is edge-connected} to {the ballistic central-$N$} in the long-
or intermediate-junction limit, i.e. the central-$N$ linear dimension
{$L\agt\xi_{ball}$ exceeds the BCS coherence length 
  $\xi_{ball}=\frac{\hbar v_F}{\Delta}$, where $v_F$ is the Fermi
  velocity and $\Delta$ the superconducting gap.}

We denote by $S_a$ and $S_b$ the two left and right end-lines of the
superconducting loop connected to the central-$N$ at the $S_a$-$N$ and
$N$-$S_b$ interfaces respectively. As discussed in the Introduction, a
normal lead $N_B$ is connected at the bottom to the central-$N$,
forming the $N_B$-$N$ interface. The dissipative current $I_B$ {flows}
between $N_B$ and the grounded left and right $S_a$ and $S_b$, in
response to the voltage $V_B$ on $N_B$.

The choice of the gauge for the superconducting phase variables
$\varphi_a$ and $\varphi_b$ of $S_a$ and $S_b$ is the following:
\begin{eqnarray}
 \label{eq:varphia-varphib-1}
\varphi_a&\simeq&\varphi_0-\Phi/2\\
\varphi_b&\simeq&\varphi_0+\Phi/2
,
\label{eq:varphia-varphib-2}
\end{eqnarray}
with $\Phi$ the flux piercing through the superconducting loop. The
overall phase variable $\varphi_0$ remains free, and all physical
observables are independent of $\varphi_0$ because of gauge
invariance. In the above
Eqs.~(\ref{eq:varphia-varphib-1})-(\ref{eq:varphia-varphib-2}), we
assumed that the radius of the {\it external} superconducting loop is
much larger than the linear dimension $L$ of the
central-$N$. Considering a closed contour encircling the
superconducting loop, we neglect the line integral of the vector
potential in the region sustained by the central-$N$.

{We distinguish between the following two situations:}

{(i) {\it The central-$N$ has a smooth tunneling density of states
    plotted as a function of energy}, which is captured by an infinite
  2D normal metal, see Fig.~\ref{fig:infinite-planar}. Lowest-order
  Keldysh} perturbation theory in the tunneling amplitudes with the
superconductors {then produces the nonstandard microscopic process of
  the phase-AR which couples the differential conductance to the
  superconducting phase variables and to the infinite 2D normal metal
  spectral density of states and populations, see
  section~\ref{sec:lowest-order}. However,} Fabry-P\'erot oscillations
between the superconducting contacts are possible at higher-order in
tunneling with the superconductors, which is compatible with the
harmonic oscillations that we here find at the lowest order in
tunneling.

(ii) {{\it The finite-size central-$N$ is connected to the infinite
    normal metal $N_B$ and to the infinite superconductors $S_a$ and
    $S_b$.} Then, the tunneling density of states of $N$ coupled to
  $N_B$ can reveal sharp resonances emerging on top of a smaller
  smooth background. This is why we consider in
  section~\ref{sec:intrinsic1} a theory that holds for general values
  of the interface transparencies and for all possible dimensions of
  the central-$N$ in comparison with the Fermi wave-length}
{$\lambda_F$} {or the BCS coherence length} {$\xi_{ball}$.}

{The notation $N_{eff}$ stands for gathering $N$ and
  $N_B$ into a single metal that is nonhomogeneous in space, see
  Figs.~\ref{fig:thedevice}b and~\ref{fig:thedevice}c. According to
  the dimensions and contact transparencies, the spectrum of $N_{eff}$
  falls within the above-mentioned items (i) or (ii), regarding the
  absence or the presence of well-defined resonances in the corresponding
  spectral or tunneling density of states.}

Further {simplifications are carried out} about the geometry, {such as
  {reducing} the multichannel interfaces {shown on
    Fig.~\ref{fig:thedevice}b to} the single-channel ones {on}
  Fig.~\ref{fig:thedevice}c.  In order to capture the qualitative
  physics {of averaging out the oscillations at the scale of the Fermi
    wave-length} {$\lambda_F$, we will carry out} {an averaging of
    those pairs of contact points over the small linear dimension of
    both interfaces,} see for instance
  Refs.~\onlinecite{theory-CPBS12,theory-CPBS11,theory-CPBS13,Floser}
  in connection with the ferromagnet-superconductor-ferromagnet or}
{the} normal metal-superconductor-normal metal {Cooper
  pair} beam splitters.

\begin{figure}[htb]
  \includegraphics[width=.8\columnwidth]{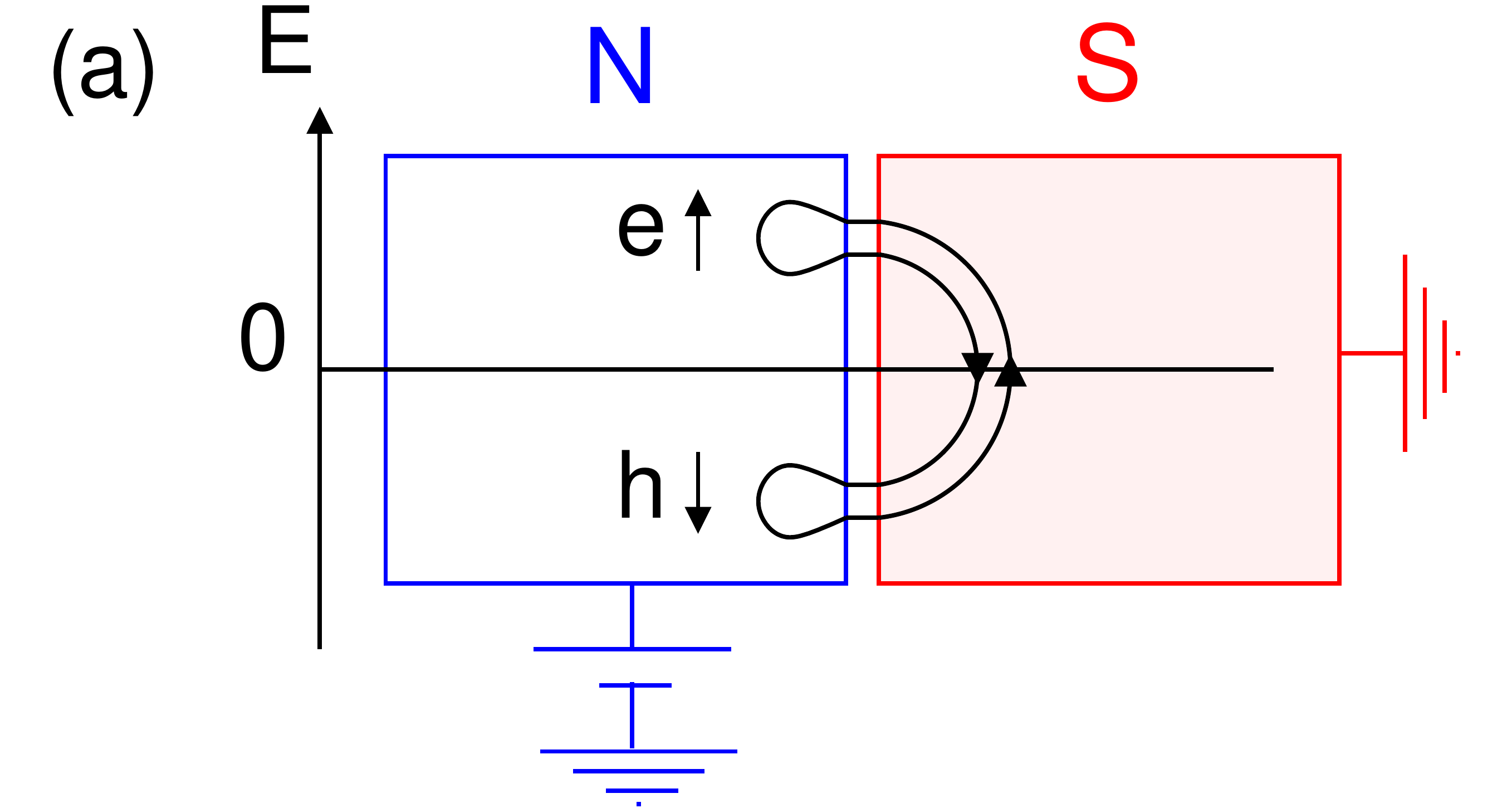}

  \includegraphics[width=.8\columnwidth]{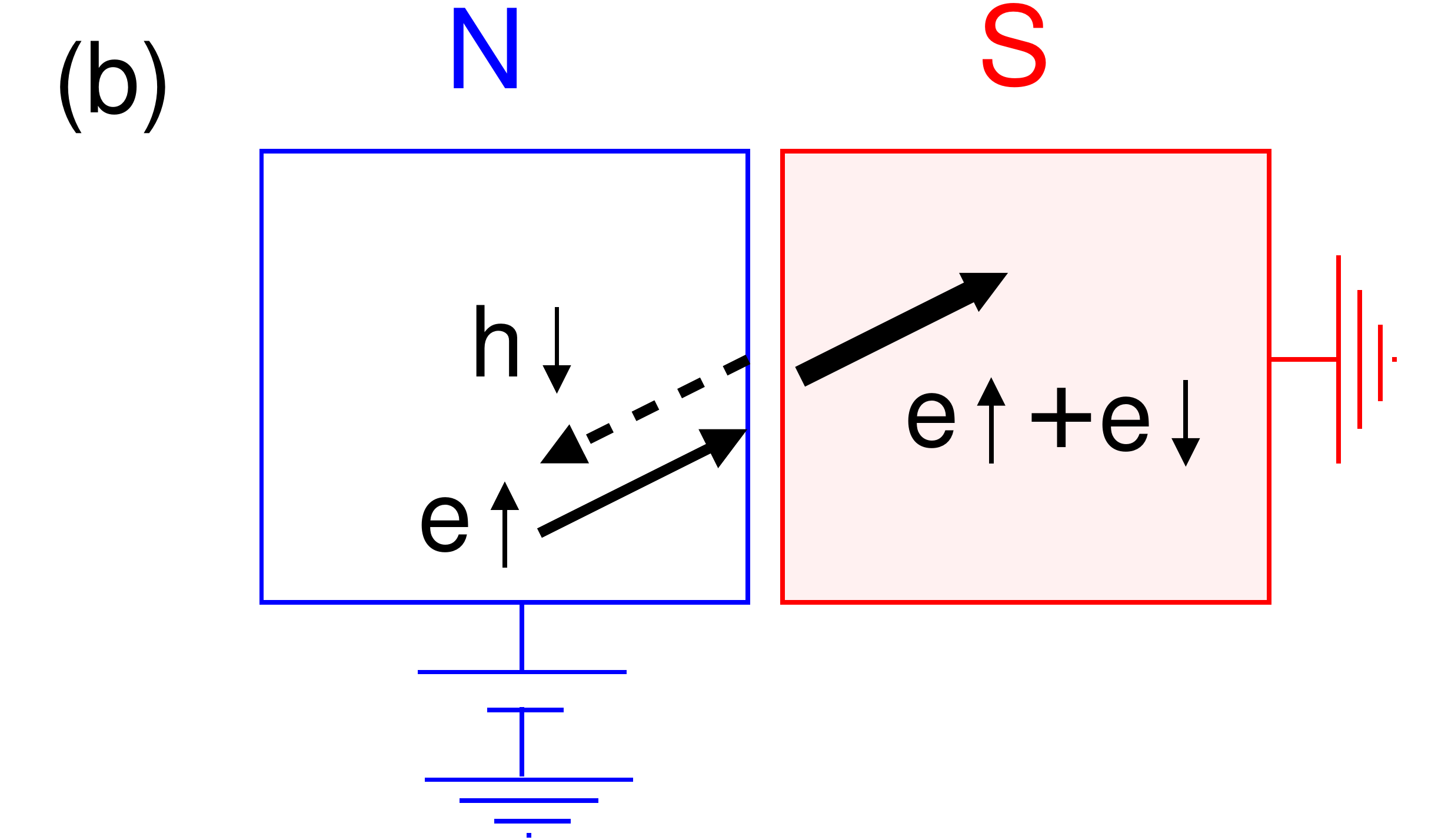}

  \caption{{{The standard process of the two-terminal Andreev
        reflection.} A spin-up electron is incoming from the $N$-side
      of the normal metal-superconductor $N$-$S$ interface, and a
      Cooper pair is transmitted into the condensate of $S$ while a
      spin-down hole is Andreev reflected in $N$. Panels a and b are
      in energy and in real space respectively.}
  \label{fig:AR}
}
\end{figure}

\begin{figure}[htb]

  \includegraphics[width=.5\textwidth]{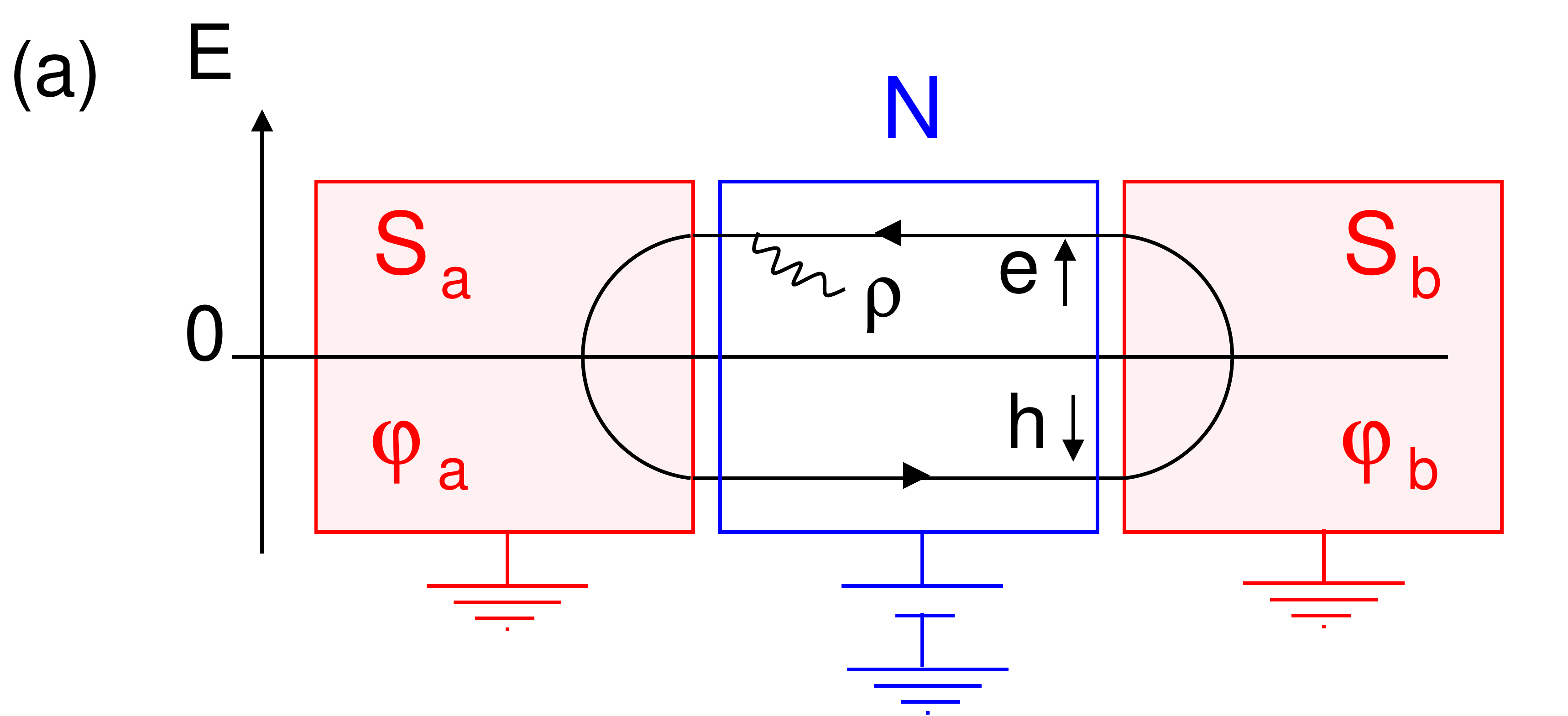}

  \includegraphics[width=.5\textwidth]{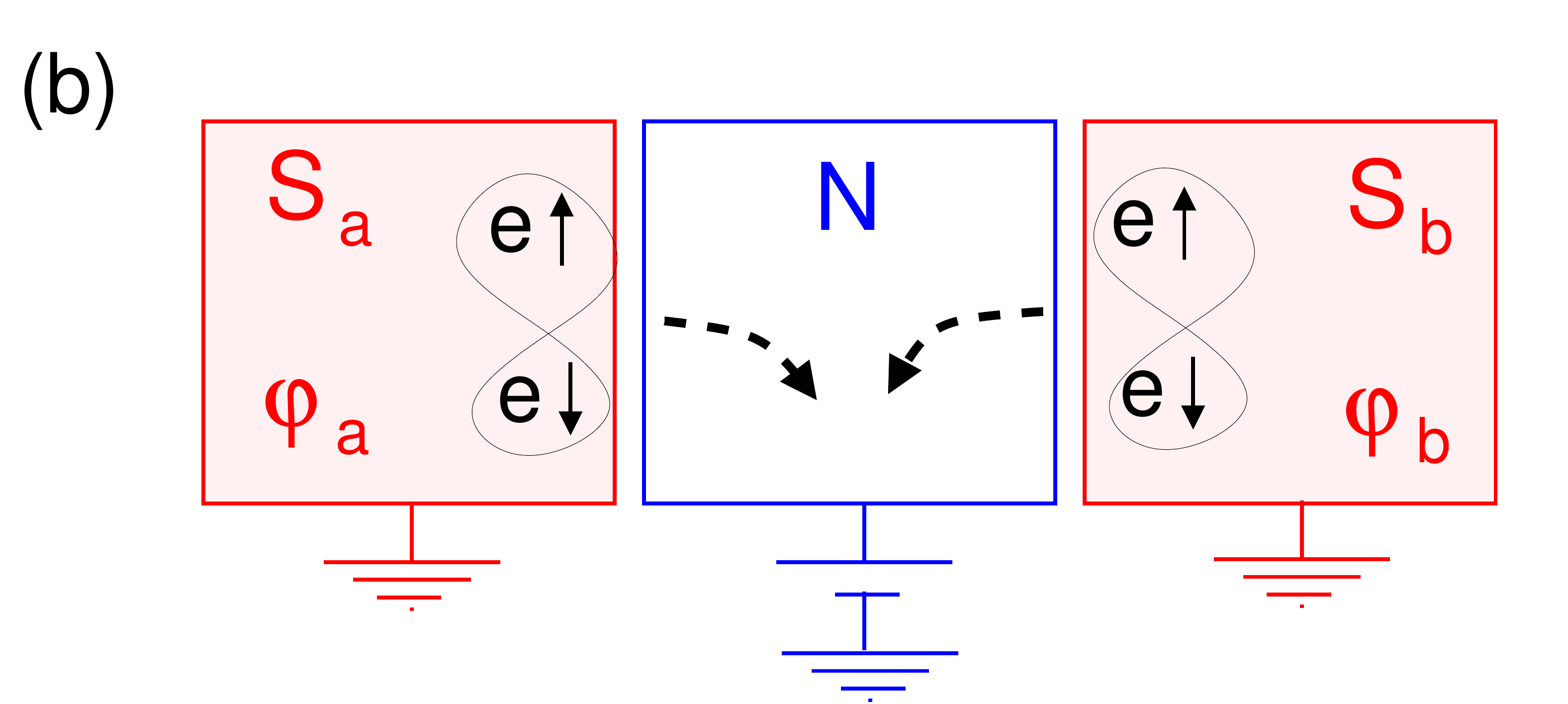}

  \caption{{{The nonstandard process of the three-terminal
        phase-Andreev reflection (phase-AR).} On panel a, the wavy
      line terminated by the symbol $\rho$ is a representation of the
      coupling to the spectral density} {of states} {and to the
      nonequilibrium Fermi surface in the central-$N$, via the
      nonlocal Keldysh Green's function. On panel b, we schematically
      represent the} {phase-AR quantum event} {in which a
      superposition of two Cooper pairs enters $N$ from each
      superconducting terminal.}
  \label{fig:phase-AR}
}
\end{figure}

\subsection{Methods}
\label{sec:methods}
All technical details of the methods are gathered in the
Appendix~\ref{app:methods}. Appendix~\ref{subsec:N} presents the form
of the advanced and retarded Green's
functions. Appendix~\ref{sec:large-gap} presents the large-gap
approximation suitable to capture a small bias voltage $V_B$ in the
{normal-superconducting Andreev interferometer} shown in
Fig.~\ref{fig:thedevice}a. Appendix~\ref{subsec:assumptions-popus}
presents the assumptions about the population and
Appendix~\ref{sec:thecurrent} explains how the current is calculated
from the {Keldysh Green's functions. Now, we present}
our results in the following.

{
\section{Perturbative transport theory and the phase-AR}
\label{sec:lowest-order}
}

{In this section, we present the first step of the theoretical
  framework. We assume the absence of resonances in the tunneling
  density of states with vanishingly small transmission into the
  superconducting leads,} and calculate the differential conductance
$dI_B/dV_B$ defined by Eq.~(\ref{eq:conductance}) as a function of the
bias voltage $V_B$ applied to the normal lead $N_B$, where $I_B$
{is the current flowing in $N_B$. The microscopic
  process of the phase-AR} is presented in
subsection~\ref{sec:phase-AR}. Subsection~\ref{subsec:general} deals
with a general model for $N_{eff}$, whether or not {its spectral
  density of states sustains resonances in the absence of coupling to
  the superconductors}, see Fig.~\ref{fig:thedevice}b,
Fig.~\ref{fig:thedevice}c, Fig.~\ref{fig:thedevice}f and
{Fig.~\ref{fig:infinite-planar}. We assume} in
subsection~\ref{subsec:infinite} an infinite 2D normal metal for the
central-$N$, {sustaining a tunneling density of states with a smooth
  energy dependence}, see Fig.~\ref{fig:infinite-planar}.

\subsection{Phase-Andreev reflection (phase-AR)}
\label{sec:phase-AR}

In this subsection, we physically introduce the process of the
phase-AR by which a normal current {in $N_B$} is converted into a
{phase-sensitive supercurrent} in the loop, {while coupling to the
  central-$N$ nonequilibrium populations}, see
Fig.~\ref{fig:thedevice}a for the
{normal-superconducting Andreev interferometer. Such
  conversion} {between a normal and supercurrent} is already
operational in the two-terminal Andreev reflection \cite{Andreev,BTK}
at a normal metal-insulator-superconductor ($NIS$) interface, see
Figs.~\ref{fig:AR}a and b. A spin-up electron incoming on the $N$ side
is Andreev reflected as a spin-down hole while a Cooper pair is
transmitted into the superconductor $S$ as evanescent quasiparticles
which eventually join the superconducting condensate
\cite{Andreev,BTK} as Cooper pair.

\begin{figure*}[htb]
  \includegraphics[width=.8\textwidth]{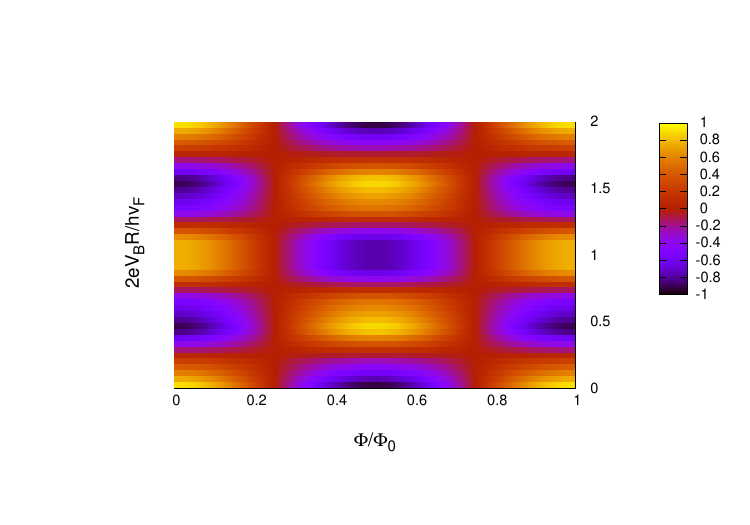}
  \caption{{The calculated checkerboard pattern of the
      normal-superconducting Andreev interferometer dimensionless
      differential conductance $g_0\left(\frac{\Phi}{\Phi_0},\frac{e
        V_B R}{\pi \hbar v_F}\right)\propto d I_B / dV_B$ plotted as a
      function of the reduced magnetic flux} {$\frac{\Phi}{\Phi_0}$
      (on the $x$-axis) and the reduced {effective Fermi energy}
      $y=\frac{2e V_B R}{h v_F}$ (on the $y$-axis).} We obtain
    alternations between {\it inversion} and {\it noninversion}, where
    $g_0$ is larger at the flux {$\frac{\Phi}{\Phi_0}=\frac{1}{2}$
      than at $\frac{\Phi}{\Phi_0}=0$} for the inversion, or the
    reverse for {the noninversion}.
    \label{fig:colorplot}
    }
\end{figure*}

In the three-terminal devices, we microscopically consider that a
spin-up electron is Andreev reflected at one interface and the
resulting spin-down hole is coherently transferred through ABS to the
other interface where the spin-down hole is backscattered into a
spin-up electron, see Figs.~\ref{fig:phase-AR}a and b. We propose to
use the wording {\it phase-AR} for the process shown in
Figs.~\ref{fig:phase-AR}a and b in order to reflect the interplay
between phase-sensitivity and the coupled spin-up electron {to}
spin-down hole current conversions typical of AR. We argue that,
microscopically, this density-phase {and} {nonequilibrium}
{population-phase couplings produce} a transient coherent
superposition between Cooper pairs (CPs) at each of the $S_aIN_a$ and
$N_bIS_b$ interfaces, e.g. a superposition of the following type:
{$\left(|CP\,a\rangle \pm \exp\left(\frac{2i\pi\Phi}{\Phi_0}\right)
  |CP\,b \rangle\right)/\sqrt{2}$.}
{Fig.~\ref{fig:phase-AR}b shows the phase-AR, where a
  superposition of Cooper pairs from each superconducting terminal
  cooperatively transports charge into the normal lead $N_B$, however
  in a two-step process in the case of sequential
  tunneling. Noncoherent transport of Cooper pairs from each
  superconducting interface is via Andreev reflection, which produces
  current that is linear in the bias voltage but insensitive on the
  superconducting phase variable of each interface. The physical
  observable of the current cannot depend on a single
  nongauge-invariant superconducting phase variable. The Andreev
  processes involving a single superconducting-normal metal interface
  are thus independent on the magnetic flux $\Phi$ through the
  superconducting loop. Still, the central-$N$ Fermi surface also
  depends on the magnetic flux $\Phi$ at higher-order}
{in tunneling, and such $\Phi$-sensitivity necessarily
  involves microscopic processes coupling to both superconducting
  interfaces. Those noncoherent Andreev processes are expected to
  contribute for a constant to the conductance, that is independent on
  the flux and bias voltage. In the following, we will not include
  this trivial global shift in the conductance spectra such as the
  forthcoming Fig.~\ref{fig:colorplot}.}

Now, we discuss the parity of the differential conductance $d I_B/d
V_B$ upon changing the signs of the voltage $V_B$ and the phase
difference $\varphi_b-\varphi_a\simeq\Phi$. Let us first assume a
tunnel contact between $N_B$ and the central-$N$. Then, the
differential conductance $d I_B/d V_B$ is proportional to the
{spectral density of states in the central-$N$ connected to $S_a$ and
  $S_b$.  The latter} is even in the phase difference
$\varphi_b-\varphi_a\simeq\Phi$ and $dI_B/dV_B$ is also even in the
magnetic flux $\Phi\simeq\varphi_b-\varphi_a$. The observation that
$I_B$ changes sign if both $V_B$ and $\Phi \simeq \varphi_b-\varphi_a$
are simultaneously reversed implies that $I_B$ is odd in the voltage
$V_B$ and $d I_B/dV_B$ is even in $V_B$.

{In addition, let us first consider the supercurrent
  circulating around the loop, and necessarily being transmitted from
  $S_a$ to $S_b$ or from $S_b$ to $S_a$ across the
  central-$N$. Current conservation implies that both $S_a$-$N$ and
  $S_b$-$N$ interfaces transmit opposite supercurrents, which thus
  have to be antisymmetric if $S_a$ and $S_b$ are interchanged. The
  supercurrent flowing from $S_a$ to $S_b$ or from $S_b$ to $S_a$ then
  has to be odd in the difference between the superconducting phase
  variables $\varphi_a$ and $\varphi_b$ of $S_a$ and $S_b$ and, at low
  interface transparency, it takes the standard form of the
  $\sim\sin\left(\frac{2\pi\Phi}{\Phi_0}\right)$ current-phase
  relation of the two-terminal Josephson effect, with $\Phi\simeq
  \varphi_b-\varphi_a$. A similar symmetry argument reveals that the
  current flowing from $(S_a,S_b)$ to $N_B$ has to be even in
  $\varphi_b-\varphi_a$, thus behaving like $\sim
  \cos\left(\frac{2\pi\Phi}{\Phi_0}\right)$ at the lowest order in
  tunneling.}

\subsection{Expression of the differential conductance for
  tunnel contacts with the superconductors}
\label{subsec:general}

In this subsection, we obtain a simple expression for the differential
conductance $dI_B/dV_B$ in the limit {of small interface
  transparencies. The dissipative spectral current} transmitted into
$S_a$ [see Eq.~(\ref{eq:I1-model-A})] is expanded to the order-$t_a^2
t_b^2$:
\begin{eqnarray}
 \label{eq:ici2}
&& \hat{t}_{\alpha,a}\delta \hat{G}_{a,\alpha}^{+,-}(\omega)\simeq\\
 \nonumber
&&\hat{t}_{\alpha,a}\hat{g}^R_{a,a}
 \hat{t}_{a,\alpha} \hat{g}^R_{\alpha,\beta}(\omega)
 \hat{t}_{\beta,b} \hat{g}^R_{b,b} \hat{t}_{b,\beta} \delta
 \hat{g}^{+,-}_{\beta,\alpha}(\omega)\\
&+& \hat{t}_{\alpha,a}\hat{g}^R_{a,a}
 \hat{t}_{a,\alpha} \delta \hat{g}^{+,-}_{\alpha,\beta}(\omega)
 \hat{t}_{\beta,b} \hat{g}^A_{b,b}
 \hat{t}_{b,\beta}\hat{g}^A_{\beta,\alpha}(\omega)
 \nonumber
 ,
\end{eqnarray}
and
\begin{eqnarray}
&&  \hat{t}_{a,\alpha}\delta \hat{G}_{\alpha,a}^{+,-}(\omega)\simeq\\
  \nonumber
&&\hat{t}_{a,\alpha} \hat{g}^R_{\alpha,\beta}(\omega) \hat{t}_{\beta,b}
\hat{g}^R_{b,b} \hat{t}_{b,\beta} \delta
\hat{g}^{+,-}_{\beta,\alpha}(\omega) \hat{t}_{\alpha,a}
\hat{g}^A_{a,a} \\&+&\hat{t}_{a,\alpha} \delta
\hat{g}^{+,-}_{\alpha,\beta}(\omega) \hat{t}_{\beta,b} \hat{g}^A_{b,b}
\hat{t}_{b,\beta} \hat{g}^A_{\beta,\alpha}(\omega) \hat{t}_{\alpha,a}
\hat{g}^A_{a,a} .\nonumber
\end{eqnarray}
The $\{\mu_{N,q}\}$-sensitive component of the spectral current takes the
following form at the order-$t_a^2 t_b^2$:
{
\begin{eqnarray}
&& I_{a,\alpha}(\{\mu_{N,q}\},\{Z_q\},\omega)
  \\\nonumber&=&\mbox{Nambu-trace}\left\{\hat{\tau}_3 \left[
    \hat{t}_{\alpha,a}(\ref{eq:(1)}) -\hat{t}_{a,\alpha}
    (\ref{eq:(1)-prime})\right]\right\}\\\nonumber&\equiv&
  I_0(\{\mu_{N,q}\},\{Z_q\},\omega) +
  I_1\left(\{\mu_{N,q}\},\{Z_q\},\frac{\Phi}{\Phi_0},\omega\right) ,
\end{eqnarray}
where} $Z_q$ is the jump labeled by $q$ at the energy $\mu_{N,q}$ in
the nonequilibrium distribution function of the central-$N$, see
Appendix~\ref{subsec:assumptions-popus}.  We separate between the
{reduced magnetic flux-$\frac{\Phi}{\Phi_0}$-insensitive} term
$I_0(\{\mu_{N,q}\},\{Z_q\},\omega)$ of order-$t_a^4$ and the {reduced
  magnetic flux-$\frac{\Phi}{\Phi_0}$-sensitive
  $I_1\left(\{\mu_{N,q}\},\{Z_q\},\frac{\Phi}{\Phi_0},\omega\right)$}
coupling both interfaces,
i.e. {$I_1\left(\{\mu_{N,q}\},\{Z_q\},\frac{\Phi}{\Phi_0},\omega\right)$}
is of order-$t_a^2 t_b^2$: {
\begin{eqnarray}
 \label{eq:I1-Phi-1}
 && I_1\left(\{\mu_{N,q}\},\{Z_q\},\frac{\Phi}{\Phi_0},\omega\right)\\
 \nonumber
&& \simeq\mbox{Nambu-trace}\left\{\hat{\tau}_3 \left[
  \hat{t}_{\alpha,a}\hat{g}^R_{a,a}
  \hat{t}_{a,\alpha} \hat{g}^R_{\alpha,\beta}(\omega)\times\right.\right.\\
  \nonumber
  &&\left.\left.
 \hat{t}_{\beta,b} \hat{g}^R_{b,b} \hat{t}_{b,\beta} \delta
 \hat{g}^{+,-}_{\beta,\alpha}(\omega)\right.\right.\\
 \nonumber
&+& \left.\left.\hat{t}_{\alpha,a}\hat{g}^R_{a,a}
 \hat{t}_{a,\alpha} \delta \hat{g}^{+,-}_{\alpha,\beta}(\omega)
 \hat{t}_{\beta,b} \hat{g}^A_{b,b}
 \hat{t}_{b,\beta}\hat{g}^A_{\beta,\alpha}(\omega)\right.\right.\\
 \nonumber
 &&-
 \hat{t}_{a,\alpha}
\hat{g}^R_{\alpha,\beta}(\omega) \hat{t}_{\beta,b} \hat{g}^R_{b,b} \hat{t}_{b,\beta} \delta \hat{g}^{+,-}_{\beta,\alpha}(\omega)
\hat{t}_{\alpha,a} \hat{g}^A_{a,a}\\
\nonumber
&-&\left.\left.\hat{t}_{a,\alpha}
\delta \hat{g}^{+,-}_{\alpha,\beta}(\omega) \hat{t}_{\beta,b}
\hat{g}^A_{b,b} \hat{t}_{b,\beta} \hat{g}^A_{\beta,\alpha}(\omega)
\hat{t}_{\alpha,a} \hat{g}^A_{a,a}\right]\right\}
 .
\end{eqnarray}}
We {now} make the Nambu labels explicit in the above
Eq.~(\ref{eq:I1-Phi-1}), and consider voltage or energy scales that
are small compared to the superconducting gap, see
Appendix~\ref{sec:large-gap}. Within this large-gap approximation, the
spectral current {$\tilde{I}_1\left(\{\mu_{N,q}\}, \{Z_q\},
  \frac{\Phi}{\Phi_0}, \omega\right)$} takes the following form at the
order-$t_a^2t_b^2$: {
\begin{eqnarray}
 \label{eq:I1-A1}
 && \tilde{I}_1\left(\{\mu_{N,q}\},\{Z_q\},\frac{
   \Phi}{\Phi_0},\omega\right)\simeq\\ \nonumber && 2 \frac{t_a^2
   t_b^2}{W^2}e^{\frac{2i\pi\Phi}{\Phi_0}}
 \left[\hat{g}^{R,2,2}_{\alpha,\beta}(\omega)\delta\hat{g}^{+,-,1,1}_{\beta,\alpha}(\omega)\right.\\
   &&\left.\nonumber
   + \delta
   \hat{g}^{+,-,2,2}_{\alpha,\beta}(\omega)\hat{g}_{\beta,\alpha}^{A,1,1}(\omega)
   \right]\\ \nonumber &-& 2 \frac{t_a^2
   t_b^2}{W^2}e^{-\frac{2i\pi\Phi}{\Phi_0}}
 \left[\hat{g}^{R,1,1}_{\alpha,\beta}(\omega)\delta\hat{g}^{+,-,2,2}_{\beta,\alpha}(\omega)\right.\\
   \nonumber
&&   + \left.\delta
   \hat{g}^{+,-,1,1}_{\alpha,\beta}(\omega)\hat{g}_{\beta,\alpha}^{A,2,2}(\omega)
   \right],
\end{eqnarray}}
where the superscripts ``1'' and ``2'' refer to the {\it spin-up
  electron} and to the {\it spin-down hole} Nambu components
respectively.  We further assume that the interfaces have transverse
dimension that is much smaller than the distance between them. This
{simplifying} assumption allows parameterizing the geometry with the
single variable $R\equiv R_{\alpha,\beta}$ for the separation between
the {contacts.}  A generalization to extended interfaces is possible,
see the discussion concluding the above
subsection~\ref{sec:geometry}. For instance, one option is to sum over
all the pairs of the tight-binding sites at both interfaces in the
tunnel limit. At higher transparency, the Dyson matrices can
{numerically} be inverted, with entries in the set of
the tight-binding sites making the contacts, see for instance
Ref.~\onlinecite{theory-CPBS12}.

\subsection{Infinite planar geometry}
\label{subsec:infinite}

In this subsection, we calculate the differential conductance $d
I_B/dV_B$ of the phase-AR within lowest-order perturbation theory in
the tunneling amplitudes, and in the infinite planar geometry
{of Fig.~\ref{fig:infinite-planar}. We also assume
  that} {a two-step distribution function \cite{Saclay} appears in the
  central-$N$, due to the interplay between the distribution functions
  of the grounded superconducting loop, and those of the infinite
  normal metal $N_B$ biased at the voltage $V_B$.} {It turns out that
  the step at zero energy associated with the grounded loop is not
  sensitive on the voltage $V_B$ and everything is as if $dI_B/dV_B$
  would solely be controlled by the other step at the energy $V_B$.
  This is why our} {calculations are also valid if the distribution
  function is characterized by the single step obtained for strong
  electron-phonon coupling in the central-$N$.}

{{\it Further physical comments about the central-$N$ distribution
    functions: }}{A diffusive quantum wire was connected in
  Ref.~\onlinecite{Saclay} to the left and right normal leads $N_a$
  and $N_b$ biased at the voltages $\pm V/2$}{,
  respectively}{. Several types of distribution functions can be
  probed by the superconducting tunneling spectroscopy of such}
{nonequilibrium} {diffusive quantum wire \cite{Saclay}. For instance,
  a double-step distribution function is theoretically obtained in the
  absence of electron-phonon coupling or electron-electron
  interaction. The nonequilibrium voltage-$V$ biasing condition on the
  diffusive wire then produces a Fermi surface that consists of the
  two pieces having the local Fermi energies $\pm \delta \mu_N$ and
  the steps $Z=1/2$ in their distribution functions. Conversely, the
  energy relaxation produced by a possibly strong electron-phonon
  coupling in the quantum wire produces a distribution function
  characterized by the single-step nonequilibrium local effective
  Fermi energy $\delta \mu_N$} {and by the step $Z=1$ in the
  corresponding distribution function at this energy.}  {Finally,
  strong electron-electron interactions produce rounded distribution
  functions characterized by the nonequilibrium effective electronic
  temperate~$T_{eff}$} {and by the single effective local Fermi energy
  $\delta \mu_N$.}

{Now, we come back to our Andreev
  interferometer. Under the condition of a single effective
  $V_B$-sensitive step,} we deduce from Eq.~(\ref{eq:I1-A1}) the
following expression for the spectral differential conductance
$\partial\tilde{I}_1/\partial\mu_{N,p}(\omega)$ at the order-$t_a^2
t_b^2$: {
\begin{eqnarray}
 \label{eq:spectral-current-2D-debut}
&&\frac{\partial
   \tilde{I}_1\left(\{\mu_{N,q}\},\{Z_q\},\frac{\Phi}{\Phi_0},\omega\right)}{\partial\mu_{N,p}}
 \simeq\\&&\nonumber 8Z_p \frac{t_a^2 t_b^2}{W^4\sqrt{k_h R}\sqrt{k_e
     R}}\cos\left(\frac{2\pi\Phi}{\Phi_0}\right)
 \times\\ &&\cos\left(k_h R-\frac{\pi}{4}\right) \cos\left(k_e
 R-\frac{\pi}{4}\right)\nonumber\\&&\times
 \left[\delta(\omega-\mu_{N,p})+\delta(\omega+\mu_{N,p})\right]
 .\nonumber
\end{eqnarray}
}

Approximating the slowly varying $\sqrt{k_h R}\sqrt{k_e R}$ as $k_F R$
in the prefactor and integrating over the energy $\omega$ leads to the
following leading-order term in the expansion of $\partial
\tilde{I}_1/\partial\mu_{N,p}(\omega)$ as a function of the hopping amplitudes
$t_a$ and $t_b$:
{
\begin{eqnarray}
  &&\frac{\partial\tilde{I}_1\left(\{\mu_{N,q}\},\{Z_q\},\frac{\Phi}{\Phi_0}\right)}{\partial\mu_{N,p}}=\\ \nonumber
  && \int d\omega
  \frac{\partial\tilde{I}_1\left(\{\mu_{N,q}\},\{Z_q\},\frac{
      \Phi}{\Phi_0},\omega\right)}{\partial\mu_{N,p}}\\ \nonumber &\simeq&16
  Z_p\frac{t_a^2 t_b^2}{W^4(k_F R)}\cos\left(\frac{2\pi\Phi}{\Phi_0}\right)\\&&\times\nonumber
  \cos\left[\left(k_F-\frac{\delta \mu_{N,p}}{\hbar v_F}\right)
    R-\frac{\pi}{4}\right] \times\\ &&
  \cos\left[\left(k_F+\frac{\delta\mu_{N,p}}{\hbar v_F}\right)
    R-\frac{\pi}{4}\right].  \nonumber
\end{eqnarray}}

We note that $\delta \mu_{N,p} R / \hbar v_F$ and $k_F R$ encode the
slow and fast oscillations respectively. In a standard procedure
\cite{theory-CPBS11,theory-CPBS12,theory-CPBS13,Floser}, we average out
those $(k_F R)$-oscillations in order to mimic multichannel interfaces
according to $\langle\langle \cos(x+A)\cos(x-A)\rangle\rangle= (\cos
2A)/2$, namely,
\begin{eqnarray}
 \label{eq:DD1}
&&\langle\langle \cos(x+A)\cos(x-A)\rangle\rangle
 =\\\nonumber&&\langle\langle
 \left(\cos x \cos A - \sin x \sin A\right)
 \left(\cos x \cos A + \sin x \sin A\right)\rangle\rangle=\\\nonumber
 && \cos^2 A \langle\langle\cos^2 x\rangle\rangle - \sin^2 A \langle\langle\sin^2 x\rangle\rangle
=\frac{1}{2} \cos(2A)
,
\end{eqnarray}
where the slow and fast variables are denoted by $A=\frac{\delta
  \mu_{N,p} R_{1,2}}{\hbar v_F}$ and $x=k_F R_{1,2}-\frac{\pi}{4}$
respectively. We obtain the following expression for the differential
conductance $\partial I_1/\partial\mu_{N,p}$ at the
order-$t_a^2t_b^2$: {
\begin{eqnarray}
 \label{eq:result}
&&\frac{\partial\langle\langle
 \tilde{I}_1\left(\{\delta\mu_{N,q}\},\frac{\Phi}{\Phi_0}\right)\rangle\rangle}
   {\partial(\delta\mu_{N,p})} =\\\nonumber&& \langle\langle \int d\omega
   \frac{\partial\tilde{I}_1\left(\{\mu_{N,q}\},\{Z_q\},\frac{\Phi}{\Phi_0},\omega\right)}{\partial\mu_{N,p}}
   \rangle\rangle\simeq \\&&8Z_p\frac{t_a^2 t_b^2}{W^4(k_F R)}\cos\left(\frac{2 \pi\Phi}{\Phi_0}\right)
   \cos\left(\frac{2\delta \mu_{N,p} R}{\hbar v_F}\right) .
   \nonumber
\end{eqnarray}}
In order to illustrate this Eq.~(\ref{eq:result}), we consider a
simple $(S_a,\,S_b,\,N_B)$ {normal-superconducting Andreev
  interferometer} where $(S_a,\,S_b)$ are grounded and $N_B$ is biased
at the voltage $V_B$. {As it is mentioned above, we
  are left with a single effective $V_B$-sensitive step} of height $Z$
at the step-energy {$\delta\mu_N=eV_B$} in the distribution function
of the central-$N$. Eq.~(\ref{eq:result}) then simplifies as follows:
{
\begin{eqnarray}
 \label{eq:result2}
\frac{\partial\langle\langle\tilde{I}_1\left(V_B,\frac{\Phi}{\Phi_0}\right)\rangle\rangle}{\partial V_B}
\simeq8Z\frac{t_a^2 t_b^2}{W^4(k_F
 R)}g_0\left(\frac{\Phi}{\Phi_0},\frac{e V_B R}{\pi \hbar v_F}\right)
,
\end{eqnarray}
where} the dimensionless differential conductance
$g_0\left(\frac{\Phi}{\Phi_0},\frac{e V_B R}{\pi \hbar v_F}\right)$ is
the following: {
\begin{eqnarray}
    g_0\left(\frac{\Phi}{\Phi_0},\frac{e V_B R}{\pi \hbar
      v_F}\right)=\cos\left(\frac{2\pi \Phi}{\Phi_0}\right)
    \cos\left(\frac{2eV_B R}{\hbar v_F}\right)
    \label{eq:EE2}
    .
\end{eqnarray}}
{This Eq.~(\ref{eq:EE2}) for $g_0\left(\frac{\Phi}{\Phi_0},\frac{e V_B
    R}{\pi \hbar v_F}\right)$ is one of the main results of the paper,
  and it concludes the calculations of the first step of the
  theoretical framework.  Eq.~(\ref{eq:EE2}) for $g_0$ will be
  complemented in the forthcoming
  subsection~\ref{sec:pheno-full-ladders} with the physically-motivated
  approximation of Eq.~(\ref{eq:gt0}) for $g'''_0$, which accounts for
  the discreteness of the spectrum.}

Fig.~\ref{fig:colorplot} shows the resulting checkerboard pattern {of}
the differential conductance $g_0\left(\frac{\Phi}{\Phi_0},\frac{e V_B
  R}{\pi \hbar v_F}\right)$ as a function of {the reduced magnetic
  flux $\frac{\Phi}{\Phi_0}$ (on the $x$-axis) and the reduced voltage
  $\frac{2 e V_B R}{h v_F}$ (on the $y$-axis).} In comparison with
Ref.~\onlinecite{Huang2022}, we also conclude {that a pattern of
  multiple inversion-noninversion coherent alternations emerges}. By
{\it noninversion}, we mean the conventional behavior that $g_0$ is
larger at $\frac{\Phi}{\Phi_0}=0$ than at
$\frac{\Phi}{\Phi_0}=\frac{1}{2}$ while {\it inversion} means that
$g_0$ is smaller at $\frac{\Phi}{\Phi_0}=0$ than at
$\frac{\Phi}{\Phi_0}=\frac{1}{2}$.

\begin{figure}[htb]
  \includegraphics[width=.8\columnwidth]{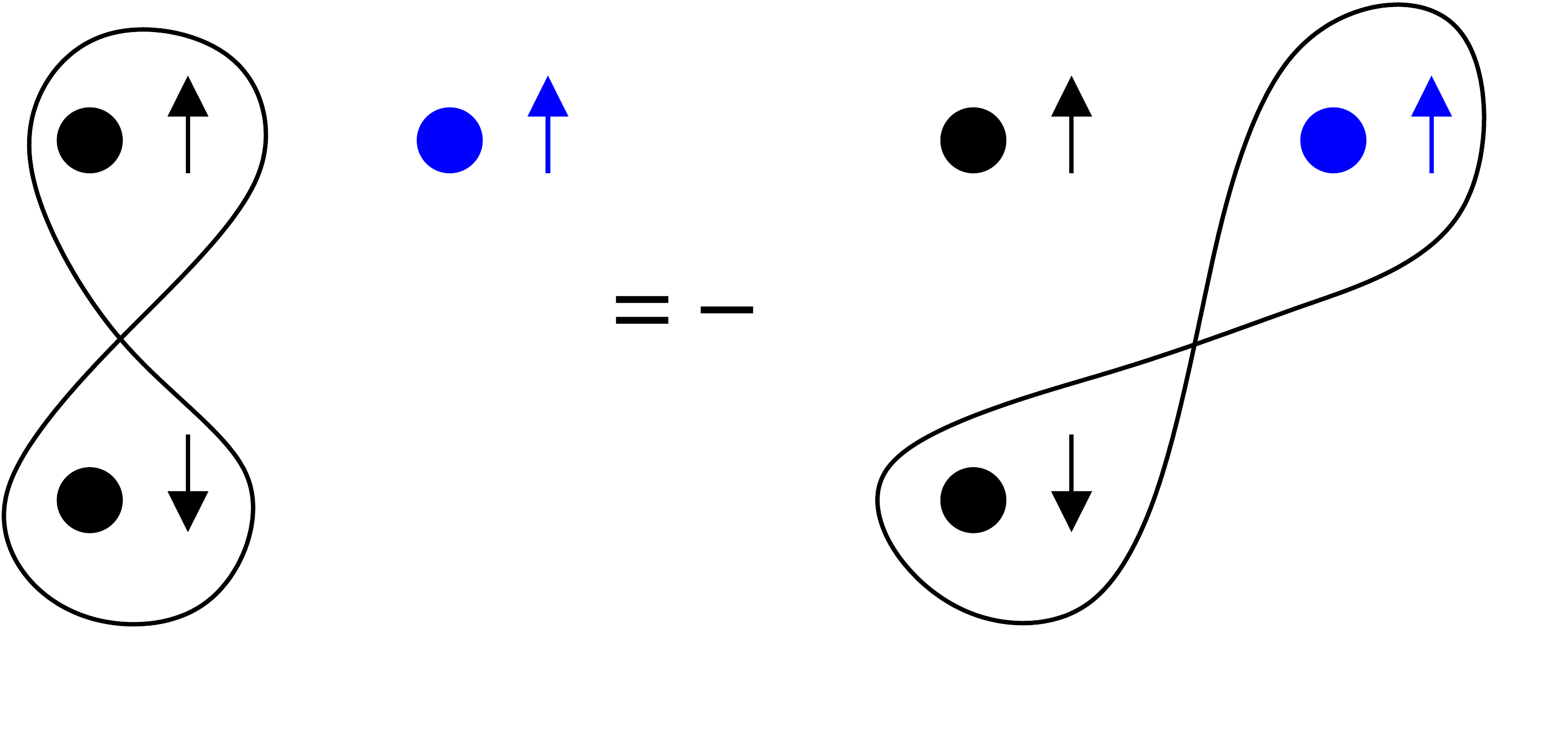}
  \caption{{The exchange between a pair and a single-particle
      electronic state.}  The anticommutation relations are given by
    Eqs.~(\ref{eq:exchange-1})-(\ref{eq:exchange-2}). The l.h.s. of
    the figure represents the paired $c_{x,\uparrow}^+
    c_{x,\downarrow}^+$ and the single-particle
    $c_{y,\uparrow}^+$. The r.h.s. represents the rearrangement as the
    single-particle $c_{x,\uparrow}^+$ and the paired
    $c_{y,\uparrow}^+c_{x,\downarrow}^+$. {Note the {\it minus} sign
      in front of the r.h.s., which reflects
      Eq.~(\ref{eq:exchange-2}).}
\label{fig:exchange}
}
\end{figure}

\subsection{Further physical remarks}
\label{sec:further-physical-remarks}
In this subsection, we present {several physical}
remarks as an interpretation of the above Eq.~(\ref{eq:EE2}).

{\it Exchange between the ABS and the normal single-particle
  background:} Let us consider an Andreev pair entering the
central-$N$, which consists of a spin-up electron and
{the} time-reversed spin-down hole. We denote this
quantum state by $\left(c_{x,\uparrow}^+ c_{x,\downarrow}^+\right)$,
where $c_{x,\sigma}^+$ creates a spin-$\sigma$ electron at the
location $x$. We also add another {unpaired} spin-up electron at the
location $y$, corresponding to $c_{y,\uparrow}^+$. Then, the fermionic
anticommutation relations lead to
\begin{eqnarray}
 \label{eq:exchange-1}
 \left(c_{x,\uparrow}^+
 c_{x,\downarrow}^+\right) c_{y,\uparrow}^+&=&
 c_{x,\uparrow}^+
 c_{x,\downarrow}^+c_{y,\uparrow}^+\\
 &=&-c_{x,\uparrow}^+ c_{y,\uparrow}^+c_{x,\downarrow}^+\\
 &=& -c_{x,\uparrow}^+ \left(c_{y,\uparrow}^+c_{x,\downarrow}^+\right)
 \label{eq:exchange-2}
 ,
\end{eqnarray}
where the pair $\left(c_{x,\uparrow}^+ c_{x,\downarrow}^+\right)$ in
Eq.~(\ref{eq:exchange-1}) exchanges its spin-up component with the
single-particle electronic state $c_{y,\uparrow}^+$, leading to
Eq.~(\ref{eq:exchange-2}). In the final state, a spin-up electron
$c_{x,\uparrow}^+$ is {separated, and the pair $\left(
  c_{y,\uparrow}^+ c_{x,\downarrow}^+ \right)$ is formed}.  {We
  conclude that a {\it minus sign} is produced if a pair is exchanged
  with a single electron, see Fig.~\ref{fig:exchange}. The coupling to
  the {electronic density and populations} via the
  Keldysh Green's function is nonequilibrium, which explains the
  noninversion at $V_B=0$ on Fig.~\ref{fig:colorplot} and the
  threshold voltage to produce the inversion.}

{{\it Comment on $g_0\left(\frac{\Phi}{\Phi_0},\frac{e V_B R}{\pi
      \hbar v_F}\right)=0$ for all {$\frac{\Phi}{\Phi_0}$} at the
    periodic values of $V_B$ on Fig.~\ref{fig:colorplot}:} We note
  that, at low energy compared to the superconducting gap $\Delta$ and
  for a long-junction, the corresponding spectrum
  \cite{Ishii,Kulik,Bagwell} features symmetry lines in the (phase,
  energy) parameter plane, which are parallel to the $x$-axis and
  located in between each pair of the consecutive energy levels. Then,
  the influence of the energy levels located} {\it at energies above a
  given inter-level symmetry line} {cancels with that of the levels
  located below.}  {It is then natural that, in our calculations, the
  differential conductance $g_0\left(\frac{\Phi}{\Phi_0},\frac{e V_B
    R}{\pi \hbar v_F}\right)=0$ is vanishingly small on the symmetry
  lines. However, we work in the limit of low energy compared the
  superconducting gap $\Delta$, which is justified by the voltage
  scales $|eV_B|\alt\Delta/10$ that are used both in the Penn State
  \cite{PSE} and in the Harvard \cite{Huang2022} group experiments.}
{Restoring a finite superconducting gap and making the calculations at
  a finite energy is expected to break} {the symmetry of the
  checkerboard pattern in Fig.~\ref{fig:colorplot}. Those remarks
  suggest that the oscillatory checkerboard pattern of the conductance
  $d I_B/dV_B$ in Fig.~\ref{fig:colorplot} {is related
    to the spectrum} of a long Josephson junction
  \cite{Ishii,Bagwell,Kulik}. Direct evidence for the connection
  between the differential conductance spectra and the spectra of ABS
  in the long-junction limit \cite{Ishii,Kulik,Bagwell} is provided in
  the forthcoming
  section~\ref{sec:pheno-full-ladders}. {Increasing
    the line-width broadening allows crossing-over from the
    forthcoming Fig.~\ref{fig:colorplot}a (revealing the discrete-like
    sharp resonances with small line-width) to the forthcoming
    Fig.~\ref{fig:colorplot}d (revealing at large line-width
    broadening the same oscillatory checkerboard pattern as the above
    Fig.~\ref{fig:gt0}). This issue is further discussed in the
    forthcoming subsections~\ref{sec:periodic-compression}
    and~\ref{sec:final}, and we will conclude that the conductance
    maps generally follow the ABS spectra throughout this cross-over.}

{{\it Voltage and magnetic flux dependence:}} The
voltage-$V_B$-sensitive alternations can be viewed as a simple
consequence of the oscillatory $\exp(\frac{2iE\tau}{\hbar})$, where
the bias voltage energy is $E = e V_B$, with $\tau=\frac{R}{v_F}$ the
delay for ballistic propagation between the $S_a$-$N_\alpha$ and the
$N_\beta$-$S_b$ interfaces. {In addition, the} phase-AR current
transmitted from $N_B$ towards $(S_a,\,S_b)$ is proportional to {$\sim
  \cos\left(\frac{2\pi\Phi}{\Phi_0}\right)$}, see the above
Eq.~(\ref{eq:EE2}), instead of {$\sim
  \sin\left(\frac{2\pi\Phi}{\Phi_0}\right)$} for the equilibrium
DC-Josephson current transmitted between $S_a$ and $S_b$, see also the
remarks at the end of the above
subsection~\ref{sec:phase-AR}. Eq.~(\ref{eq:EE2}) also reveals the
{undamped $\cos\left(\frac{2ER}{\hbar v_F}\right)$ alternations
  typical of the ballistic limit.}

{{\it Variations in the gate voltage $V_g$ are
    inequivalent to variations in the nonequilibrium Fermi surface:}
  Two options are generally possible: (i) Varying the gate voltage
  $V_g$, which produces a global shift $W_g$ of the entire dispersion
  relation [see the above Eq.~(\ref{eq:Wg})], leaving unchanged the
  chemical potentials of the leads. (ii) Varying the bias voltage
  $V_B$ of the lead $N_B$, which changes the energies of the steps in
  the infinite 2D normal metal distribution function. While we are
  here primarily focused on this second possibility, it is known that
  the energy level spacing of the spectrum of a long Josephson
  junction \cite{Kulik,Ishii,Bagwell} scales proportionally to the
  Fermi velocity} {$v_F$}{. This is
  why changing the gate voltage leaves the differential conductance
  $dI_B/dV_B$ unchanged as long as the effects of the band curvature
  are negligibly small, i.e.  as long as the Fermi velocity $v_F$ can
  be approximated as independent on the value of the gate voltage
  $V_g$. This feature of the model is in a qualitative agreement with
  the experimental data of the Penn State group \cite{PSE}.}

{{\it Comparison to the diffusive limit:} In the diffusive limit, the
  elastic mean free path is large compared to the Fermi wave-length
  {$\lambda_F$} but small compared to the
  {linear} dimension of the device.  It was
  established
  \cite{Pfeffer2014,Wilhelm,Nakano,Zaitsev,Kadigrobov,Volkov,Nazarov1,Nazarov2,Yip,Belzig,Zaikin1,Zaikin2,Zaikin3,Zaikin4}
  that, in the diffusive limit, the differential conductance shows
  damped oscillations as a function of the bias voltage $V_B$, which
  are controlled by the Thouless energy{, see also the
    experimental Refs.~\onlinecite{Klapwijk,Margineda}, the second one
    being recent.} This contrasts with the absence of damping in
  Eq.~(\ref{eq:EE2}) as a function of the bias voltage.}

{{\it Analogy with a normal-metallic loop: }In order to further
  understand and confirm the results of the Keldysh Green's function
  calculation, we identify the following ingredients in the model of
  the $(S_a,S_b,N_B)$ Andreev interferometer: (i) The coupling to the
  superconducting phase differences between the two superconducting
  leads, (ii) Constructive interference in real space between the
  {1D} Andreev tubes, (iii) The finite size of the
  central normal metal and (iv) The tunnel coupling to the
  leads. Similar ingredients are} {also there in the} {model of
  normal-metallic loop taken from Ref.~\onlinecite{Landauer-Buttiker},
  see also Fig.~\ref{fig:loop}a: (i) The coupling to the phase of the
  magnetic flux $\Phi$, (ii) {A 1D geometry,} (iii)
  The finite length of the normal loop, and (iv) The tunnel junction
  that interrupts the loop.}
\begin{figure}[htb]
  \begin{minipage}{\columnwidth}
    \begin{minipage}{.57\textwidth}
    \includegraphics[width=\textwidth]{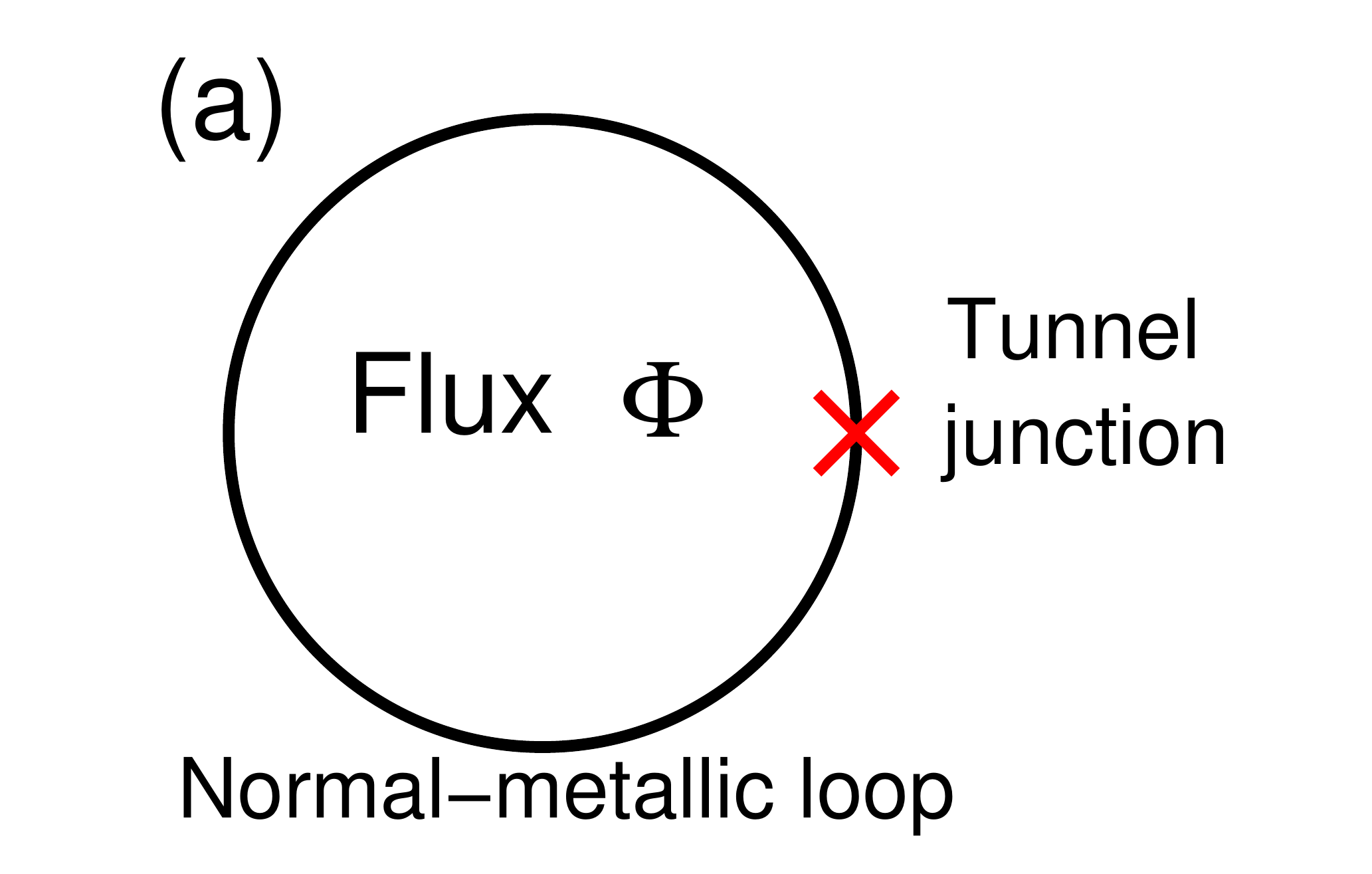}\end{minipage}\begin{minipage}{.42\textwidth}
      \includegraphics[width=\textwidth]{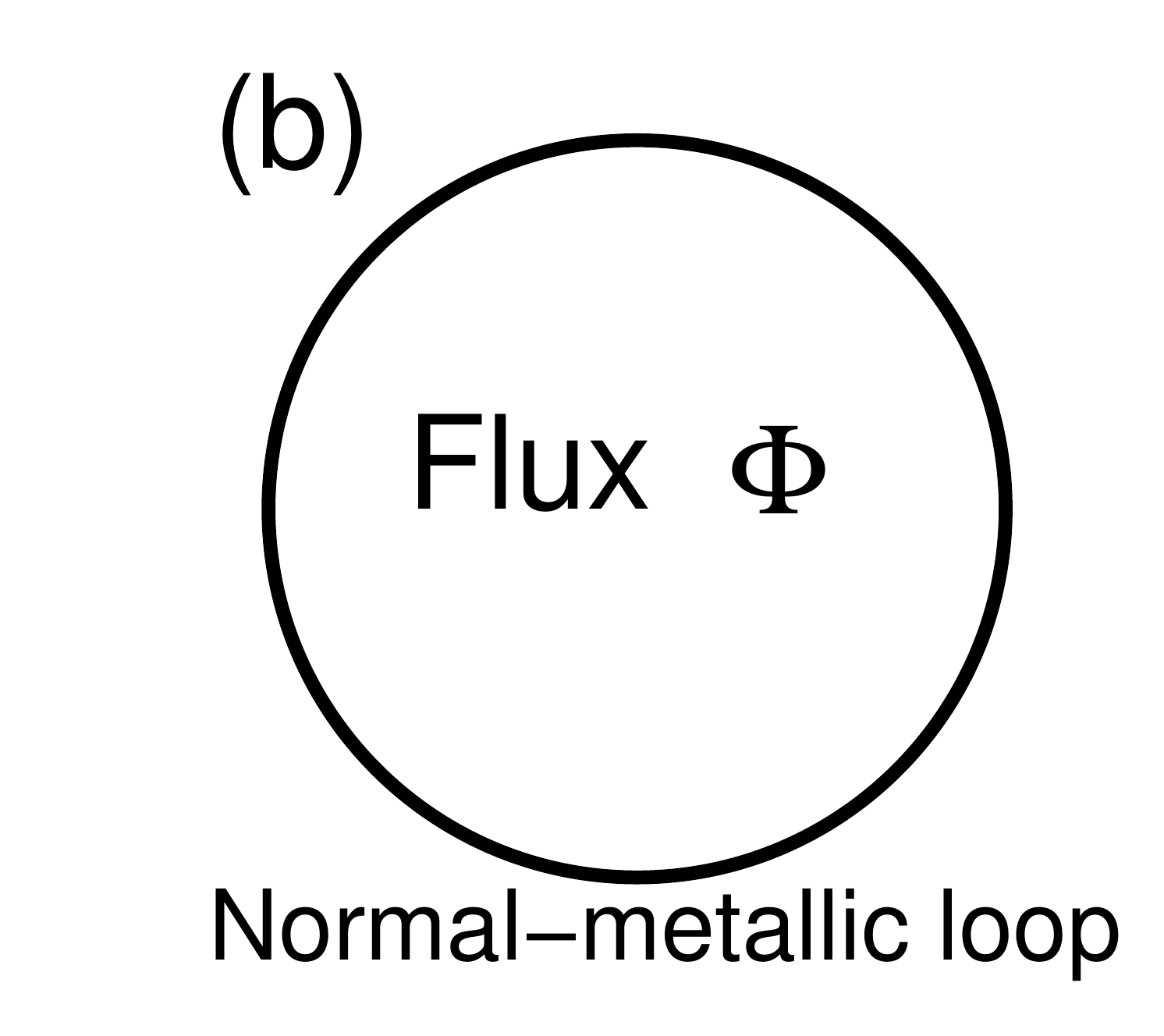}\end{minipage}
    \end{minipage}
  \caption{{The analogous normal-metallic loop,
      interrupted by a tunnel junction (a) or without a tunnel
      junction (b).}
\label{fig:loop}
}
\end{figure}
{The normal loop Hamiltonian is simply given by
\begin{eqnarray}
  {\cal H}_{loop}&=&-\sum_{\langle i,j\rangle} \left(T|i\rangle
  \langle j|+T^*|j\rangle \langle i| \right)\\&-& \left(T_0|i_0\rangle
  \langle j_0|+T_0^*|j_0\rangle \langle i_0| \right) ,
\end{eqnarray}
where the summation in the first term {runs} over all
of the pairs $\langle i,j \rangle$ of the neighboring tight-binding
sites coupled by the hopping amplitude $T$, at the exception of the
$\langle i_0,j_0\rangle$ pair having the much smaller hopping
amplitude $T_0$. The first-order correction to the energy is the
following: \begin{eqnarray} \delta E_n=- T_0 \psi_n^*(i_0) \psi_n(j_0)
  - T_0^* \psi_n^*(j_0) \psi_n(i_0),
  \label{eq:E-T0}
  \end{eqnarray}
where $\psi_n(i_0)$ and $\psi_n(j_0)$ are the complex-valued
wave-functions {$\psi_n(\theta)=\varphi_n(\theta) \exp\left(\frac{i
    \theta \Phi }{ \Phi_0} \right)$.} The notation $\varphi_n(\theta)$
denotes the real-valued wave-functions of the standing waves on the
open chain with $T_0=0$, and $\theta$ is the angle defining the point
$x$ on the circle. Then, the product $\varphi_n(i_0) \varphi_n(j_0)$
changes sign according to the alternation between the even and odd
number of nodes in the open-chain wave-function. This produces the
alternating positive and negative first-order corrections to the
energy, which are here considered as being analogous to spectrum of a
long Josephson junction \cite{Kulik,Ishii,Bagwell}. We thus obtain an
analogy to the oscillatory checkerboard pattern in
Fig.~\ref{fig:colorplot}. We also note that {Eq.~(\ref{eq:E-T0}) leads
  to
\begin{eqnarray}
    \delta E_n&=&- \left(U_0 + U_0^*\right)\varphi_n(i_0)
    \varphi_n(j_0)\\ &=& - 2 |T_0|
    \cos\left(\frac{2\pi\Phi}{\Phi_0}\right) \varphi_n(i_0)
    \varphi_n(j_0) ,
  \end{eqnarray}}
where $U_0=T_0\exp\left(\frac{2i\pi(1-1/n)\Phi}{\Phi_0}\right) = |T_0|
\exp\left( \frac{2 i \pi \Phi}{\Phi_0} \right)$, with $n$ the number
of the tight-binding sites making the loop. This implies that the
$\cos\left(\frac{2\pi\Phi}{\Phi_0}\right)$ term factors out, as it is
the case for $g_0\left(\frac{\Phi}{\Phi_0},\frac{e V_B R}{\pi \hbar
  v_F}\right)$ in our Keldysh Green's function calculation, see the
above Eq.~(\ref{eq:EE2}). We also note that the loop without the weak
link, see Fig.~\ref{fig:loop}b, is analogous to the fully developed
resonances discussed in the forthcoming section~\ref{sec:intrinsic1}.}

{{\it Expression of the current $I_B$: }Let us assume
  for simplicity a single step in the distribution function, being due
  to a strong electron-phonon coupling on the central-$N$. Then, we
  express that the current transmitted into $N$ is vanishingly small:
  \begin{eqnarray}
    \label{eq:A1}
  0=I_N&\equiv&(G_a+G_b) \delta \mu_N + G_B (\delta\mu_N - V_B)
  \\&&+G_{a,b} \delta\mu_N \cos\left(\frac{2\pi\Phi}{\Phi_0}\right) ,
  \nonumber
\end{eqnarray}
where} $G_a$ and $G_b$ are the Andreev reflection differential
conductances of each $S_a$-$N$ and $N$-$S_b$ interfaces, $G_B$ is the
differential conductance of the $N$-$N_B$ interface
{and $G_{a,b} \mu_N
  \cos\left(\frac{2\pi\Phi}{\Phi_0}\right)$ is the contribution of the
  phase-AR to the current entering the central-$N$.} We deduce the
expression of the normal metal Fermi energy {
  \begin{equation}
    \label{eq:A2}
  \delta \mu_N=\frac{G_B}{G_a+G_b+ G_B +G_{a,b} \cos\left(\frac{2\pi\Phi}{\Phi_0}\right)} V_B
  .
\end{equation}
The current transmitted into $N_B$ is the following:}
{
\begin{eqnarray}
  I_B&=&G_B(V_B-\delta\mu_N)\\ &=& G_B\frac{G_a+G_b+G_{a,b}
    \cos\left(\frac{2\pi\Phi}{\Phi_0}\right)}{G_a+G_b+ G_B +G_{a,b}
    \cos\left(\frac{2\pi\Phi}{\Phi_0}\right)} V_B
  \label{eq:I-simple}
  ,
\end{eqnarray}
and} {the $G_{a,b}$ contribution to Eq.~(\ref{eq:I-simple}) oscillates
  with the reduced flux $\frac{\Phi}{\Phi_0}$ and the voltage $V_B$ in
  a way that is similar to Fig.~\ref{fig:colorplot}.}

{
  \section{Nonperturbative transport theory}
\label{sec:intrinsic1}
}

{In this section, we establish the second step in our theoretical
  framework, namely, we nonperturbatively} {calculate
  the conductance $d I_B/d V_B$. A general device} can be modeled by a
finite tight-binding lattice for the central-$N$, connected by
transmission amplitudes to the grounded infinite superconducting leads
$S_a$ and $S_b$, and to the normal lead $N_B$ biased at the voltage
$V_B$, see section~\ref{sec:H-geom-method}. The overall physical
behavior depends on how the typical device linear dimension $L$
compares to the Fermi wave-length $\lambda_F$ and to the BCS coherence
length $\xi_{ball}$, such that $\lambda_F\ll\xi_{ball}$ under usual
conditions. This defines three physical domains of the typical device
dimension $L$: $L\alt \lambda_F$ (quantum dots), $\lambda_F \ll L \ll
\xi_{ball}$ (the short-junction limit) and $\xi_{ball}\ll L$ (the long
junction limit). In addition, the spectrum is discrete-like at small
$L$, and continuous-like at large $L$. The wording {\it discrete-like}
refers to well-defined resonances in the tunneling density of states
of the central-$N$, on top of much smaller continuous
background. Conversely, {\it continuous-like} has the meaning of weak
resonances in comparison with a large continuous background in the
energy-dependence of the tunneling density of states of the
central-$N$. Those different physical regimes of the device dimensions
can be handled with different approaches and methods. For instance,
the semi-classical {theory} \cite{Kulik} holds in the
long-junction limit $L\gg \xi_{ball}$ but {the semi-classical
  quantization is usually not developed} {to address the
  short-junction limit $\lambda_F\ll L \ll \xi_{ball}$ or the limit
  $L\alt \lambda_F$ of a quantum dot. By contrast, the calculations
  presented in this section are fully general and they apply to all
  regimes of the device dimensions, and independently on whether the
  spectrum is discrete-like or continuous-like.}

{One of the goals of this section is to simplify the
  expansion of the current given by
  Eqs.~(\ref{eq:DEBUT})-(\ref{eq:O2-prime-4}) in
  Appendix~\ref{app:thecurrent}. We will obtain the forthcoming
  Eqs.~(\ref{eq:Ia-1-1}) and~(\ref{eq:I(2)-debut}) for the conductance
  $d I_B/d V_B$ in the regime of sequential tunneling. Those
  expressions have the same status as Eq.~(18) in
  Ref.~\onlinecite{Cuevas}. Namely, we will infer from the forthcoming
  Eqs.~(\ref{eq:Ia-1-1}) and~(\ref{eq:I(2)-debut}) that the phase-AR
  couples the spectral density of states and the nonequilibrium
  populations to the pair amplitude in the superconducting lead
  $S_a$. This contrasts with standard Andreev reflection which
  involves the spin-up electron and spin-down hole spectral density of
  states in the initial and final states, see for instance
  Ref.~\onlinecite{Cuevas}.}

{This section is organized as follows. Subsection~\ref{sec:graph}
  presents {the} graph-theoretical arguments that justify the form of
  the Green's functions that are used for the central-$N$ coupled to
  the infinite superconducting and normal leads. The ideas of
  sequential tunneling and elastic cotunneling are presented in
  subsection~\ref{sec:elco}. The next
  subsection~\ref{sec:different-types} introduces the presentation of
  the nonperturbative calculation of the resonances. The one- and
  two-particle resonances are discussed in subsections~\ref{sec:1part}
  and~\ref{sec:2parts} respectively, within the dominant elastic
  cotunneling channel. The unitary limit is discussed in
  subsection~\ref{sec:unitary-ST}, still within elastic
  cotunneling. Subsection~\ref{subsec:further-calculations} discusses
  the regime of elastic cotunneling and the corresponding unitary
  limit is discussed in subsection~\ref{sec:unitary-EC}.  It is
  further suggested in subsection~\ref{sec:distinction} that
  sequential tunneling dominates over elastic cotunneling at low or
  moderate values interface transparencies, and if the interfaces
  contain many channels, so as to perform averaging over the
  oscillations at the scale of the Fermi wave-length}
{$\lambda_F$}. {The details of the analytical results are presented as
  supplemental Appendices.}
  
\begin{figure*}[htb]
  \begin{minipage}{.7\textwidth}
  \begin{minipage}{.7\textwidth}
  \begin{minipage}{\textwidth}
    \begin{minipage}{.49\textwidth}
      \includegraphics[width=\textwidth]{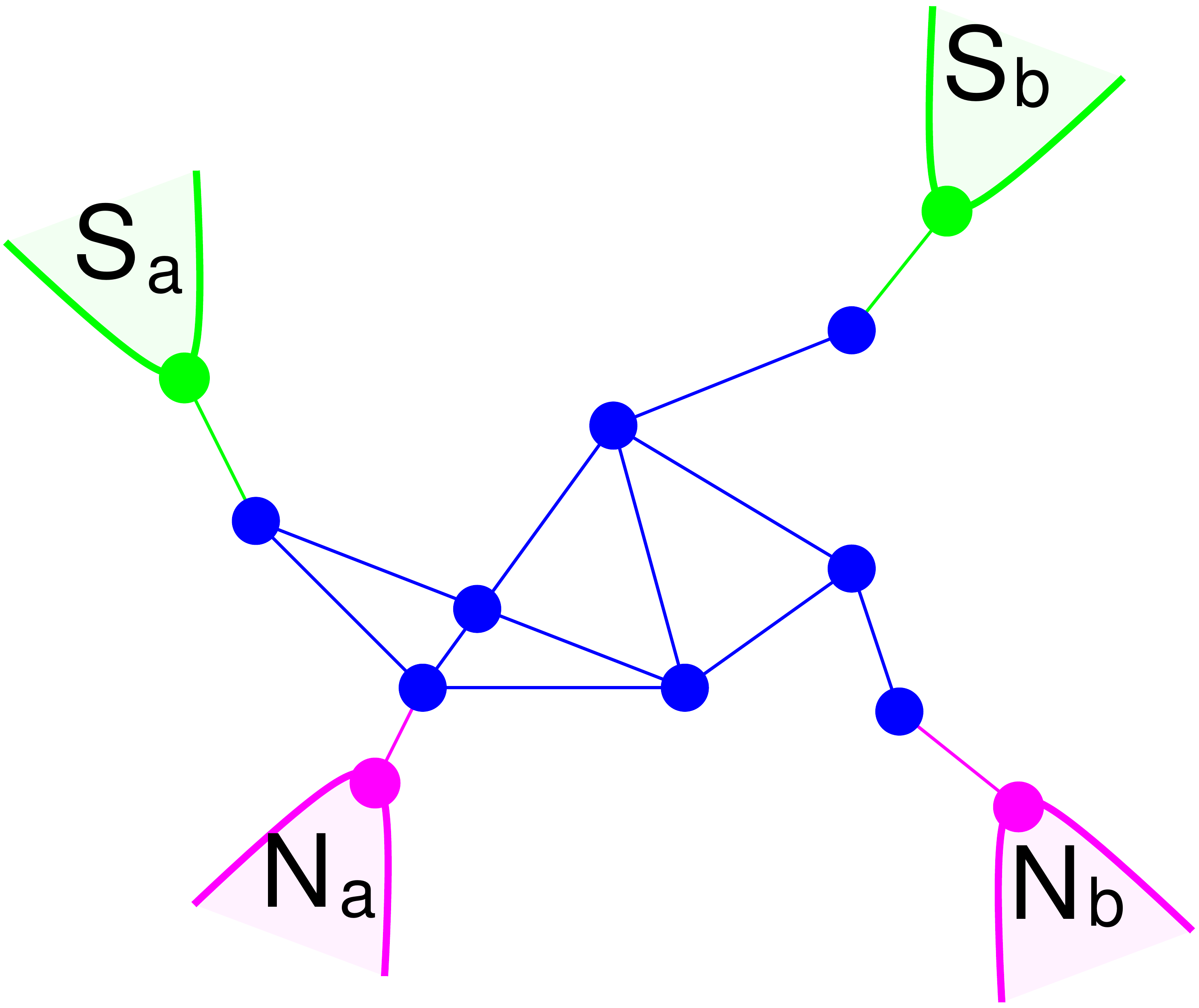}\end{minipage}\begin{minipage}{.49\textwidth}
      \begin{minipage}{.2\textwidth}
        {\bf \huge $\rightarrow$}
      \end{minipage}\begin{minipage}{.79\textwidth}
        \includegraphics[width=.8\textwidth]{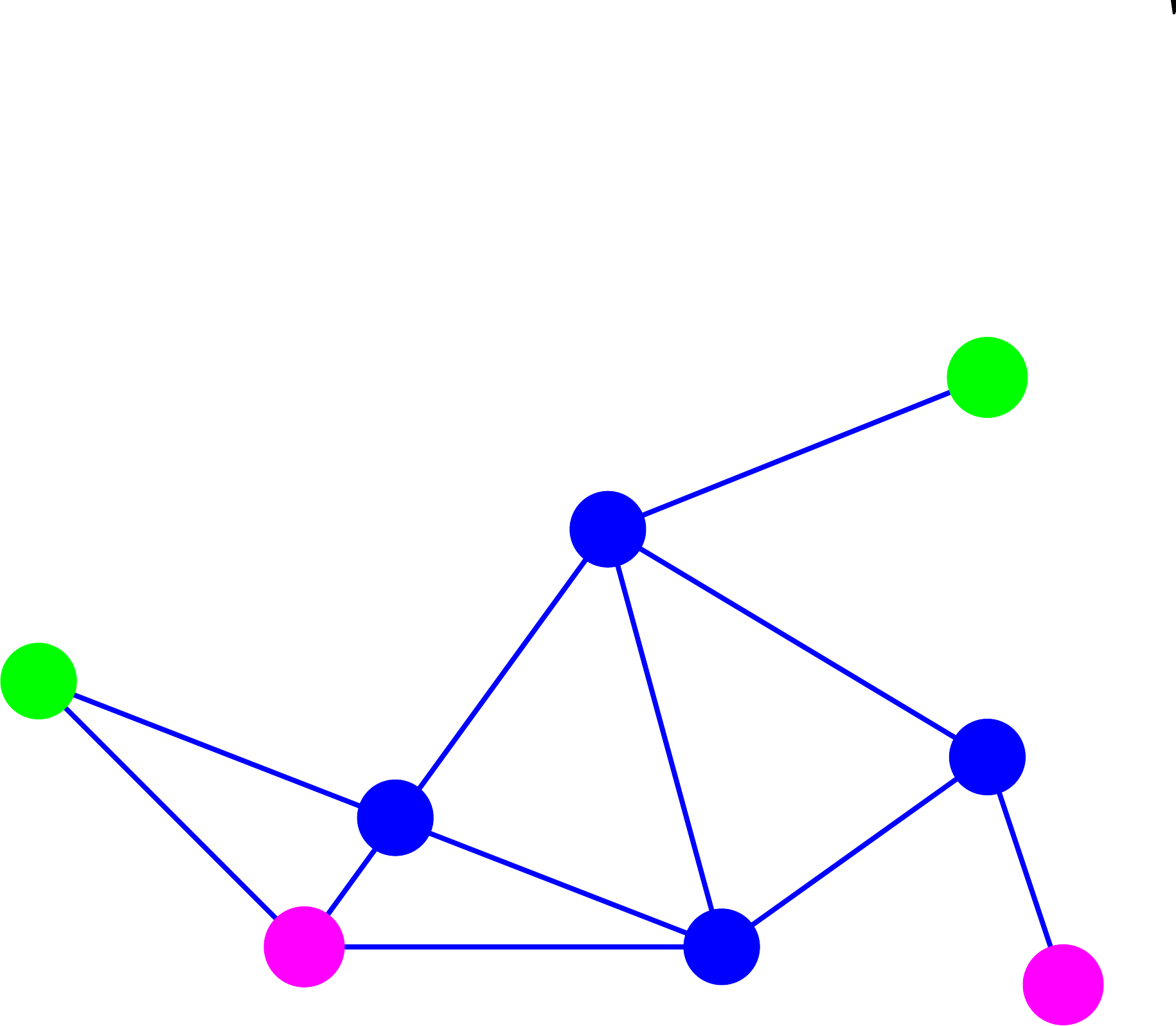}\end{minipage}
      \end{minipage}
  \end{minipage}
  \begin{minipage}{\textwidth}
    \begin{minipage}{.49\textwidth}
      \centerline{\large ${\cal H}$}\end{minipage}\begin{minipage}{.49\textwidth}
      \centerline{\large ${\cal H}_{eff}$}\end{minipage}
    \end{minipage}
  \end{minipage}
  \end{minipage}
  \begin{minipage}{.29\textwidth}
  \caption{{{A tight-binding graph locally connected to two infinite
        superconducting and two infinite normal leads.}  The initial
      Hermitian Hamiltonian ${\cal H}$ is transformed into the
      effective non-Hermitian Hamiltonian ${\cal H}_{eff}$.}
    \label{fig:graph1}
    }
    \end{minipage}
\end{figure*}
\begin{figure*}[htb]
  \begin{minipage}{.7\textwidth}
  \begin{minipage}{.7\textwidth}
  \begin{minipage}{\columnwidth}
    \begin{minipage}{.49\textwidth}
      \includegraphics[width=\textwidth]{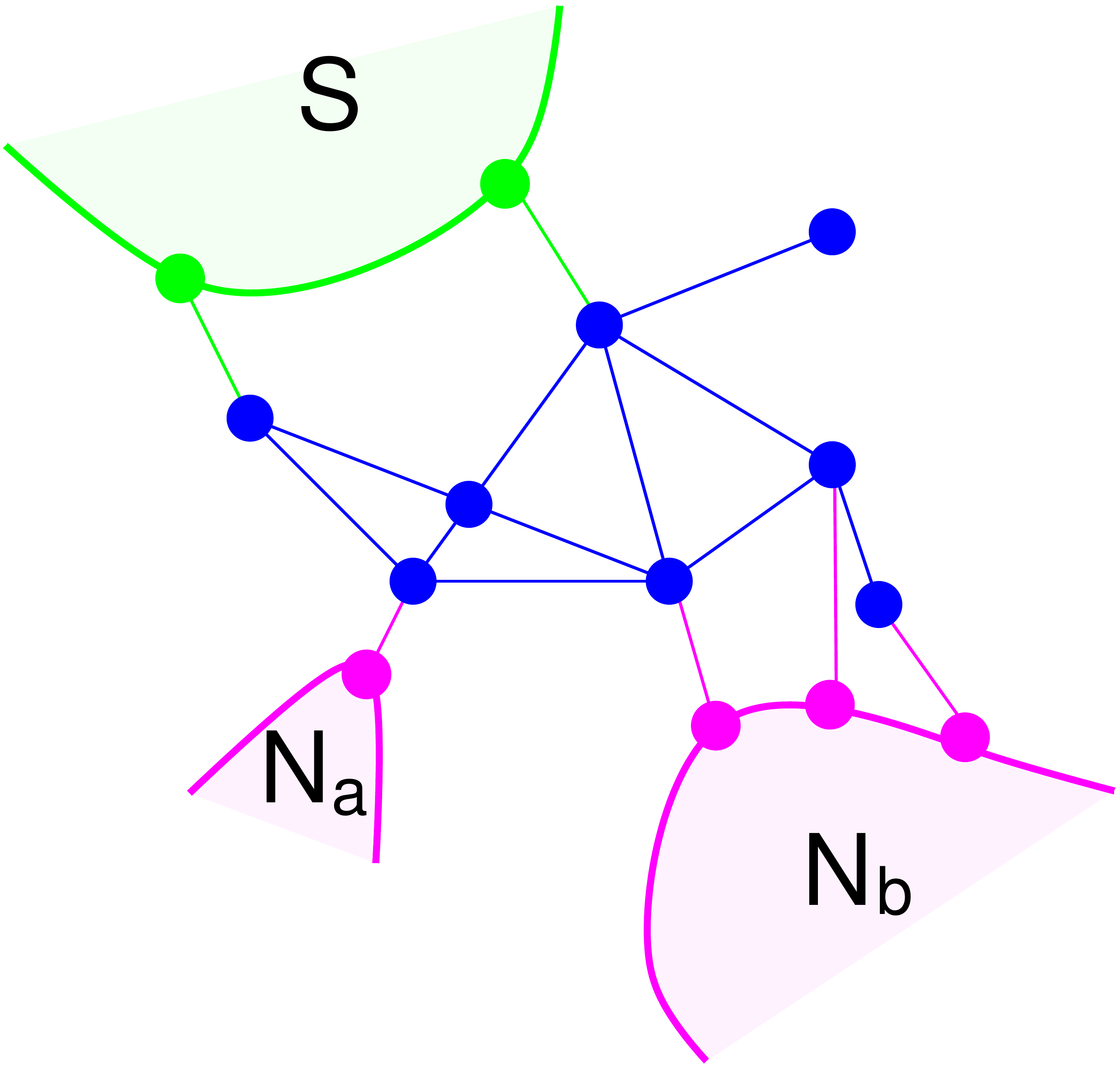}\end{minipage}\begin{minipage}{.49\textwidth}
      \begin{minipage}{.2\textwidth}
        {\bf \huge $\rightarrow$}
      \end{minipage}\begin{minipage}{.79\textwidth}
        \vspace*{-1.75cm}
        \includegraphics[width=.8\textwidth]{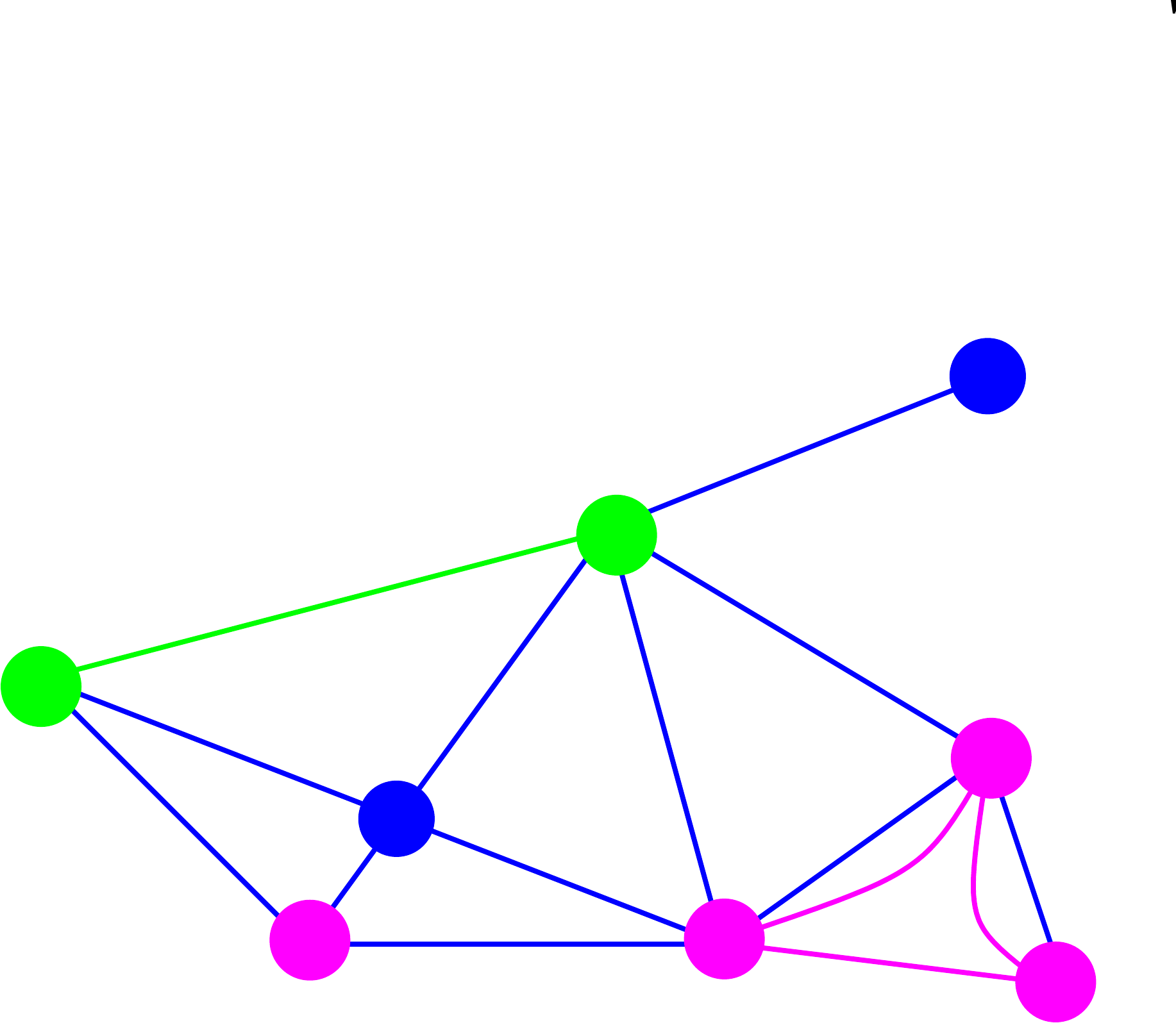}\end{minipage}
      \end{minipage}
  \end{minipage}
  \begin{minipage}{\textwidth}
    \begin{minipage}{.49\textwidth}
      \centerline{\large ${\cal H}$}\end{minipage}\begin{minipage}{.49\textwidth}
      \centerline{\large ${\cal H}_{eff}$}\end{minipage}
    \end{minipage}
  \end{minipage}
  \end{minipage}
  \begin{minipage}{.29\textwidth}
  \caption{{The change in connectivity upon including
      more extended interfaces and longer range couplings in the
      infinite superconducting and normal leads.}
    \label{fig:graph2}
  }
  \end{minipage}
\end{figure*}

{
  \subsection{Expression of the Green's functions
    deduced from graph-theory arguments}
  \label{sec:graph}
}

{In this subsection, we present and demonstrate the
  form of the Green's functions. In the absence of coupling to the
  superconductors, the bare Green's functions of the central-$N$
  connected to $N_B$ have the following spectral representation:
\begin{eqnarray}
 \label{eq:residu-debut-bare}
 \hat{g}^A_{\alpha_k,\alpha_l}(\omega)&=& \sum_\mu \frac{\hat{\cal R}_{\alpha_k,\alpha_l}^{'(\mu)}}
 {\omega-\zeta'_\mu-i \eta'_\mu}
\end{eqnarray}
and
\begin{eqnarray}
 \hat{g}^R_{\alpha_k,\alpha_l}(\omega)&=& \sum_\mu \frac{\hat{\cal R}_{\alpha_k,\alpha_l}^{'(\mu)}}
 {\omega-\zeta'_\mu+i\eta'_\mu}
   \label{eq:residu-fin-bare}
 ,
\end{eqnarray}
where $\zeta'_\mu$ {and $\eta'_\mu$} are the real and
imaginary parts of the poles, respectively, and ${\cal R}'$ is the
matrix of the residues.}

{Conversely, the fully dressed advanced and retarded
  Green's functions have poles at the ABS energies $\Omega_\lambda\pm
  i \eta_\lambda$. The fully dressed Green's functions have the
  following spectral representation:
\begin{eqnarray}
 \label{eq:residu-debut}
 \hat{G}^A_{\alpha_k,\alpha_l}(\omega)&=& \sum_\lambda \frac{\hat{\cal R}_{\alpha_k,\alpha_l}^{(\lambda)}}
 {\omega-\Omega_\lambda-i\eta_\lambda}
\end{eqnarray}
and
\begin{eqnarray}
  \hat{G}^R_{\alpha_k,\alpha_l}(\omega)&=& \sum_\lambda \frac{\hat{\cal R}_{\alpha_k,\alpha_l}^{(\lambda)}}
 {\omega-\Omega_\lambda+i\eta_\lambda}
   \label{eq:residu-fin}
 ,
\end{eqnarray}
where $\hat{\cal R}$ is the matrix of the residues.}

{Now, we demonstrate
  Eqs.~(\ref{eq:residu-debut-bare})-(\ref{eq:residu-fin-bare}) and
  Eqs.~(\ref{eq:residu-debut})-(\ref{eq:residu-fin})}{. We assume}
{that the central-$N$ consists of a general tight-binding graph
  connected to the normal and superconducting leads, see
  Figs.~\ref{fig:graph1} and~\ref{fig:graph2}. We implement the
  large-gap approximation, see the remarks in the above
  subsection~\ref{sec:further-physical-remarks}. The large-gap
  approximation is combined to the energy-independent Green's
  functions in the normal leads, which, physically, have the meaning
  of small-area interfaces. Within those approximations, the connected
  structure can be captured by an effective Hamiltonian
  \cite{Melin-Winkelmann-Danneau} which takes the following form on
  the basis of the graph or the lead tight-binding sites ($G$ or $L$
  respectively) :
  \begin{equation}
    {\cal H}=\left(\begin{array}{cc} {\cal H}_{G,G}&{\cal H}_{G,L}\\
      {\cal H}_{L,G}&H_{L,L}\end{array}\right)
    ,
  \end{equation}
  with ${\cal H}_L$ the Hamiltonian of the infinite leads and ${\cal
    H}_G$ the Hamiltonian of the finite tight-binding graph. Then, we
  deduce the following} {set} {of
  linear equations for the eigenvalues and eigenstates of the
  Hermitian Hamiltonian ${\cal H}$ of the entire device, including the
  tight-binding graph and the leads:
  \begin{eqnarray}
    {\cal H}_{G,G} \psi_G + {\cal H}_{G,L} \psi_L &=& E \psi_G\\
    {\cal H}_{L,G} \psi_G + {\cal H}_{L,L} \psi_L &=& E \psi_L
    ,
  \end{eqnarray}
  which yields ${\cal H}_{eff}(E) \psi_G = E \psi_G$ with the effective
  Hamiltonian
  \begin{equation}
   {\cal H}_{eff}(E)= {\cal H}_{G,G}+{\cal H}_{G,L} \left(E-{\cal H}_{L,L}\right)^{-1}
    {\cal H}_{L,G}
    .
  \end{equation}
  Within the two assumptions mentioned above on the superconducting
  and normal leads, the resolvent $\left(E-{\cal H}_{L,L}\right)^{-1}$
  and the effective Hamiltonian ${\cal H}_{eff}\equiv {\cal
    H}_{eff}(E)$ are both energy-independent.}

{Figs.~\ref{fig:graph1} and~\ref{fig:graph2}
  graphically show how the non-Hermitian effective Hamiltonian ${\cal
    H}_{eff}$ is deduced from the Hermitian ${\cal H}$. The
  connectivity of the tight-binding graph remains unchanged in
  Fig.~\ref{fig:graph1} because of the absence of nonlocality in the
  leads. However, the matrix elements of the Hamiltonian are changed
  on the tight-binding sites making the contacts, which is shown by
  the colorcode in Figs.~\ref{fig:graph1} and~\ref{fig:graph2}.}
{In the initial tight-binding model,}
{the quasiparticles can escape towards infinity into
  the lead, which produces the non-Hermitian ${\cal H}_{eff}$ on the
  finite effective graph. Fig.~\ref{fig:graph2} additionally
  illustrates how the connectivity of the tight-binding graph changes
  when ${\cal H}$ is transformed into ${\cal H}_{eff}$ in the presence
  of nonlocality in the leads. }

{We conclude that the above
  Eqs.~(\ref{eq:residu-debut-bare})-(\ref{eq:residu-fin-bare}) and
  Eqs.~(\ref{eq:residu-debut})-(\ref{eq:residu-fin}) are simply
  deduced as the spectral representation of the corresponding
  non-Hermitian Hamiltonians, where {$\Omega_\lambda$
    and $\eta_\lambda$ are} the real and imaginary parts of the
  complex-valued energies of ${\cal H}_{eff}$.}

{
\subsection{Sequential tunneling and elastic cotunneling}
\label{sec:elco}}

{In this section, we evaluate how the electronic populations couple to
  the {phase-sensitive} current. {In
    analogy with the wording used for Coulomb blockaded quantum dots,
    we distinguish} between the limiting cases of {\it dominant
    elastic cotunneling} and {\it dominant sequential tunneling}.}

{{\it Current conservation:} For simplicity, we start with a
  zero-dimensional (0D) quantum dot with broadened energy
  levels. {The finite width of the energy levels is
    produced either by introducing an intrinsic imaginary part of the
    energy, see Ref.~\onlinecite{Melin2017}, or by coupling the
    noninteracting quantum dot to an extrinsic normal metallic lead
    extending to infinity.}  We assume that this 0D quantum dot $x$
  with broadened energy levels is connected to the grounded left and
  right superconducting leads $S_a$ and $S_b$ respectively, and to the
  normal lead $N_B$ biased at the voltage $V_B$, see
  Fig.~\ref{fig:thedevice}f. Then,} {using the Keldysh Green's
  functions presented in Appendix~\ref{app:methods},} {we find the
  following expression for $I_a+I_b$, the sum of the currents
  transmitted into $S_a$ and $S_b$:
\begin{eqnarray}
  I_a+I_b&=& \mbox{Nambu-trace}\left\{\left(\hat{\Gamma}_{x,x}^A
  \hat{\tau}_3 -\hat{\tau}_3
  \hat{\Gamma}_{x,x}^R\right)\hat{G}^{+,-}_{x,x} \right.\\&+&\left.
  \hat{\Gamma}_{x,x}^{+,-} \hat{\tau}_3 \hat{G}^R_{x,x}
  -\hat{\tau}_3 \hat{\Gamma}_{x,x}^{+,-} \hat{G}^A_{x,x}\right\}
  ,
  \nonumber
\end{eqnarray}
with
\begin{eqnarray}
  \hat{G}_{x,x}^{+,-}&=&
  \left(\hat{I}+\hat{G}{x,x} \hat{\Gamma}^R_{x,x}\right) \hat{g}^{+,-}_{x,x}
  \left(\hat{I}+\hat{\Gamma}^A_{x,x} \hat{G}^A_{x,x} \right)\\
  &+&\hat{G}^R_{x,x} \hat{\Gamma}^{+,-}_{x,x} \hat{G}^A_{x,x}
  ,
  \nonumber
\end{eqnarray}
where we use the following notations
\begin{eqnarray}
  \hat{\Gamma}_{x,x}^A &=& \hat{t}_{x,a} \hat{g}^A_{a,a} \hat{t}_{a,x}
  + \hat{t}_{x,b} \hat{g}^A_{b,b} \hat{t}_{b,x}\\
  \hat{\Gamma}_{x,x}^R &=& \hat{t}_{x,a} \hat{g}^R_{a,a} \hat{t}_{a,x}
  + \hat{t}_{x,b} \hat{g}^R_{b,b} \hat{t}_{b,x}\\
  \hat{\Gamma}_{x,x}^{+,-} &=& \hat{t}_{x,a} \hat{g}^{+,-}_{a,a} \hat{t}_{a,x}
  + \hat{t}_{x,b} \hat{g}^{+,-}_{b,b} \hat{t}_{b,x}
  ,
\end{eqnarray}
and the Keldysh Green's function is defined as
$\hat{g}^{+,-}_{x,x}=n_F(\omega) \left[g^A_{x,x} -g^R_{x,x}\right]$.
Using the identities
\begin{eqnarray}
  \hat{\Gamma}_{x,x}^R \hat{G}_{x,x}^R &=& \left(g^R_{x,x}\right)^{-1} \hat{G}^R_{x,x} - \hat{I}\\
  \hat{G}_{x,x}^A \hat{\Gamma}_{x,x}^A &=& \hat{G}^A_{x,x} \left(g^A_{x,x}\right)^{-1} - \hat{I}
  ,
\end{eqnarray}
we obtain a simplified expression for $I_a+I_b$:
{\begin{eqnarray}
    \label{eq:term-gpmxx}
  &&  I_a+I_b = \mbox{Nambu-trace}\left\{\right.\\
  \nonumber
  &&\left(\hat{\Gamma}^A_{x,x} \hat{\tau}_3 - \hat{\tau}_3
  \hat{\Gamma}_{x,x}^R \right) \left(\hat{I} + \hat{G}^R_{x,x}
  \hat{\Gamma}_{x,x}^R\right) \hat{g}^{+,-}_{x,x} \left(\hat{I} +
  \hat{\Gamma}_{x,x}^A \hat{G}_{x,x}^A\right) \\
  &+& \left. \hat{\tau}_3 \hat{G}^R_{x,x} \hat{\Gamma}_{x,x}^{+,-}
  \hat{G}_{x,x}^A \left[ \left(\hat{g}^A_{x,x} \right)^{-1} -
    \left(\hat{g}^R_{x,x} \right)^{-1}\right] \right\} .
  \nonumber
\end{eqnarray}}
{In this example, the} populations $n_F(\omega)$ of
the central {quantum dot} at the energy $\omega$ are
calculated from imposing the vanishingly small current $I_a+I_b\equiv
0$, which makes $n_F(\omega)$ acquire steps at each of the {external}
normal reservoir chemical potential.}

{The following calculations will be carried out in the following limiting
  cases: First, in the} {\it regime of elastic cotunneling}{, the
  steps in {the distribution function} {$n_F(\omega)$ contribute to}
  the differential conductance $d I_B/d V_B$ according to {the first
    term in the r.h.s. of} Eq.~(\ref{eq:term-gpmxx}). We will also
  consider the {\it regime of elastic cotunneling} {where the second
    term in the r.h.s. of Eq.~(\ref{eq:term-gpmxx}) is much larger
    than the first one.}

{{\it Multichannel interfaces:} Considering next a
  multilevel quantum dot connected to the collection of the leads
  $\{L_p\}$, we deduce from Appendix~\ref{app:methods} that the
  spectral current flowing at the $D_p$-$L_p$ interface is given by
\begin{equation}
I_p=\mbox{Nambu-trace}\{\hat{\tau}_3 \hat{t}_{L_p,D_p}
\hat{G}^{+,-}_{D_p,L_p} - \hat{\tau}_3 \hat{t}_{D_p,L_p}
\hat{G}^{+,-}_{L_p,D_p}\} ,
\end{equation}
where the bare advanced and retarded Green's functions are given by
\begin{eqnarray}
  \hat{G}^A_{D_p,L_p}&=& \hat{G}^A_{D_p,D_p} \hat{t}_{D_p,L_p} \hat{g}^A_{L_p,L_p}\\
  \hat{G}^R_{D_p,L_p}&=& \hat{G}^R_{D_p,D_p} \hat{t}_{D_p,L_p} \hat{g}^R_{L_p,L_p}
\end{eqnarray}
and the Keldysh Green's function is given by
\begin{eqnarray}
  \hat{G}^{+,-}_{D_p,L_p}&=& \hat{G}^{+,-}_{D_p,D_p} \hat{t}_{D_p,L_p} \hat{g}^A_{L_p,L_p}
  + \hat{G}^{R}_{D_p,D_p} \hat{t}_{D_p,L_p} \hat{g}^{+,-}_{L_p,L_p}
  .
\end{eqnarray}
We reach the same conclusions as above regarding sequential tunneling
and elastic cotunneling because the single 0D tight-binding site $x$ in
the above calculations now becomes the vector-valued $x\equiv
\left(D_1,\,...,\,D_n\right)$.}

{
  \subsection{The different types of resonances in the regime
    of sequential tunneling}\label{sec:different-types}}

{The current can uniquely be decomposed into the
  contributions of the regular terms, and the one- and two-particle
  resonances. The corresponding computational algorithm for making the
  decomposition is detailed in Appendix~\ref{app:thecurrent}. The
  algorithm expresses the spectral current as a function of the fully
  dressed advanced and retarded Green's functions $G^A$ and $G^R$ of
  the central-$N$. The starting point of the calculation is the
  collection of the bare Green's functions of the central-$N$
  connected to the infinite normal lead $N_B$, in the absence of
  coupling to the superconducting leads $S_a$ and $S_b$. Similarly to
  Eq.~(18) in Ref.~\onlinecite{Cuevas}, the expanded expression of the
  current given by Eqs.~(\ref{eq:DEBUT})-(\ref{eq:O2-prime-4})
  decomposes into a collection of terms that are regular (terms of
  order zero), or involve a single of $G^A$ or $G^R$ (terms of order
  one for the one-particle resonances), or involve the product $G^A
  G^R$ (terms of order two for the two-particle resonances). Those
  different terms are interpreted as the quasiequilibrium (i.e. the
  terms of orders zero or one) or the truly nonequilibrium (i.e. the
  terms of order two) contributions to the current.}

{The current $I_\alpha$ through the lead $\alpha$ is
  then written as $I^{(\alpha)}_{tot} =
  I_0^{(\alpha)}+I_{1\,part}^{(\alpha)} + I_{2\,part}^{(\alpha)}$,
  where $I_0^{(\alpha)}$ denotes the regular term, and
  $I_{1,\,part}^{(\alpha)}$ and $I_{2\,part}^{(\alpha)}$ respectively
  denote the one- and two-particle currents transmitted into the lead
  $\alpha$. Each of those terms depends on all of the device control
  parameters, and, in addition, on the entire energy dependence of the
  nonequilibrium distribution function. The latter can be approximated as
  being parameterized by the steps of height $Z_p$ at the energies
  $\mu_{N,p}$ in the central-$N$} {nonequilibrium}
{distribution function, such that $\sum_p Z_p<1$.}

{Within this approximation for the distribution function, the current
  conservation implies $\sum_\alpha
  I^{(\alpha)}_{tot}(\{\mu_{N,p}^*\},\{Z_p^*\})=0$ at the
  self-consistent nonequilibrium Fermi surface
  $\mu_{N,p}\equiv\mu_{N,p}^*$ and $Z_p\equiv Z_p^*$. However, we also
  find that no specific constraint is put on separate current
  conservation for each of the regular terms, or the one- or
  two-particle contributions. This argument weakens the physical
  meaning of separately considering the regular and the one- and
  two-particle terms. However, in the following, we still separately
  discuss each sector, which is justified by the uniqueness of this
  decomposition, as it is mentioned above.}

{
\subsection{One-particle resonances within sequential tunneling}
\label{sec:1part}}

In this subsection, {we assume the dominant {sequential} tunneling
  channel and} address {the} emergence of the one-particle resonances
in the expression of the current, on the basis of the calculation that
is detailed in Appendix~\ref{app:thecurrent}. We assume that the
central-$N$ is connected to the {infinite}
superconductors $S_a$, $S_b$ and to the {infinite}
normal lead $N_B$ by interfaces having arbitrary transmission.

In Appendix~\ref{app:thecurrent}, we find eight terms containing a
single fully dressed Green's function, see Eqs.~(\ref{eq:O1-1}),
(\ref{eq:O1-2}), (\ref{eq:O1-3}), (\ref{eq:O1-4}) and
Eqs. (\ref{eq:O1-1-bis}), (\ref{eq:O1-2-bis}), (\ref{eq:O1-3-bis}),
(\ref{eq:O1-4-bis}). Using the cyclic properties of the trace, the
summation over those eight terms takes the following compact form:
{
\begin{eqnarray}
  \nonumber
&& I_a^{(1)}\left(\frac{\Phi}{\Phi_0},\omega\right)= \mbox{Nambu-trace}\left\{\sum_k \hat{t}_{\alpha,a}
  \left[\hat{\tau}_3 \hat{g}^R_{a,a}
    - \hat{g}^A_{a,a} \hat{\tau}_3 \right] \times\right.\\
  \nonumber
&&\left.  \hat{t}_{a,\alpha} \hat{G}^R_{\alpha,\alpha_k}(\omega) \hat{t}_{\alpha_k,a_k}
  \hat{g}^R_{a_k,a_k} \hat{t}_{a_k,\alpha_k} \delta \hat{g}^{+,-}_{\alpha_k,\alpha}(\omega)\right\}\\
  \nonumber
  &+&\mbox{Nambu-trace}\left\{\sum_k \hat{t}_{\alpha,a}
  \left[\hat{\tau}_3 \hat{g}^R_{a,a}
    - \hat{g}^A_{a,a} \hat{\tau}_3 \right] \hat{t}_{a,\alpha} \times\right.\\
&&\left.  \delta \hat{g}^{+,-}_{\alpha,\alpha_k}(\omega) \hat{t}_{\alpha_k,a_k} \hat{g}^A_{a_k,a_k}
  \hat{t}_{a_k,\alpha_k} \hat{G}^A_{\alpha_k,\alpha}(\omega)\right\}
 \label{eq:Ia-1-1}
  .
\end{eqnarray}}
{In the first term of Eq.~(\ref{eq:Ia-1-1}), the fully
  dressed Green's function $\hat{G}^R_{\alpha,\alpha_k}(\omega)$
  couples the analogous superconducting pair amplitude
  \begin{equation}
    \label{eq:pair-amplitude}
    \hat{\tau}_3 \hat{g}^R_{a,a}-\hat{g}^A_{a,a}\hat{\tau}_3\equiv 2
    \hat{\tau}_3 \hat{g}_{a,a}
  \end{equation}
  to the spin-up electron and spin-down hole density and populations
  appearing in the Keldysh Green's function
  $\hat{g}^{+,-}_{\alpha_k,\alpha}$. Eq.~(\ref{eq:pair-amplitude}) was
  evaluated in the large-gap approximation. This expression contrasts
  with the Andreev reflection term in Eq.~(18) of
  Ref.~\onlinecite{Cuevas}, in the sense that the above
  Eq.~(\ref{eq:Ia-1-1}) involves the density-pair amplitude and
  population-pair amplitude couplings, instead of the initial and
  final states given by the spin-up electron and spin-down hole
  densities in Ref.~\onlinecite{Cuevas}.}

{{\it Consistency with the Keldysh calculation of the
    phase-AR presented in the above section~\ref{sec:lowest-order}:}}
First coming back to the tunnel limit $t_{a,\alpha},\,t_{b,\beta}\ll
W$, we replace the fully dressed Green's functions
$\hat{G}_{\alpha_k,\alpha_l}$ by the bare ones
$\hat{g}_{\alpha_k,\alpha_l}$ and expand the Nambu labels according to
{
\begin{eqnarray}
 \label{eq:Ia-1-1-bis}
 && \tilde{I}_a^{(1)}\left(\frac{\Phi}{\Phi_0},\omega\right)=\\
 \nonumber
 &&2\hat{t}_{\alpha,a}^{1,1}
  \hat{g}^{R,1,2}_{a,a}\hat{t}_{a,\alpha}^{2,2}
  \hat{g}^{R,2,2}_{\alpha,\beta}(\omega) \hat{t}_{\beta,b}^{2,2}
  \hat{g}^{R,2,1}_{b,b} \hat{t}_{b,\beta}^{1,1} \delta \hat{g}^{+,-,1,1}_{\beta,\alpha}(\omega)\\
  \nonumber
  &-&2\hat{t}_{\alpha,a}^{2,2}
  \hat{g}^{R,2,1}_{a,a}\hat{t}_{a,\alpha}^{1,1}
  \hat{g}^{R,1,1}_{\alpha,\beta}(\omega) \hat{t}_{\beta,b}^{1,1}
  \hat{g}^{R,1,2}_{b,b} \hat{t}_{b,\beta}^{2,2} \delta \hat{g}^{+,-,2,2}_{\beta,\alpha}(\omega)\\
  \nonumber
  &+&2 \hat{t}_{\alpha,a}^{1,1} \hat{g}^{R,1,2}_{a,a} \hat{t}_{a,\alpha}^{2,2}
  \delta \hat{g}^{+,-,2,2}_{\alpha,\beta}(\omega) \hat{t}_{\beta,b}^{2,2} \hat{g}^{A,2,1}_{b,b}
  \hat{t}_{b,\beta}^{1,1} \hat{g}^{A,1,1}_{\beta,\alpha}(\omega)\\
  \nonumber
  &-&2 \hat{t}_{\alpha,a}^{2,2} \hat{g}^{R,2,1}_{a,a} \hat{t}_{a,\alpha}^{1,1}
  \delta \hat{g}^{+,-,1,1}_{\alpha,\beta}(\omega) \hat{t}_{\beta,b}^{1,1} \hat{g}^{A,1,2}_{b,b}
  \hat{t}_{b,\beta}^{2,2} \hat{g}^{A,2,2}_{\beta,\alpha}(\omega)
  .
\end{eqnarray}}
This Eq.~(\ref{eq:Ia-1-1-bis}) implies the above
Eq.~(\ref{eq:I1-A1}). Thus, the phase-AR process evaluated in the
above section~\ref{sec:lowest-order} is entirely within this
one-particle {sector. Now,} {we make
  the one-particle resonances explicit} and insert the fully dressed
Green's functions $\hat{G}^A_{\alpha_k,\alpha_l}(\omega)$ and
{ $\hat{G}^R_{\alpha_k,\alpha_l}(\omega)$ into
  Eq.~(\ref{eq:Ia-1-1}).
  Eqs.~(\ref{eq:residu-debut-bare})-(\ref{eq:residu-fin-bare})} yield
the following expression for the derivative of the Keldysh Green's
function with respect to $\mu_{N,p}$:
\begin{eqnarray}
 \label{eq:residu-bare}
&& \frac{\partial}{\partial
   \mu_{N,p}}\hat{g}^{+,-}_{\alpha_k,\alpha_l}(\omega)\\&=& 2i\pi Z_p
 \hat{\tau}_3 \delta\left(\omega-\mu_{N,p}\hat{\tau}_3\right) \sum_\mu
 \hat{\cal R}_{\alpha_k,\alpha_l}^{'(\mu)}
 \delta_{\eta'_\mu}\left(\omega-\zeta'_\mu\right)
 ,
 \nonumber
\end{eqnarray}
where $\delta_{\eta'_\mu}$ is the Dirac $\delta$-function broadened by
the line-width $\eta'_\mu$.

Now, we insert Eqs.~(\ref{eq:residu-debut})-(\ref{eq:residu-bare})
into {Eq.}~(\ref{eq:Ia-1-1}). Then,
$\partial\tilde{I}_1/\partial \mu_{N,p}$ takes the following form:
{
 \begin{eqnarray}
   \label{eq:res-final1}
&&\frac{\partial}{\partial \mu_{N,p}} \int d\omega
   {I}_a^{(1)}\left(\frac{\Phi}{\Phi_0},\omega\right) =\\\nonumber
   &&4i\pi Z_p \mbox{Nambu-trace} \left\{
\sum_{\lambda,\mu} \hat{t}_{\alpha,a} \hat{\tau}_3 \hat{g}_{a,a}
\hat{t}_{a,\alpha} \hat{\cal R}^{(\lambda)}_{\alpha,\beta}
\hat{t}_{\beta,b} \hat{g}_{b,b} \hat{t}_{b,\beta}\times\right.\\
\nonumber
&&\left.\hat{\tau}_3
\hat{\cal R}^{'(\mu)}_{\beta,\alpha} \int
d\omega\frac{\delta\left(\omega-
  \mu_{N,p}\hat{\tau}_3^{\beta,\alpha}\right) \delta_{\eta'_\mu}
  \left(\omega-\zeta'_\mu\right)
}{\omega-\Omega_\lambda+i\eta_\lambda}\right\}\\
\nonumber
&+& 4i\pi Z_p \mbox{Nambu-trace} \left\{\sum_{\lambda,\mu} \hat{t}_{\alpha,a}
\hat{\tau}_3 \hat{g}_{a,a} \hat{t}_{a,\alpha} \hat{\cal
  R}^{'(\mu)}_{\alpha,\beta} \hat{t}_{\beta,b} \hat{g}_{b,b}
\hat{t}_{b,\beta}\times\right.
\\&&\left.\hat{\tau}_3 \hat{\cal R}^{(\lambda)}_{\beta,\alpha} \int
d\omega\frac{
  \delta\left(\omega-\mu_{N,p}\hat{\tau}_3^{\alpha,\beta}
  \right)\delta_{\eta'_\mu}\left(\omega-\zeta'_\mu\right)}
{\omega-\Omega_\lambda-i\eta_\lambda} \right\}
,
\nonumber
 \end{eqnarray}}
where we denoted by $g_{a,a} \equiv g^A_{a,a}=g^R_{a,a}$ and $g_{b,b}
\equiv g^A_{b,b}=g^R_{b,b}$ the local superconducting Green's function
of $S_a$ and $S_b$ in the large-gap approximation {and
  we used the above Eq.~(\ref{eq:pair-amplitude}).} The superscript on
the matrix $\hat{\tau}_3$ refers to where it has been inserted in the
chain of the Green's function{, i.e. at the location
  of the Keldysh Green's function $\hat{g}^{+,-}$.}

{{\it Conclusions:} The above energy-integral given by
  Eq.~(\ref{eq:res-final1}) is} sensitive to the values of the complex
energies $\zeta'_\mu\pm i \eta'_\mu$ of the central-$N$ connected to
$N_B$, and to the values of the fully dressed ABS energies
$\Omega_\lambda\pm i \eta_\lambda$ in the presence of all couplings to
the superconducting and normal leads $S_a$, $S_b$ and $N_B$
respectively. {We additionally deduce from Eq.~(\ref{eq:res-final1})
  that the harmonic dependence of Eq.~(\ref{eq:EE2}) on both the
  reduced magnetic flux $\frac{\Phi}{\Phi_0}$ and the bias voltage
  $V_B$ now becomes peaked around the values $\Omega_\lambda$ of the
  spectrum of a long Josephson junction
  \cite{Kulik,Ishii,Bagwell}. This supports one of the main claims of
  the paper, namely, the conductance spectrum of a three-terminal
  Andreev interferometer in the long-junction limit follows the
  spectrum of a long Josephson junction
  \cite{Kulik,Ishii,Bagwell}. Numerical calculations supporting this
  claim are presented in the forthcoming
  Sec.~\ref{sec:pheno-full-ladders}.}

{
\subsection{Two-particle resonances within sequential tunneling}
\label{sec:2parts}}

In this subsection, {we assume the dominant sequential tunneling
  channel and} discuss the emergence {of two-particle
  resonances. We obtain in Appendix~\ref{app:thecurrent}} eight terms
corresponding to the second-order contribution
{$I_a^{(2)}\left(\frac{\Phi}{\Phi_0},\omega\right)$} to the spectral
current {$I_a\left(\frac{\Phi}{\Phi_0},\omega\right)$}, see
Eqs.~(\ref{eq:O2-1}), (\ref{eq:O2-2}), (\ref{eq:O2-3}),
(\ref{eq:O2-4}) and Eqs.~(\ref{eq:O2-prime-1}), (\ref{eq:O2-prime-2}),
(\ref{eq:O2-prime-3}), (\ref{eq:O2-prime-4}). Using the cyclic
properties of the trace, those contributions are summarized as
follows: {
\begin{eqnarray}
  \nonumber
&& I_a^{(2)}\left(\frac{\Phi}{\Phi_0},\omega\right)= \mbox{Nambu-trace} \left\{
 \sum_{k,l}\hat{t}_{\alpha,a} \left[\hat{\tau}_3 \hat{g}^R_{a,a}
   - \hat{g}^A_{a,a} \hat{\tau}_3 \right] \hat{t}_{a,\alpha} \times \right.\\
 \nonumber
&& \hat{G}^R_{\alpha,\alpha_k}(\omega) \hat{t}_{\alpha_k,a_k}
 \hat{g}^R_{a_k,a_k} \hat{t}_{a_k,\alpha_k} \delta
 \hat{g}^{+,-}_{\alpha_k,\alpha_l}(\omega) \times\\
 && \left.\hat{t}_{\alpha_l,a_l}
 \hat{g}^A_{a_l,a_l} \hat{t}_{a_l,\alpha_l}
 \hat{G}^A_{\alpha_l,\alpha}(\omega)\right\}
 ,
  \label{eq:I(2)-debut}
\end{eqnarray}}
where Eq.~(\ref{eq:I(2)-debut}) generally holds for arbitrary values
of the tunneling amplitudes $t_a$ and $t_b$ connecting by single
channel contacts the central normal metal to the superconducting leads
$S_a$ and $S_b$. {We note that
  Eq.~(\ref{eq:I(2)-debut}) also involves a coupling between the
  spectral density of states in the central-$N$ and the pair amplitude
  is $S_a$.}

Now, we insert into Eq.~(\ref{eq:I(2)-debut}) the spectral
decomposition given by
Eqs.~(\ref{eq:residu-debut-bare})-(\ref{eq:residu-fin}):
{
  \begin{eqnarray}
  \nonumber &&\frac{\partial}{\partial \mu_{N,p}} \int d\omega
            {I}_a^{(2)}\left(\frac{\Phi}{\Phi_0},\omega\right) =
            \\ \nonumber && 4i\pi Z_p \mbox{Nambu-trace}\left\{
            \sum_{k,l}\sum_{\lambda,\lambda',\mu}\hat{t}_{\alpha,a}
            \hat{\tau}_3\right.\\ \nonumber &&\left.\hat{g}_{a,a}
            \hat{t}_{a,\alpha} \hat{\cal
              R}^{(\lambda)}_{\alpha,\alpha_k} \hat{t}_{\alpha_k,a_k}
            \hat{g}_{a_k,a_k} \hat{t}_{a_k,\alpha_k} \hat{\tau}_3
            \hat{\cal R}_{\alpha_k,\alpha_l}^{'(\mu)}
            \hat{t}_{\alpha_l,a_l} \hat{g}_{a_l,a_l}
            \hat{t}_{a_l,\alpha_l} \hat{\cal
              R}^{(\lambda')}_{\alpha_l,\alpha} \times\right.
            \\ &&\left.\int d\omega
            \frac{\delta\left(\omega-\mu_{N,p}\hat{\tau}_3^{\alpha_k,\alpha_l}\right)
              \delta_{\eta'_\mu}\left(\omega-\zeta'_\mu\right)}
                 {\left(\omega-\Omega_\lambda+i\eta_\lambda\right)
                   \left(\omega-\Omega_{\lambda'}-i\eta_{\lambda'}\right)}
                 \right\},
\label{eq:resB}
\end{eqnarray}}
where we used Eq.~(\ref{eq:pair-amplitude}) in the large-gap
{approximation.}  As it was the case for the
one-particle resonances, see Eq.~(\ref{eq:res-final1}), we find that
the two-particle contribution to the current depends on the resonance
energies $\Omega_\lambda$, $\Omega_{\lambda'}$ and $\zeta'_\mu$, see
Eq.~(\ref{eq:resB}). This contribution is maximal if $\Omega_\lambda$
and $\Omega_\lambda'$ are within the line-width broadening
$\delta_{eff}$:
\begin{equation}
  \label{eq:resonance-condition}
  \Omega_\lambda=\Omega_{\lambda'} + {\cal O}\left(\delta_{eff}\right)
  .
\end{equation}

{{\it Physical remark: }It is generally true that conservation laws
  yield energy level that can freely cross each other. The possible
  symmetries such as perfect wave-guide geometry or bipartite
  Bogoliubov-de Gennes Hamiltonian can produce genuine energy level
  crossings between the ABS. However, this requires perfect control
  over the microscopic details of the device geometry, which is not
  within the midterm experimental agenda. Thus, we practically
  consider the absence of degeneracies in the ABS spectrum yielding
  $\lambda=\lambda'$ as the only option for the sharpest resonances in
  Eq.~(\ref{eq:resB}) and in Eq.~(\ref{eq:resonance-condition}).} {As
  above, we deduce from Eq.~(\ref{eq:resB}) that the conductance
  $dI_B/dV_B$ is peaked on the real parts $\Omega_\lambda$ of the
  energies of the spectrum of a long Josephson junction
  \cite{Kulik,Ishii,Bagwell}. {A further discussion of
    the presence or absence of energy level repulsion in the ABS
    spectra is presented in the forthcoming sections~\ref{sec:final},
    \ref{sec:conjecture} and Appendix~\ref{app:billiards}.}}

{
\subsection{Unitary limit for dominant sequential tunneling}
\label{sec:unitary-ST}}
{We note here that the above divergences in the current response are
  regularized by self-consistently calculating the nonequilibrium
  Fermi surface. For instance, the simple toy model of
  Eqs.~(\ref{eq:A1})-(\ref{eq:I-simple}) produces the unitary current
  $I_B= G_B V_B$ in the limit of diverging {$G_{a,b}
    \cos\left(\frac{2\pi\Phi}{\Phi_0}\right)\rightarrow \pm \infty$.}}

{
\subsection{Further calculations for the two-particle
    resonances within elastic cotunneling}
\label{subsec:further-calculations}}

In this subsection, {we {now} assume the dominant
  elastic cotunneling channel. We} provide a complementary point of
view {on the phase-AR current} and directly evaluate the current $I_B$
flowing through the lead $N_B$:
\begin{equation}
  I_B=\mbox{Nambu-trace} \left\{\hat{\tau}_3
  \left[ \hat{\Sigma}_{N_B,N} \hat{G}^{+,-}_{N,N_B}
    -\hat{\Sigma}_{N,N_B} \hat{G}^{+,-}_{N_B,N} \right] \right\}
  .
\end{equation}
We find the following for each of the Nambu components of
$\hat{\Sigma}_{N_B,N} \hat{G}^{+,-}_{N,N_B}$ and $\hat{\Sigma}_{N,N_B}
\hat{G}^{+,-}_{N_B,N}$:
\begin{eqnarray}
  \label{eq:(A1)}
&&  \left(\hat{\Sigma}_{N_B,N} \hat{G}^{+,-}_{N,N_B}\right)^{1,1}=
  \Sigma_{N_B,N}^{1,1} G^{R,1,1}_{N,N} \Sigma_{N,N_B}^{1,1} g^{+,-,1,1}_{N_B,N_B}\\
  \label{eq:(A2)}
  &+& \Sigma_{N_B,N}^{1,1} G^{R,1,1}_{N,N} \Sigma_{N,N_B}^{1,1} g^{+,-,1,1}_{N_B,N_B}
  \Sigma_{N_B,N}^{1,1} G^{A,1,1}_{N,N} \Sigma_{N,N_B}^{1,1} g^{A,1,1}_{N_B,N_B}\\
  \label{eq:(A3)}
  &+& \Sigma_{N_B,N}^{1,1} G^{R,1,2}_{N,N} \Sigma_{N,N_B}^{2,2} g^{+,-,2,2}_{N_B,N_B}
  \Sigma_{N_B,N}^{2,2} G^{A,2,1}_{N,N} \Sigma_{N,N_B}^{1,1} g^{A,1,1}_{N_B,N_B},
\end{eqnarray}
\begin{eqnarray}
  \label{eq:(B1)}
&&  \left(\hat{\Sigma}_{N_B,N} \hat{G}^{+,-}_{N,N_B}\right)^{2,2}=
  \Sigma_{N_B,N}^{2,2} G^{R,2,2}_{N,N} \Sigma_{N,N_B}^{2,2} g^{+,-,2,2}_{N_B,N_B}\\
  \label{eq:(B2)}
  &+& \Sigma_{N_B,N}^{2,2} G^{R,2,2}_{N,N} \Sigma_{N,N_B}^{2,2} g^{+,-,2,2}_{N_B,N_B}
  \Sigma_{N_B,N}^{2,2} G^{A,2,2}_{N,N} \Sigma_{N,N_B}^{2,2} g^{A,2,2}_{N_B,N_B}\\
  \label{eq:(B3)}
  &+& \Sigma_{N_B,N}^{2,2} G^{R,2,1}_{N,N} \Sigma_{N,N_B}^{1,1} g^{+,-,1,1}_{N_B,N_B}
  \Sigma_{N_B,N}^{1,1} G^{A,1,2}_{N,N} \Sigma_{N,N_B}^{2,2} g^{A,1,1}_{N_B,N_B},
\end{eqnarray}
\begin{eqnarray}
  \label{eq:(C1)}
&&  \left(\hat{\Sigma}_{N,N_B}\hat{G}^{+,-}_{N_B,N}\right)^{1,1}=
  \Sigma_{N,N_B}^{1,1} g^{+,-,1,1}_{N_B,N_B} \Sigma_{N_B,N}^{1,1} G^{A,1,1}_{N,N}  \\
  \label{eq:(C2)}
  &+& \Sigma_{N,N_B}^{1,1} g^{R,1,1}_{N_B,N_B} \Sigma_{N_B,N}^{1,1} G^{R,1,1}_{N,N}
  \Sigma_{N,N_B}^{1,1} g^{+,-,1,1}_{N_B,N_B} \Sigma_{N_B,N}^{1,1} G^{A,1,1}_{N,N}\\
  \label{eq:(C3)}
  &+& \Sigma_{N,N_B}^{1,1} g^{R,1,1}_{N_B,N_B} \Sigma_{N_B,N}^{1,1} G^{R,1,2}_{N,N}
  \Sigma_{N,N_B}^{2,2} g^{+,-,2,2}_{N_B,N_B} \Sigma_{N_B,N}^{2,2} G^{A,2,1}_{N,N},
\end{eqnarray}
and
\begin{eqnarray}
  \label{eq:(D1)}
&& \left(\hat{\Sigma}_{N,N_B}\hat{G}^{+,-}_{N_B,N}\right)^{2,2}=
  \Sigma_{N,N_B}^{2,2} g^{+,-,2,2}_{N_B,N_B} \Sigma_{N_B,N}^{2,2}
  G^{A,2,2}_{N,N} \\
  \label{eq:(D2)}
  &+& \Sigma_{N,N_B}^{2,2} g^{R,2,2}_{N_B,N_B} \Sigma_{N_B,N}^{2,2}
  G^{R,2,2}_{N,N} \Sigma_{N,N_B}^{2,2} g^{+,-,2,2}_{N_B,N_B}
  \Sigma_{N_B,N}^{2,2} G^{A,2,2}_{N,N}\\
  \label{eq:(D3)}
  &+& \Sigma_{N,N_B}^{2,2} g^{R,2,2}_{N_B,N_B} \Sigma_{N_B,N}^{2,2}
  G^{R,2,1}_{N,N} \Sigma_{N,N_B}^{1,1} g^{+,-,1,1}_{N_B,N_B}
  \Sigma_{N_B,N}^{1,1} G^{A,1,2}_{N,N}.
\end{eqnarray}
We next assume that the $N_B$-$N$ interface has moderately small
transparency. Considering low energy in comparison to the
superconducting gap, we perform the following approximation in
Eqs.~(\ref{eq:(A1)})-(\ref{eq:(D3)}):
\begin{eqnarray}
  \label{eq:truncation1}
  G_{N,N}^{1,1}&\simeq& G_{N,N}^{(1),1,1}\\\nonumber &+&
  G_{N,N}^{(1),1,1} \Sigma_{N,S}^{1,1} g_{S,S}^{1,2}
  \Sigma_{S,N}^{2,2} G_{N,N}^{(1),2,2} \Sigma_{N,S}^{2,2}
  g_{S,S}^{2,1} \Sigma_{S,N}^{1,1}
  G_{N,N}^{(1),1,1}\\ G_{N,N}^{1,2}&\simeq& G_{N,N}^{(1),1,1}
  \Sigma_{N,S}^{1,1} g_{S,S}^{1,2} \Sigma_{S,N}^{2,2}
  G_{N,N}^{(1),2,2} ,
  \label{eq:truncation2}
\end{eqnarray}
where
\begin{eqnarray}
 \hat{G}^{(1),A}_{\alpha_k,\alpha_l}(\omega)&=& \sum_\mu
 \frac{\hat{\cal R}_{\alpha_k,\alpha_l}^{''(\mu)}}
      {\omega-\zeta''_\mu-i \eta''_\mu}
\end{eqnarray}
and
\begin{eqnarray}
 \hat{G}^{(1),R}_{\alpha_k,\alpha_l}(\omega)&=& \sum_\mu
 \frac{\hat{\cal R}_{\alpha_k,\alpha_l}^{''(\mu)}}
      {\omega-\zeta''_\mu+i\eta''_\mu}
\end{eqnarray}
are the Green's function in the absence of coupling
{to $N_B$,} and where $\zeta''_\mu$ and $\eta''_\mu$
are the corresponding ABS energies and line-width broadening. In the
large-gap approximation, we obtain the following expression for the
difference between Eq.~(\ref{eq:(A3)}) and Eq.~(\ref{eq:(C3)}):
\begin{eqnarray}
  \nonumber
  (\ref{eq:(A3)})-(\ref{eq:(C3)})&=&
  \mbox{Nambu-trace}\left\{\hat{\tau}_3^{1,1} \Sigma_{N,N_B}^{1,1}
  \left[g^{A,1,1}_{N_B,N_B}-g^{R,1,1}_{N_B,N_B}\right]\times\right.\\
  \label{eq:result1}
&&\left.  \Sigma_{N_B,N}^{1,1} G^{(1),1,1,R}_{N,N} \Sigma_{N,S}^{1,1}
  g^{1,2}_{S,S} \Sigma_{S,N}^{2,2}
  G^{(1),2,2,R}_{N,N}\times\right.\\&&\left. \Sigma_{N,N_B}^{2,2}
  g^{+,-,2,2}_{N_B,N_B} \Sigma_{N_B,N}^{2,2} G^{(1),2,2,A}_{N,N}
  \Sigma_{N,S}^{2,2} G^{(1),1,1,A}_{N,N}\right\}, \nonumber
\end{eqnarray}
and a similar equation holds for the difference between
Eq.~(\ref{eq:(B3)}) and Eq.~(\ref{eq:(D3)}). The remaining term given
by Eqs.~(\ref{eq:(A1)})-(\ref{eq:(A2)}),
Eqs.~(\ref{eq:(B1)})-(\ref{eq:(B2)}),
Eqs.~(\ref{eq:(C1)})-(\ref{eq:(C2)}) and
Eqs.~(\ref{eq:(D1)})-(\ref{eq:(D2)}) contribute to the current in the
``$1,1$'' electron-electron channel in response to the ``$1,1$''
component of the Keldysh Green's function $g^{+,-,1,1}$. Those
quasiparticle terms contribute to coupling the effective Fermi surface
in the central-$N$ to the bias voltage $V_B$ if the spectral density
of states in $N$ is nonvanishingly small.

{
\subsection{Unitary limit for elastic cotunneling}
\label{sec:unitary-EC}}

{In Eqs.~(\ref{eq:truncation1})-(\ref{eq:truncation2}), $G^{(1)}\sim
  \frac{1}{\eta}$ in the limit of small $\eta$ at resonance and the
  product $G^{1,1,R} G^{1,1,A} G^{2,2,R} G^{2,2,A}$ then scales like
  $\frac{1}{\eta^4}$. The products
  $\Sigma_{N,N_B}^{1,1}\left[g^{A,1,1}_{N_B,N_B}-g^{R,1,1}_{N_B,N_B}\right]
  \Sigma_{N_N,N}^{1,1}$ and $\Sigma_{N,N_B}^{2,2}
  g^{+,-,2,2}_{N_B,N_B} \Sigma_{N_B,N}^{2,2}$ in the numerator both
  behave like $\sim \eta$.  This makes Eq.~(\ref{eq:result1}) scale
  like $\sim \frac{1}{\eta^2}$ in the narrow energy window $\delta
  \omega \sim \eta$.  We conclude that the energy integral of
  Eq.~(\ref{eq:result1}) scales like $\sim \frac{\delta
    \omega}{\eta^2} \sim \frac{1}{\eta}$, which is diverging. However,
  Eq.~(\ref{eq:truncation1}) takes the following form at the next
  order:
\begin{eqnarray}
  \label{eq:truncation1-bis}
  G_{N,N}^{1,1}&\simeq& G_{N,N}^{(1),1,1}\\\nonumber &+&
  G_{N,N}^{(1),1,1} \Sigma_{N,S}^{1,1} g_{S,S}^{1,2}
  \Sigma_{S,N}^{2,2} G_{N,N}^{(1),2,2} \Sigma_{N,S}^{2,2}
  g_{S,S}^{2,1} \Sigma_{S,N}^{1,1}
  G_{N,N}^{(1),1,1}\\\nonumber
&+&G_{N,N}^{(1),1,1} \Sigma_{N,S}^{1,1} g_{S,S}^{1,2}
  \Sigma_{S,N}^{2,2} G_{N,N}^{(1),2,2} \Sigma_{N,S}^{2,2}
  g_{S,S}^{2,1} \Sigma_{S,N}^{1,1}\times\\\nonumber
&&  G_{N,N}^{(1),1,1}\Sigma_{N,S}^{1,1} g_{S,S}^{1,2}
  \Sigma_{S,N}^{2,2} G_{N,N}^{(1),2,2} \Sigma_{N,S}^{2,2}
  g_{S,S}^{2,1} \Sigma_{S,N}^{1,1}
  G_{N,N}^{(1),1,1}
  ,
\end{eqnarray}
from what we deduce that the fully dressed $G_{N,N}$ is obtained from
resuming a geometric series of the $G^{(1)}$ terms having the
following form:
\begin{eqnarray}
  G_{N,N}^{1,1}&\simeq& \frac{1}{\eta}\left[\lambda_0+\frac{\lambda_1}{\eta^2} + \frac{\lambda_2}{\eta^4} + ...\right]
  ,
\end{eqnarray}
where the $\lambda_n$ scale like $\lambda_n\sim \eta^0$. The entire
series for $G_{N,N}^{1,1}$ resums as a number of order $\eta$. We
conclude that the physical differential conductance is vanishingly
small if the} {real part of the energy exactly
  matches that of one of the resonances.}

{{\it Physical remark:} At this point, we refer to the
  Blonder-Tinkham-Klapwijk (BTK) model \cite{BTK} of a
  normal-superconducting weak link at energies close to the
  superconducting gap $E\approx \Delta$. A singularity is then
  obtained at $E=\Delta$ at the lowest order in a large-$Z$ expansion
  of the conductance, where $Z$ parameterizes the barrier at the
  interface (the larger $Z$, the smaller current). However, BTK
  calculate the differential conductance to all orders in $1/Z$, and the
  differential conductance is unitary at the gap edge $E=\Delta$, with
  a sharp maximum if $Z\agt 1$.} {Regarding the considered
  three-terminal $(S_a,S_b,N_B)$ device, we conclude that this analogy
  with the BTK solution shows that the unitary limit is obtained at
  energies that are slightly detuned from the real part of the
  resonance energies in $G^{(1)}$.}

\begin{figure}[htb]
  \includegraphics[width=\columnwidth]{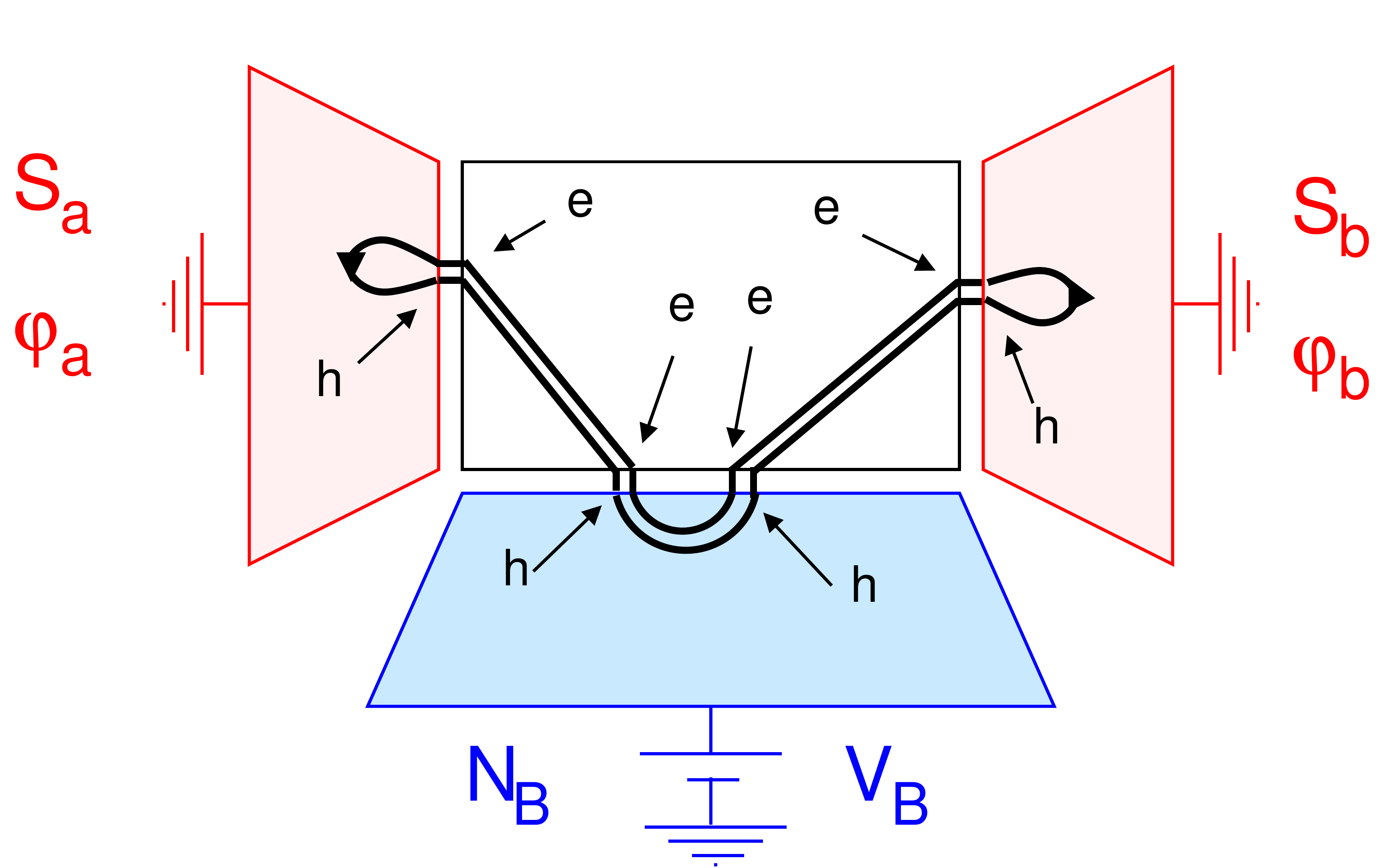}
  \caption{{{The lowest-order elastic cotunneling diagram.}  This
      higher-order Keldysh diagram is of order four in the tunneling
      amplitude between the central-$N$ and the infinite bottom
      $N_B$. The corresponding differential conductance scales like
      $(T_N)^2$ in the limit of a small tunable interface transparency
      $T_N$ between the central-$N$ and the bottom normal lead $N_B$.}
\label{fig:QPC}
}
\end{figure}

{
  \subsection{Can we distinguish between elastic cotunneling and
    sequential tunneling?}\label{sec:distinction}}

{The conjectured emergence of the unitary limit for both sequential
  tunneling and elastic cotunneling, see the above
  subsections~\ref{sec:unitary-ST} and~\ref{sec:unitary-EC},
  respectively, suggests that both channels yield similar signatures
  in the conductance $dI_B/dV_B$ plotted as a function of the bias
  voltage, namely, they both yield resonances corresponding to maxima
  in $dI_B/dV_B$ at the energies of the spectrum of a long Josephson
  junction \cite{Kulik,Ishii,Bagwell}. A difficulty is then to
  experimentally provide a way to tell whether the experimental signal
  is mostly dominated by the sequential tunneling or
  {by} the elastic cotunneling channel.}

{Let us now consider an experiment that could possibly answer this
  question, for a metallic device having dimensions that are large
  compared to the Fermi wave-length} {$\lambda_F$}{. Namely, we
  envision tuning by a quantum point contact (QPC) the transparency of
  the contact with $N_B$. The device differential conductance is
  envisioned to be plotted as a function of the transparency of this
  tunable $N$-QPC-$N_B$ contact.}

{The} {differential conductance $d
  I_B/dV_B$ would scale like $dI_B/dV_B\sim T_N$ if sequential
  tunneling would dominate. Conversely, the dominant sequential
  tunneling process would result in $dI_B/dV_B\sim\left(T_N\right)^2$,
  see lowest-order diagram of the process in Fig.~\ref{fig:QPC}, which
  intersects four times the interface with $N_B$. In those
  expressions, the notation $T_N$ refers to the}
{small or moderately small}
{normal-state transmission coefficient of the QPC
  making the $N$-$N_B$ interface.}

{{\it Conclusion: } The scaling of the elastic
  cotunneling and the sequential tunneling channels with $T_N$
  suggests that sequential tunneling dominates over elastic
  cotunneling if the device dimensions are large compared to the Fermi
  wave-length} {$\lambda_F$}{, and if
  the contacts contain a number of channels that is sufficiently large
  to ensure averaging over the oscillations at the scale of the Fermi
  wave-length} {$\lambda_F$}{. In
  addition, the range of small to intermediate values of the interface
  transparencies is experimentally relevant.}

{In the next section~\ref{sec:num-dot}, we combine the calculations of
  sections~\ref{sec:lowest-order} and~\ref{sec:intrinsic1} to propose
  the physically-motivated approximations that yield simple
  expressions for the dimensionless conductance, which will be used to
  produce the figures.}
  
\begin{figure*}[htb]
  \begin{minipage}{.9\textwidth}
    \includegraphics[width=.49\textwidth]{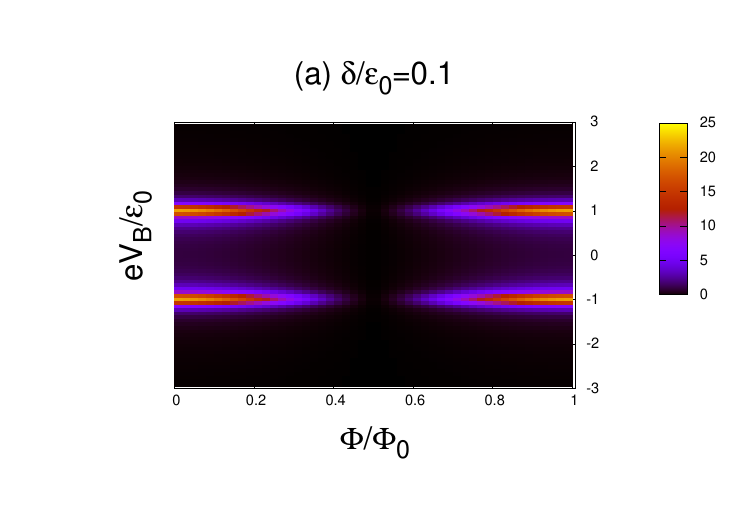}
    \includegraphics[width=.49\textwidth]{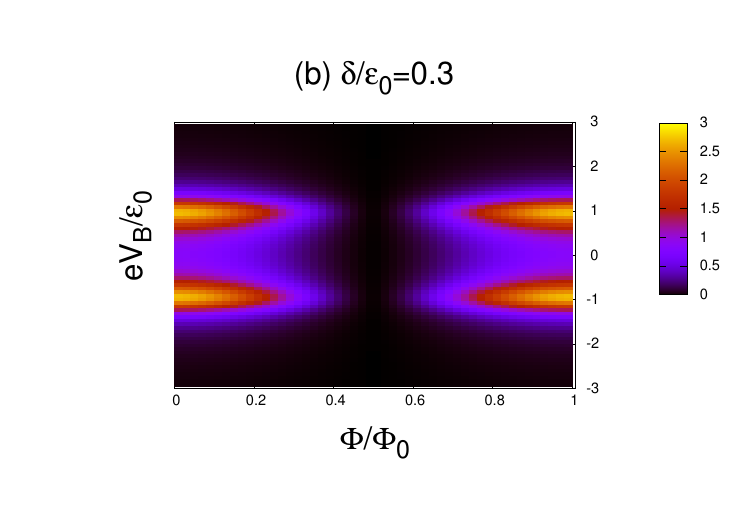} \end{minipage}
  
  \begin{minipage}{.9\textwidth}
    \includegraphics[width=.49\textwidth]{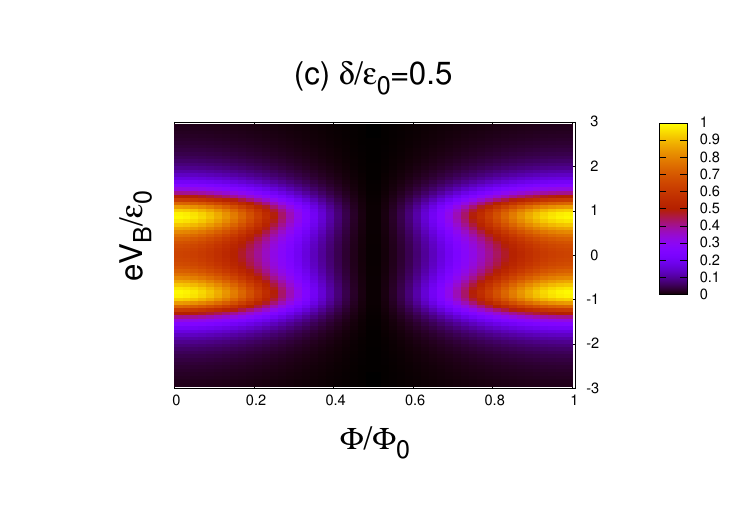}
    \includegraphics[width=.49\textwidth]{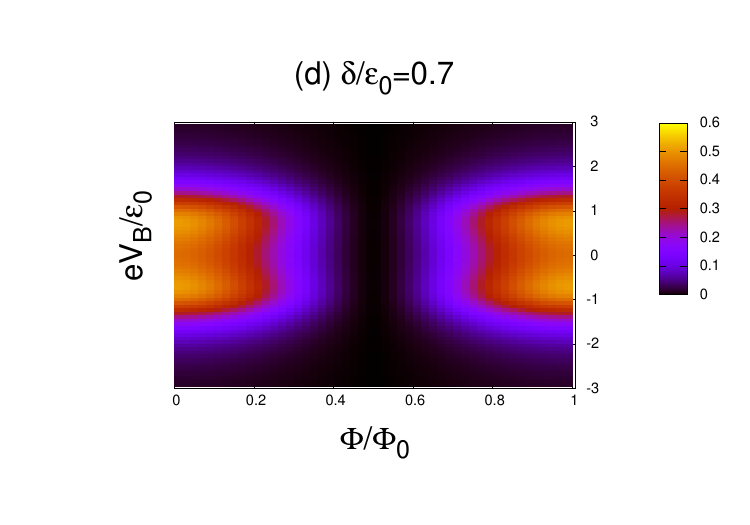} \end{minipage}
  
  \caption{{{The dimensionless differential conductance
        $g'_0\left(\frac{\Phi}{\Phi_0}, \frac{eV_B}{\epsilon_0},
        \frac{\delta}{\epsilon_0}\right)$ of a quantum dot connected
        to two superconducting and to a normal lead in the Andreev
        limit, see Eq.~(\ref{eq:gp0}).} The contact with the normal
      lead $N_B$ has transparency that is much larger than with the
      superconducting leads $S_a$ and $S_b$. The real parts of the two
      ABS energies are given by $\pm \epsilon_0$ for all
      $\frac{\Phi}{\Phi_0}$. The dimensionless differential
      conductance is shown in color in the plane of the parameters
      $\left(\frac{\Phi}{\Phi_0},\frac{eV_B}{\epsilon_0}\right)$. The
      four panels correspond to the following values of the
      dimensionless line-width broadening:
      $\frac{\delta}{\epsilon_0}=0.1$ (panel a),
      $\frac{\delta}{\epsilon_0}=0.3$ (panel b),
      $\frac{\delta}{\epsilon_0}=0.5$ (panel c),
      $\frac{\delta}{\epsilon_0}=0.7$ (panel d). The
      $\frac{\Phi}{\Phi_0}$-sensitivity of
      $g'_0\left(\frac{\Phi}{\Phi_0}, \frac{eV_B}{\epsilon_0},
      \frac{\delta}{\epsilon_0}\right)$ obtained on all panels
      illustrates the interplay between the destructive interference
      at $\frac{\Phi}{\Phi_0}=\frac{1}{2}$, and the
      $\frac{\Phi}{\Phi_0}$-sensitive effect of the line-width
      broadening.}
\label{fig:dIB/dVB-dot-large-transparency}
}
\end{figure*}

\begin{figure*}[htb]
  \begin{minipage}{.9\textwidth}
    \includegraphics[width=.49\textwidth]{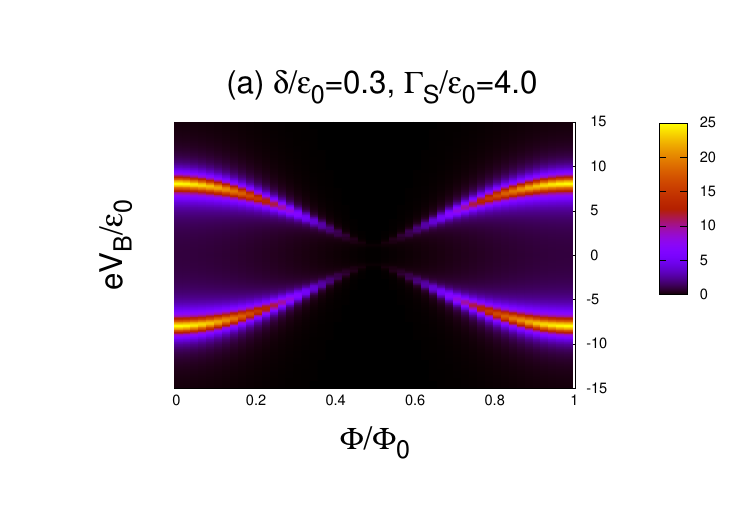}
    \includegraphics[width=.49\textwidth]{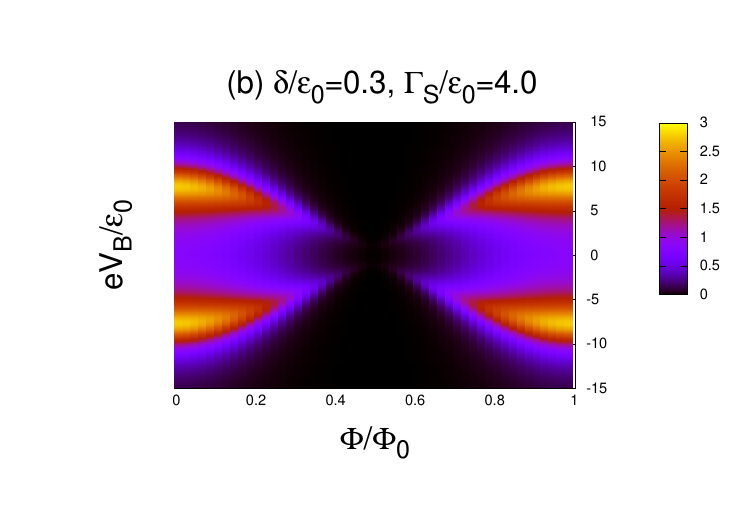} \end{minipage}
  
  \begin{minipage}{.9\textwidth}
    \includegraphics[width=.49\textwidth]{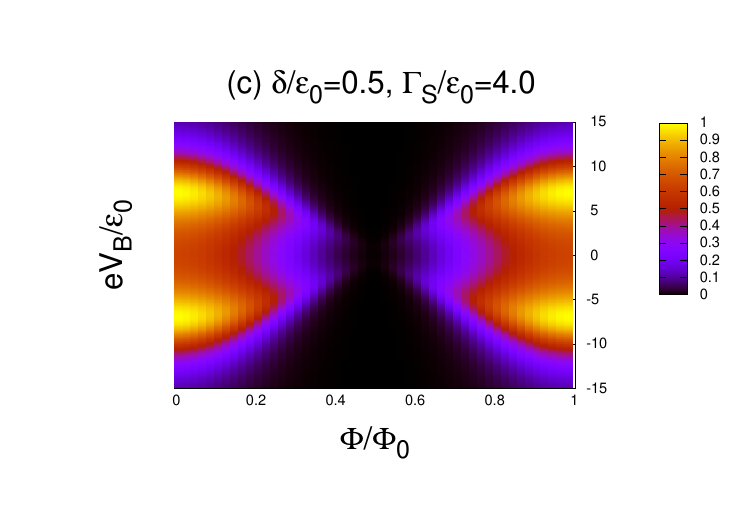}
    \includegraphics[width=.49\textwidth]{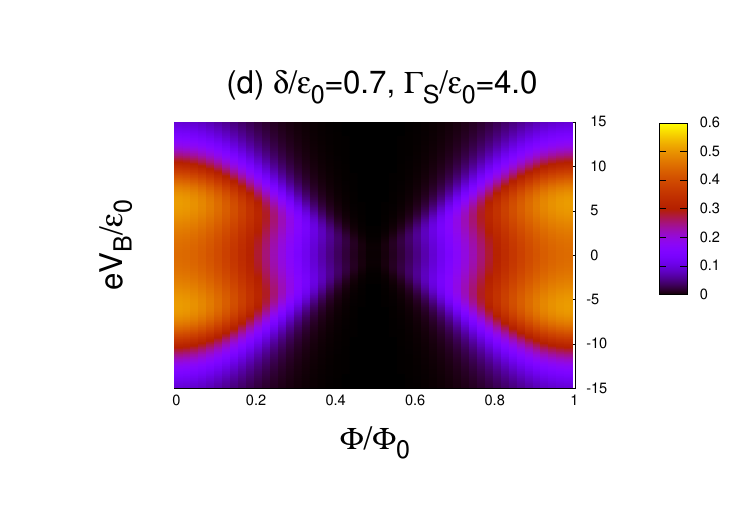} \end{minipage}
  
  \caption{{The physically-approximated dimensionless
      differential conductance $g''_0\left(\frac{\Phi}{\Phi_0},
      \frac{eV_B}{\epsilon_0}, \frac{\Gamma_S}{\epsilon_0},
      \frac{\delta}{\epsilon_0}\right)$ of a quantum dot connected to
      two superconducting and a normal lead in the Andreev limit, see
      Eq.~(\ref{eq:gs0}). As in the above
      Fig.~\ref{fig:dIB/dVB-dot-large-transparency}, the four panels
      correspond to the following values of the dimensionless
      line-width broadening: $\frac{\delta}{\epsilon_0}=0.1$ (panel
      a), $\frac{\delta}{\epsilon_0}=0.3$ (panel b),
      $\frac{\delta}{\epsilon_0}=0.5$ (panel c),
      $\frac{\delta}{\epsilon_0}=0.7$ (panel d). The finite line-width
      broadening can have a drastic effect in smearing the features of
      the ABS energy-phase dispersions.
\label{fig:dIB/dVB-dot-large-transparency-2}
}}
\end{figure*}

{
  \section{Physically-motivated approximations of transport theory}
  \label{sec:num-dot}}

{Now, we start the third step in our theoretical
  framework, namely, we provide the physically-motivated approximations
  for the conductance $d I_B/d V_B$ that are used to produce the
  numerical results. We start the discussion in
  subsection~\ref{sec:example-0D} with the simple example of a 0D
  quantum dot, and next proceed in
  subsection~\ref{sec:pheno-full-ladders} with an Andreev
  interferometer in the long-junction limit
  \cite{Kulik,Ishii,Bagwell}.}

{
  \subsection{Physically-motivated approximations for a 0D quantum dot}
  \label{sec:example-0D}}

{ In order to present the approximations that will later be used in
  subsection~\ref{sec:pheno-full-ladders} for a long Josephson
  junction \cite{Kulik,Ishii,Bagwell}, we here start with the simple
  example of a 0D quantum dot, see Fig.~\ref{fig:thedevice}f. This
  simple 0D geometry yields a pair of ABS at opposite energies. In
  addition, we assume that the transparency is large for the interface
  with the infinite normal lead $N_B$, and small for the interface
  with the infinite superconductors $S_a$ and $S_b$. We coin this
  approximation as the {\it Andreev limit}, as opposed to the {\it
    Josephson limit} in the case of highly transparent interfaces with
  the superconductors $S_a$ and $S_b$, and tunnel interfaces with the
  normal lead $N_B$. We will next proceed with physically-motivated
  approximations for arbitrary interface transparencies.}

{Analytical calculations for this simple device in the
  Andreev limit are presented in Appendix~\ref{sec:0D}, leading to the
  following compact expression for the differential conductance $dI_B/dV_B$:
  \begin{eqnarray}
 \label{eq:final}
\frac{dI_B}{dV_B} &\simeq&8\pi^2 \gamma_B^2 \times\\ &&
\frac{\left|\Gamma_a e^{i\varphi_a} + \Gamma_b e^{i\varphi_b}
  \right|^2}{\left[\left(eV_B-\epsilon_0\right)^2+
    \left(\gamma_B\right)^2\right]\left[\left(eV_B+\epsilon_0\right)^2+
    \left(\gamma_B\right)^2\right]}
,
\nonumber
  \end{eqnarray}
  which illustrates the above two-particle resonance in
  Eq.~(\ref{eq:res-final1}) for $\Omega_\lambda=\epsilon_0$ and
  $\Omega_{\lambda'}=-\epsilon_0$. In Eq.~(\ref{eq:final}),
  $\gamma_B=(t_B)^2/W$ parameterizes the transparency of the interface
  with $N_B$, with $t_B$ the corresponding hopping amplitude and $W$
  the band-width in the leads. The transparencies of the interfaces with
  the left and right superconductors are parameterized by
  $\Gamma_{a,b}=(t_{a,b})^2/W$, where $t_{a,b}$ are the corresponding
  hopping amplitudes. The variable $\epsilon_0$ is the on-site energy
  on the quantum dot.}

\begin{figure*}[htb]
  \begin{minipage}{.9\textwidth}
    \includegraphics[width=.49\textwidth]{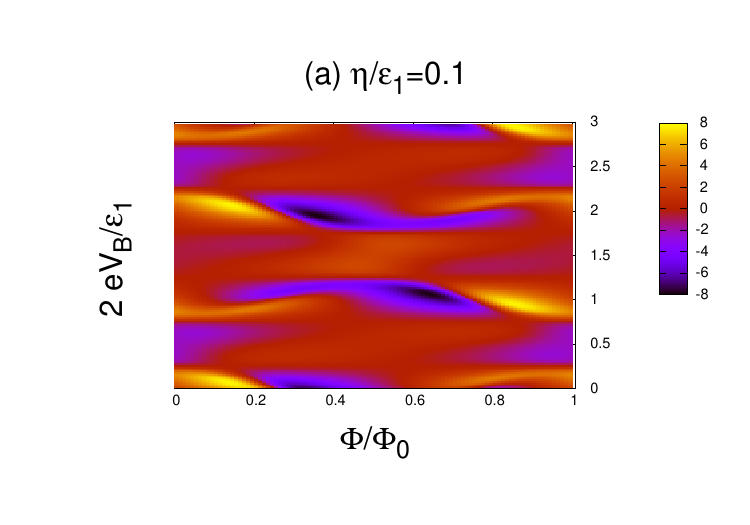}
    \includegraphics[width=.49\textwidth]{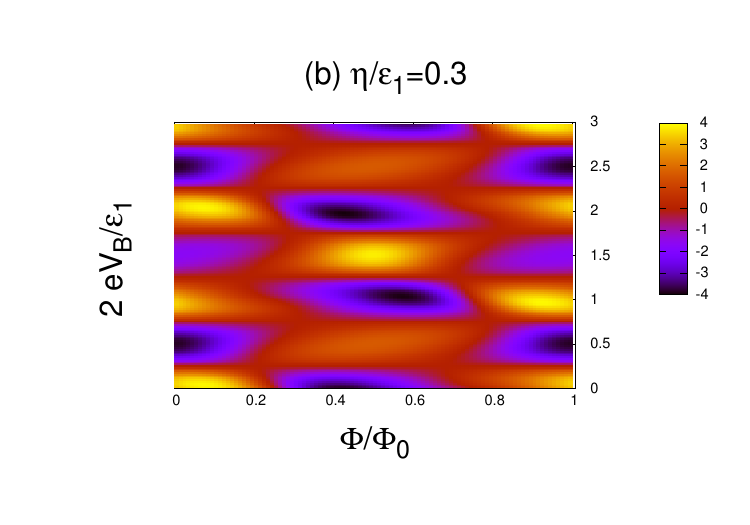} \end{minipage}
  
  \begin{minipage}{.9\textwidth}
    \includegraphics[width=.49\textwidth]{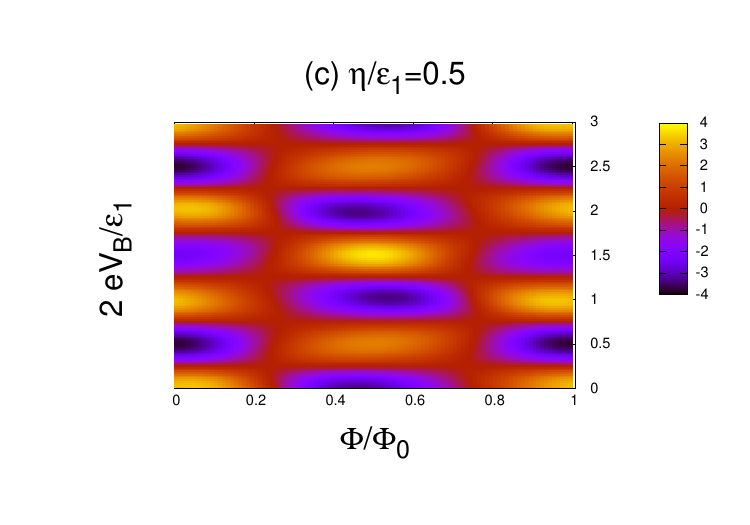}
    \includegraphics[width=.49\textwidth]{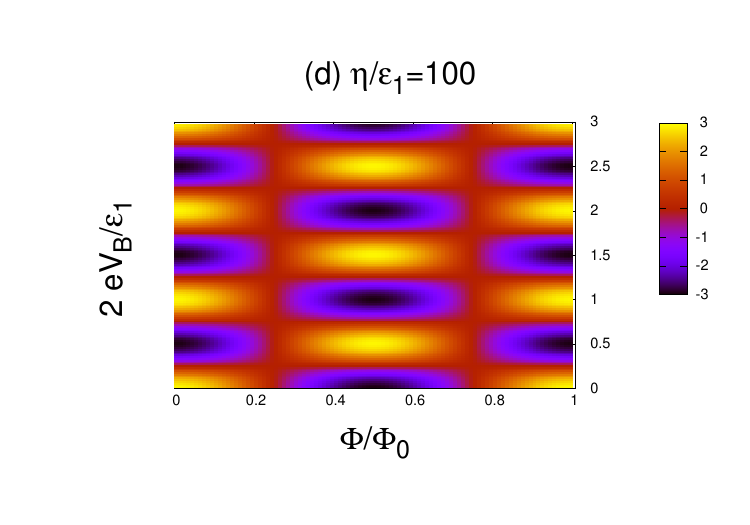} \end{minipage}
  
  \caption{{The physically-approximated dimensionless differential
      conductance $g'''_0\left(\frac{\Phi}{\Phi_0},
      \frac{eV_B}{\epsilon_1}, \frac{E_J}{\epsilon_1},
      \frac{\eta}{\epsilon_1}\right)$, see Eq.~(\ref{eq:gt0}). The
      four panels correspond to the following values of the
      dimensionless line-width broadening:
      $\frac{\eta}{\epsilon_1}=0.1$ (panel a),
      $\frac{\eta}{\epsilon_1}=0.3$ (panel b),
      $\frac{\eta}{\epsilon_1}=0.5$ (panel c),
      $\frac{\eta}{\epsilon_1}=100$ (panel d). Increasing the
      line-width broadening from $\frac{\eta}{\epsilon_1}=0.1$ on
      panel a to $\frac{\eta}{\epsilon_1}=100$ on panel d produces a
      cross-over from narrow ABS (panel a) to a regular checkerboard
      pattern similar to Fig.~\ref{fig:colorplot}. The dimensionless
      Josephson energy between the left and right superconducting
      leads is $\frac{E_J}{\epsilon_1}=0.2$.
\label{fig:gt0}
}}
\end{figure*}

{{\it Physical remarks: } (i) We find two resonances
  if $eV_B\simeq \pm \epsilon_0$, which coincide with the real parts
  of the ABS energies in the limit of small $\Gamma_a$ and $\Gamma_b$,
  see also Eq.~(\ref{eq:ABS-ener}). (ii) We note that, in
  Eq.~(\ref{eq:final}), the transparency of the interface with $N_B$ is
  parameterized by $ \gamma_B^2=t_B^4/W^2$. The scaling $dI_B/d
  V_B\sim(t_B)^4$ is typical of Andreev reflection. (iii) The Keldysh
  diagram of the lowest-order elastic cotunneling process also scales
  like $dI_B/dV_B\sim(t_B)^4$, see Fig.~\ref{fig:QPC} discussed in the
  above section~\ref{sec:distinction}.  (iv) In addition, $dI_B/dV_B$
  in Eq.~(\ref{eq:final}) features in the denominator a product of two
  terms having resonances at $e V_B\simeq \pm \epsilon_0$, which is
  compatible with the idea of the two-particle resonance.}

{We deduce from Eq.~(\ref{eq:final}) the following
  expression for the dimensionless differential conductance
  $g'_0\left(\frac{\Phi}{\Phi_0}, \frac{eV_B}{\epsilon_0},
  \frac{\delta}{\epsilon_0}\right)$ to which $dI_B/dV_B$ in
  Eq.~(\ref{eq:final}) is proportional:} {
\begin{eqnarray}
  \label{eq:gp0}
&&  g'_0\left(\frac{\Phi}{\Phi_0}, \frac{eV_B}{\epsilon_0}, \frac{\delta}{\epsilon_0}\right)
  =\\&&
  \frac{\cos^2\left(\frac{\pi\Phi}{\Phi_0}\right)}{\left[\left(\frac{eV_B}{\epsilon_0}-1\right)^2
      + \left(\frac{\delta}{\epsilon_0}\right)^2\right]
    \left[\left(\frac{eV_B}{\epsilon_0}+1\right)^2 +
      \left(\frac{\delta}{\epsilon_0}\right)^2\right]} .
  \nonumber
\end{eqnarray}}
{The colorplots in Fig.~\ref{fig:dIB/dVB-dot-large-transparency} show
  $g'_0\left(\frac{\Phi}{\Phi_0}, \frac{eV_B}{\epsilon_0},
  \frac{\delta}{\epsilon_0}\right)$ given by Eq.~(\ref{eq:gp0}) in the
  plane of the parameters $\frac{\Phi}{\Phi_0}$ and $\frac{e
    V_B}{\epsilon_0}$. The four panels in
  Fig.~\ref{fig:dIB/dVB-dot-large-transparency} show the four
  different values $\frac{\delta}{\epsilon_0}=0.1$ (panel a),
  $\frac{\delta}{\epsilon_0}=0.3$ (panel b),
  $\frac{\delta}{\epsilon_0}=0.5$ (panel cc) and
  $\frac{\delta}{\epsilon_0}=0.7$ (panel d). Eq.~(\ref{eq:gp0}) and
  Fig.~\ref{fig:dIB/dVB-dot-large-transparency} reveal a strong
  $\frac{\Phi}{\Phi_0}$-sensitivity of $g'_0\left(\frac{\Phi}{\Phi_0},
  \frac{eV_B}{\epsilon_0}\right)$ which is due to the interplay
  between the destructive interference at
  $\frac{\Phi}{\Phi_0}=\frac{1}{2}$, and the
  $\frac{\Phi}{\Phi_0}$-sensitive effect of the line-width
  broadening.}

{However, in the Andreev limit, the real part of the
  resonance energies does not disperse with $\frac{\Phi}{\Phi_0}$, see
  the above Eq.~(\ref{eq:gp0}). This is why we now address a regime of
  intermediate interface transparencies, and, as a physically-motivated
  approximation, we replace $\epsilon_0$ in Eq.~(\ref{eq:gp0}) by the
  full expression of the ABS energy given by Eq.~(\ref{eq:ABS-ener}):}
{
  \begin{widetext}
  \begin{eqnarray}
  \label{eq:gs0}
  g''_0\left(\frac{\Phi}{\Phi_0}, \frac{eV_B}{\epsilon_0},
  \frac{\Gamma}{\epsilon_0}\frac{\delta}{\epsilon_0}\right)
  = \frac{\cos^2\left(\frac{\pi\Phi}{\Phi_0}\right)}{\left[\left(\frac{eV_B}{\epsilon_0
        F\left(\frac{\Gamma_S}{\epsilon_0},\frac{\Phi}{\Phi_0}\right)}-1\right)^2
      + \left(\frac{\delta}{\epsilon_0}\right)^2\right]
    \left[\left(\frac{eV_B}{\epsilon_0
        F\left(\frac{\Gamma_S}{\epsilon_0},\frac{\Phi}{\Phi_0}\right)}+1\right)^2
        + \left(\frac{\delta}{\epsilon_0}\right)^2\right]} ,
  \end{eqnarray}
  \end{widetext}
with
\begin{equation}
  \label{eq:F}
  F\left(\frac{\Gamma_S}{\epsilon_0},\frac{\Phi}{\Phi_0}\right)
  = \sqrt{1+4 \left(\frac{\Gamma_S}{\epsilon_0}\right)^2
    \cos^2\left(\frac{\pi\Phi}{\Phi_0}\right)}
  ,
\end{equation}
where we assumed that both superconductors are symmetrically coupled
by the hopping $t_S$ to the quantum dot, with the parameter
$\Gamma_S=(t_S)^2/W$ for the corresponding interface transparency.}

{Fig.~\ref{fig:dIB/dVB-dot-large-transparency-2} shows
  the colorplots of the conductance deduced from the expression of
  $g''_0\left(\frac{\Phi}{\Phi_0}, \frac{eV_B}{\epsilon_0},
  \frac{\Gamma}{\epsilon_0}\frac{\delta}{\epsilon_0}\right)$ given by
  Eqs.~(\ref{eq:gs0})-(\ref{eq:F}). We find that increasing the
  line-width broadening according to $\frac{\delta}{\epsilon_0}=0.1$
  (panel a), $\frac{\delta}{\epsilon_0}=0.3$ (panel b),
  $\frac{\delta}{\epsilon_0}=0.5$ (panel c),
  $\frac{\delta}{\epsilon_0}=0.7$ (panel d) drastically broadens the
  narrow energy-phase dispersion of the ABS that is obtained for
  $\frac{\delta}{\epsilon_0}=0.1$ in
  Fig.~\ref{fig:dIB/dVB-dot-large-transparency-2}a.}

  \subsection{Physically-motivated approximations
    for the long-junction limit}
  \label{sec:pheno-full-ladders}
  In this subsection, we {use the same strategy as in
    the above subsection~\ref{sec:example-0D} and} calculate an
  approximation to the conductance $dI_B/d V_B$ {of a
    $(S_a,S_b,N_B)$ Andreev interferometer. The fully dressed advanced
    Green's function} {has the following spectral representation, see
    Eqs.~(\ref{eq:residu-debut})-(\ref{eq:residu-fin}):
\begin{equation}
  G_{\alpha,\beta}^{A,1,1}=
  \sum_{n_x,n_y} \langle\alpha|n_x,n_y\rangle
  \frac{1}{\omega-\Omega_{n_x,n_y}-i\eta}
  \langle n_x,n_y|\beta\rangle
  ,
\end{equation}
where $n_x$ and $n_y$ label the standing waves in the $x$ and $y$
directions respectively.  We note that, in the wave-guide geometry of
Ref.~\onlinecite{Kulik}, $\Omega_{n_x,n_y}\equiv \Omega_{n_x}$ is
independent on the transverse quantum number $n_y$. This leads to the
following 1D form of the advanced Green's function:
\begin{equation}
  \label{eq:G-albe-1D}
  G_{\alpha,\beta}^{A,1,1}=
  \sum_{n_x} \langle \alpha_x|n_x\rangle
  \frac{1}{\omega-\Omega_{n_x}-i\eta} \langle n_x|\beta_x\rangle
  .
\end{equation}}

{We deduce from Eqs.~(\ref{eq:I1-A1}),
  (\ref{eq:Ia-1-1}) and~(\ref{eq:Ia-1-1-bis}) the following
  physically-motivated approximation for the conductance $d I_B/d
  V_B$:
\begin{eqnarray}
 \label{eq:I1-A1-approx}
 && \tilde{I}_1(\{\mu_{N,q}\},\{Z_q\},\frac{
   \Phi}{\Phi_0},\omega)\simeq\\ \nonumber && 2 \frac{t_a^2
   t_b^2}{W^2}e^{\frac{2i\pi\Phi}{\Phi_0}}
 \left[\tilde{g}^{R,2,2}_{\alpha,\beta}(\omega)\delta\hat{g}^{+,-,1,1}_{\beta,\alpha}(\omega)\right.\\
   &&\left.\nonumber
   + \delta
   \hat{g}^{+,-,2,2}_{\alpha,\beta}(\omega)\tilde{g}_{\beta,\alpha}^{A,1,1}(\omega)
   \right]\\ \nonumber &-& 2 \frac{t_a^2
   t_b^2}{W^2}e^{-\frac{2i\pi\Phi}{\Phi_0}}
 \left[\tilde{g}^{R,1,1}_{\alpha,\beta}(\omega)\delta\hat{g}^{+,-,2,2}_{\beta,\alpha}(\omega)\right.\\
   \nonumber
&&   + \left.\delta
   \hat{g}^{+,-,1,1}_{\alpha,\beta}(\omega)\tilde{g}_{\beta,\alpha}^{A,2,2}(\omega)
   \right],
\end{eqnarray}
  where the $\hat{g}$s in Eq.~(\ref{eq:I1-A1}) have been dressed in
  the $\tilde{g}$s by the poles on the spectrum of a long Josephson
  junction in Eq.~(\ref{eq:I1-A1-approx}), see also the above
  Eq.~(\ref{eq:G-albe-1D}):
  \begin{eqnarray}
    \label{eq:tilde-g-1}
    \tilde{g}^{R,2,2}_{\alpha,\beta}(\omega)&=& \sum_{n_x} {}_h\langle
    \alpha_x|n_x\rangle \frac{1}{\omega+\Omega_{n_x}+i\eta} \langle
    n_x|\beta_x\rangle_h\\
\label{eq:tilde-g-1-bis}
    \tilde{g}^{A,1,1}_{\beta,\alpha}(\omega)&=&
    \sum_{n_x} {}_e\langle \beta_x|n_x\rangle
    \frac{1}{\omega-\Omega_{n_x}-i\eta} \langle
    n_x|\alpha_x\rangle_e\\
\label{eq:tilde-g-2-bis}
    \tilde{g}^{R,1,1}_{\alpha,\beta}(\omega)&=&
    \sum_{n_x} {}_e\langle \alpha_x|n_x\rangle
    \frac{1}{\omega-\Omega_{n_x}+i\eta} \langle
    n_x|\beta_x\rangle_e\\ \tilde{g}^{A,2,2}_{\beta,\alpha}(\omega)&=&
    \sum_{n_x} {}_h\langle \beta_x|n_x\rangle
    \frac{1}{\omega+\Omega_{n_x}-i\eta} \langle
    n_x|\alpha_x\rangle_h .
    \label{eq:tilde-g-2}
\end{eqnarray}
In those Eqs.~(\ref{eq:tilde-g-1})-(\ref{eq:tilde-g-2}), the
subscripts ``e'' and ``h'' refer to electrons and holes respectively.}

{We next expand
  Eqs.~(\ref{eq:tilde-g-1})-(\ref{eq:tilde-g-2}) and average out the
  fast oscillations over the Fermi wave-length. Specifically, we obtain
  \begin{widetext}
  \begin{eqnarray}
    \label{eq:TOTOA}
    (\ref{eq:tilde-g-1})&=&\frac{4iZ_p}{\sqrt{k_F R_{\alpha,\beta}}}
    \frac{t_a^2 t_b^2}{W^3} \exp\left(\frac{2i\pi\Phi}{\Phi_0}\right)
      \sum_{n_x} \frac{{}_h\langle
    \alpha_x|n_x\rangle \langle
    n_x|\beta_x\rangle_h
    \cos\left(k_e R_{\alpha,\beta}-\frac{\pi}{4}\right)}{\mu_{N_p}
        +\Omega_{n_x} + i \eta}\\
(\ref{eq:tilde-g-1-bis})&=&\frac{4iZ_p}{\sqrt{k_F R_{\alpha,\beta}}}
      \frac{t_a^2 t_b^2}{W^3} \exp\left(\frac{2i\pi\Phi}{\Phi_0}\right)
      \sum_{n_x} \frac{{}_e\langle
    \alpha_x|n_x\rangle \langle
    n_x|\beta_x\rangle_e
    \cos\left(k_h R_{\alpha,\beta}-\frac{\pi}{4}\right)}{\mu_{N_p}
        +\Omega_{n_x} + i \eta}\\
      (\ref{eq:tilde-g-2-bis})&=&-\frac{4iZ_p}{\sqrt{k_F R_{\alpha,\beta}}}
      \frac{t_a^2 t_b^2}{W^3} \exp\left(-\frac{2i\pi\Phi}{\Phi_0}\right)
      \sum_{n_x} \frac{{}_e\langle
    \alpha_x|n_x\rangle \langle
    n_x|\beta_x\rangle_e
    \cos\left(k_h R_{\alpha,\beta}-\frac{\pi}{4}\right)}{\mu_{N_p}
        +\Omega_{n_x} - i \eta}\\
      (\ref{eq:tilde-g-2})&=&-\frac{4iZ_p}{\sqrt{k_F R_{\alpha,\beta}}}
      \frac{t_a^2 t_b^2}{W^3} \exp\left(-\frac{2i\pi\Phi}{\Phi_0}\right)
      \sum_{n_x} \frac{{}_h\langle
    \alpha_x|n_x\rangle \langle
    n_x|\beta_x\rangle_h
    \cos\left(k_e R_{\alpha,\beta}-\frac{\pi}{4}\right)}{\mu_{N_p}
        +\Omega_{n_x} - i \eta}
      .
      \label{eq:TOTOB}
  \end{eqnarray}
We deduce the following expression for the current:
\begin{eqnarray}
  \nonumber
  \tilde{I}_1(\{\mu_{N,q}\},\{Z_q\},\frac{
   \Phi}{\Phi_0},\omega)&\simeq&
 -\frac{8 Z_p}{\sqrt{k_F R_{\alpha,\beta}}} \frac{t_a^2 t_b^2}{W^3}
\sum_{n_x} \left[{}_h\langle
    \alpha_x|n_x\rangle \langle
    n_x|\beta_x\rangle_h
    \cos\left(k_e R_{\alpha,\beta}-\frac{\pi}{4}\right)
    +
    {}_e\langle
    \alpha_x|n_x\rangle \langle
    n_x|\beta_x\rangle_e
    \cos\left(k_h R_{\alpha,\beta}-\frac{\pi}{4}\right)\right]\\
&&\times
\mbox{Im}\left[\frac{\exp\left(\frac{2i\pi\Phi}{\Phi_0}\right)}
  {\mu_{N_p}+\Omega_{n_x}+i\eta}\right]
.
 \label{eq:I1-A1-approx2}
\end{eqnarray}
In order to obtain simple formula that capture the qualitative
behavior, we replace $n_x\ge 1 $ by the wave-vector $k_x=2\pi
n_x'/L_x$, with $n_x'$ a positive or negative integer. Then, we find
\begin{eqnarray}
  && \left({}_h\langle \alpha_x|n'_x\rangle \langle n'_x|\beta_x\rangle_h
  \cos\left(k_e R_{\alpha,\beta}-\frac{\pi}{4}\right) + {}_e\langle
  \alpha_x|n'_x\rangle \langle n'_x|\beta_x\rangle_e \cos\left(k_h
  R_{\alpha,\beta}-\frac{\pi}{4}\right)\right) + \left(n'_x \rightarrow
  -n'_x\right)\\ &=& \frac{1}{R_{\alpha,\beta}}\left\{
  \exp\left[-i\left(k_F-\frac{\delta \omega}{\hbar
      v_F}\right)\left(x_\beta-x_\alpha\right)\right]
  \cos\left[\left(k_F+\frac{\delta\omega}{\hbar v_F}\right)
    R_{\alpha,\beta}-\frac{\pi}{4}\right]\right.
\nonumber
  \\&&+
  \exp\left[i\left(k_F+\frac{\delta \omega}{\hbar
      v_F}\right)\left(x_\beta-x_\alpha\right)\right]
  \cos\left[\left(k_F-\frac{\delta\omega}{\hbar v_F}\right)
    R_{\alpha,\beta}-\frac{\pi}{4}\right]
\nonumber
  \\&&+
  \exp\left[i\left(k_F-\frac{\delta \omega}{\hbar
      v_F}\right)\left(x_\beta-x_\alpha\right)\right]
  \cos\left[\left(k_F+\frac{\delta\omega}{\hbar v_F}\right)
    R_{\alpha,\beta}+\frac{\pi}{4}\right]
\nonumber
  \\&&+\left.
  \exp\left[-i\left(k_F+\frac{\delta \omega}{\hbar
      v_F}\right)\left(x_\beta-x_\alpha\right)\right]
  \cos\left[\left(k_F-\frac{\delta\omega}{\hbar v_F}\right)
    R_{\alpha,\beta}+\frac{\pi}{4}\right] \right\} .
  \nonumber
\end{eqnarray}
  Averaging over the fast oscillations $k_F R_{\alpha,\beta}$ leads to
  \begin{eqnarray}
  && \langle \langle \left({}_h\langle \alpha_x|n'_x\rangle \langle n'_x|\beta_x\rangle_h
  \cos\left(k_e R_{\alpha,\beta}-\frac{\pi}{4}\right) + {}_e\langle
  \alpha_x|n'_x\rangle \langle n'_x|\beta_x\rangle_e \cos\left(k_h
  R_{\alpha,\beta}-\frac{\pi}{4}\right)\right) + \left(n'_x \rightarrow
  -n'_x\right) \rangle \rangle\\&&= \frac{\sqrt{2}}{R_{\alpha,\beta}}
  \cos\left[\frac{2\delta \omega R_{\alpha,\beta}}{\hbar v_F}\right]
  \nonumber
  .
  \end{eqnarray}
We obtain the following expression for the partial derivative of the
current with respect to $\mu_{N,p}$:
  \begin{eqnarray}
    \frac{\partial\tilde{I}\left(\{\mu_{N,q}\},\{Z_q\},\frac{\Phi}{\Phi_0}\right)}
         {\partial\mu_{N,p}}=-
    \frac{2\sqrt{2}Z_p}{(k_F R_{\alpha,\beta})^{3/2}} \frac{t_a^2
      t_b^2}{W^3} \cos\left[\frac{2\delta\omega
        R_{\alpha,\beta}}{\hbar v_F} \right] \sum_{n_x}
    \mbox{Im}\left[\frac{\exp\left(\frac{2i\pi \Phi}{\Phi_0}\right)}
       {\mu_{N_p}+\Omega_{n_x}+i\eta}\right] .
  \end{eqnarray}
  We deduce the following expression for the dimensionless differential
  conductance $g'''_0\left(\frac{\Phi}{\Phi_0},
  \frac{eV_B}{\epsilon_1}, \frac{E_J}{\epsilon_1},
  \frac{\eta}{\epsilon_1}\right)$:
  \begin{eqnarray}
    \label{eq:gt0}
    g'''_0\left(\frac{\Phi}{\Phi_0}, \frac{eV_B}{\epsilon_1},
    \frac{E_J}{\epsilon_1},
    \frac{\eta}{\epsilon_1}\right)=-
    \cos\left[\frac{4eV_B}{\epsilon_1}\right]
    \sum_{n_x}\mbox{Im}\left[\frac{\exp\left(\frac{2i\pi\Phi}{\Phi_0}\right)}
      {\frac{2eV_B}{\epsilon_1} + n_x+(-)^{n_x} \frac{E_J}{\epsilon_1}
        \cos \left(\frac{2\pi\Phi}{\Phi_0}\right)
        +i\frac{\eta}{\epsilon_1}}\right] ,
  \end{eqnarray}
  \end{widetext}
  where $\epsilon_1=2\pi \hbar v_F/R_{\alpha,\beta}$ and, as a
  physically-motivated approximation, we inserted into $\Omega_{n_x}$
  {the fit $\simeq n_x \epsilon_1 + (-)^{n_x} E_J
    \cos\left(\frac{2\pi\Phi}{\Phi_0}\right)$ to the Bagwell spectra}
  \cite{Bagwell}, where $E_J$ is the Josephson energy.  The above
  {phenomenological} Eq.~(\ref{eq:gt0}) for $g'''_0$
  is one of the main results of the paper, which becomes proportional
  to $g_0$ in Eq.~(\ref{eq:EE2}) for large ratio $\eta/\epsilon_1$.}
{Fig.~\ref{fig:gt0} shows the variations of $g'''_0$ given by
  Eq.~(\ref{eq:gt0}) in the $\left(\frac{\Phi}{\Phi_0}, \frac{e
    V_B}{\epsilon_1} \right)$ plane of the parameters, with the
  dimensionless Josephson energy $\frac{E_J}{\epsilon_1}=0.2$ and with
  the following values for the dimensionless line-width broadening:
  $\frac{\eta}{\epsilon_1} = 0.1$ (panel a), $\frac{\eta}{\epsilon_1}
  = 0.3$ (panel b), $\frac{\eta}{\epsilon_1} = 0.5$ (panel c) and
  $\frac{\eta}{\epsilon_1} = 100$ (panel d). We find that the
  conductance at small $\frac{\eta}{\epsilon_1}=0.1$ (see panel a)
  reflects the energy-flux relation of the ABS. Increasing
  $\frac{\eta}{\epsilon_1}$ produces a cross-over to the oscillatory
  checkerboard pattern that was obtained in the above
  Sec.~\ref{sec:lowest-order} for an infinite 2D metal connected by
  tunnel contacts to the superconducting leads, see the above
  Fig.~\ref{fig:colorplot}.}

{
\section{Discussion and conclusions}
\label{sec:discussion-conclusion}
}

{We present final remarks in this section: discussion
  in subsection~\ref{sec:discussion} and conclusions in
  subsection~\ref{sec:conclusions}.}

{
\subsection{Discussion}
\label{sec:discussion}}

{We now complement the paper with the following
  four-fold discussion: We argue in subsection~\ref{sec:PSEandHGE}
  that our theory can explain the experimental results of the Penn
  State group \cite{PSE} on graphene-based three-terminal Andreev
  interferometers, and of the Harvard group \cite{Huang2022} on
  graphene-based four-terminal Josephson junctions. We next address in
  section~\ref{sec:periodic-compression} a semiclassical picture,
  starting from Ref.~\onlinecite{Kulik}, and from qualitative
  arguments about the spectrum of the device at high transparency. A
  discussion is presented in section~\ref{sec:final}, regarding a
  comparison between the high- and low-transparency interfaces. We
  finally proceed in section~\ref{sec:conjecture} with a conjecture
  about the spectrum of billiards connected to an arbitrary number of
  superconducting leads by highly transparent interfaces, in
  connection with the 1D classical orbits formed by the 1D Andreev
  tubes.}

{\subsubsection{Penn State and Harvard group experiments}\label{sec:PSEandHGE}}

{In this subsection, we mention the relevance of our
  theory to the recent Penn State \cite{PSE} and Harvard
  \cite{Huang2022} group experiments.}

{\it The recent Penn State group experimental preprint \cite{PSE}}
{implements} the ballistic limit of Andreev
interferometers that are realized with graphene connected to a
grounded superconducting loop pierced by the magnetic flux
$\Phi$. This grounded loop terminates at the left and right
superconducting interfaces $S_a$ and $S_b$ biased at the phases
$\varphi_a$ and $\varphi_b$ such that $\varphi_b-\varphi_a\simeq
\Phi$. The rectangular piece of graphene is additionally connected to
the normal lead $N_B$ (see Fig.~\ref{fig:thedevice}a) or}{, in other
  devices,} {to the superconducting lead $S_B$. The experimental
  preprint of the Penn State group \cite{PSE} reports the emergence of
  low-energy resonances in the all-superconducting three-terminal
  Andreev interferometer, which nontrivially disperse as a function of
  the magnetic flux $\Phi$,} {so} {as to produce alternations between
  the $0$- and $\pi$-shifted energy-phase
  relations. {As an interpretation of the Penn State
    group experiment \cite{PSE}, the periodic variations of the
    resonance energies in the conductance as a function of the bias
    voltage and magnetic flux are put in a one-to-one correspondence
    with the periodic compression of the ABS spectrum as a function of
    energy, and the way those compressed energy levels vary as a
    function of the magnetic flux $\Phi$, and can be probed by varying
    the bias voltage $V_B$. We argue in the forthcoming
    subsections~\ref{sec:periodic-compression} and~\ref{sec:final}
    that this structure of the ABS spectrum is there in both limits of
    small and high transparency interfaces.}}

 {{\it The Harvard-group experimental paper \cite{Huang2022}} provides
   further evidence for transient {multiplets} of Cooper pairs in
   graphene-based four-terminal Josephson junctions, {also} known as
   the Cooper quartets \cite{Freyn2011}, sextets, octets, {and so on,}
   which could be revealed in the DC-current, at equilibrium or in a
   device biased at the commensurate voltage differences. In the
   voltage-biased three-terminal devices, the third terminal is
   qualitatively analogous \cite{Cuevas-Pothier} to the RF source in
   the two-terminal Shapiro step experiments \cite{Shapiro}. The
   theory of the quartets
   \cite{Freyn2011,Cuevas-Pothier,Melin2016,Melin-Sotto,Melin2017,Melin2019,Doucot2020,Melin2020,Melin2020a,Melin2021,Keliri1,Keliri2,Melin2022,Melin2023,Melin2023a,Melin-Winkelmann-Danneau}
   is in agreement with a series of experiments
   \cite{Pfeffer2014,Cohen2018,Zurich,Huang2022,Pribiag,Finkelstein-recent}
   that contrast with a conservative picture in terms of the
   resistively shunted Josephson junction (RSJ) model
   \cite{Finkelstein1,Finkelstein2,Finkelstein3,Zhang1,Gupta2023,Graziano2020}.
   In addition, many other effects
   \cite{topo1,topo2,topo3,topo4,molecule1,molecule2,molecule3,molecule4,Gupta2023,Zhang2023,Matsuo2022,Pankratova2020,Graziano2020,Khan2021,Finkelstein1}
   have focused interest in the field of multiterminal Josephson
   junctions, such as the production of Weyl-point singularities and
   topologically quantized transconductance
   \cite{topo1,topo2,topo3,topo4}, the emergence of avoided crossings
   in the spectra of Andreev bound states
   \cite{molecule1,molecule2,molecule3,molecule4}, the multiterminal
   Josephson diode effect \cite{Gupta2023,Zhang2023,Finkelstein1}, and
   so on.}

{We are now demonstrating in a work in progress that a process similar
  to the here considered phase-AR also couples the three- and
  four-terminal quartet and} {split}{-quartet modes \cite{Melin2020}
  to the nonequilibrium distribution functions. This can explain the
  voltage dependence of the quartet anomaly in the Harvard group
  experiment \cite{Huang2022}. Such approach would considerably
  simplify the description with respect to including both ingredients
  of the multiple Andreev processes allowed by the time-periodic
  Hamiltonian, and the extended interfaces of the device geometry.}
{We generally argue that the nonequilibrium electronic populations can
  produce small energy or voltage scales in the all-superconducting
  ballistic multiterminal 2D devices in the long-junction limit, if
  they are simultaneously voltage- and phase-biased.}

\begin{figure}[htb]
  \includegraphics[width=\columnwidth]{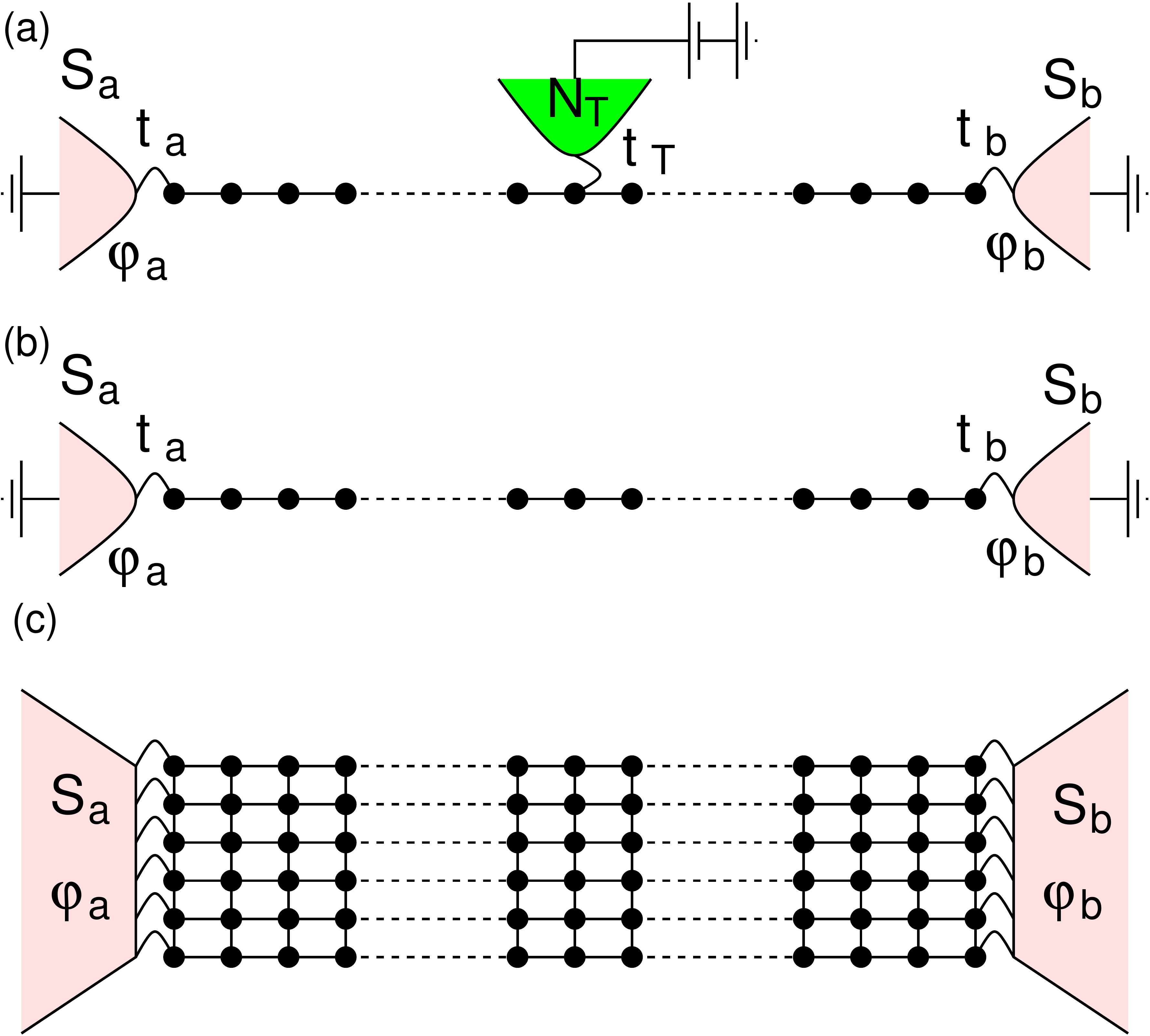}
  \caption{{A 1D Andreev interferometer connected to
      the grounded left and right superconductors $S_a$ and $S_b$
      biased at the phases $\varphi_a$ and $\varphi_b$, and to the top
      normal lead $N_T$ that is biased at the voltage $V_T$ (a); A 1D
      Josephson junction with the nondegenerate energy levels of
      Eq.~(\ref{eq:spectrum-Kulik-1D}) (panel b); A 2D Josephson
      junction (c). With periodic boundary conditions in the
      $y$-direction, each levels would approximately have $M$-fold
      degeneracy, where $M$ is the number of channels in the
      transverse direction ($M=6$ on the figure). This degeneracy is
      due to the Josephson current flow from left to right or from
      right to left along horizontal 1D Andreev tubes that are
      localized over $\approx \lambda_F$ in the vertical
      direction. Each of those tubes forms an independent sector of
      the semi-classical limit.}
\label{fig:thedevices-bis}
}
\end{figure}

\begin{figure}[htb]
  \includegraphics[width=\columnwidth]{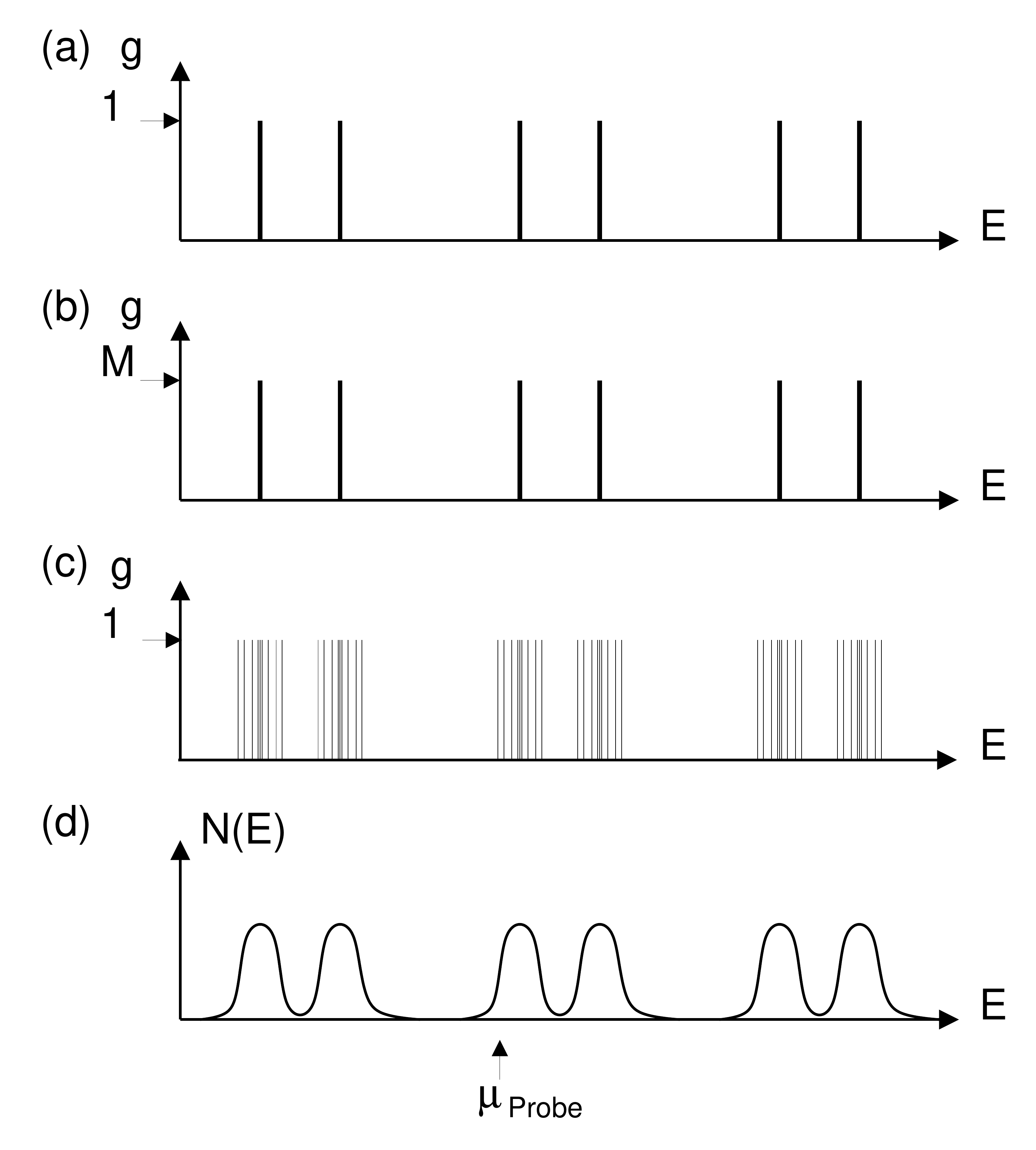}
  \caption{{Spectra of ballistic $S_a$-$N$-$S_b$
      Josephson devices in the long-junction limit and at high
      transparency: The specialization of Ref.~\onlinecite{Kulik} to
      1D, with the nondegenerate energy levels arranged according to
      Eq.~(\ref{eq:spectrum-Kulik-1D}) (panel a); The 2D model of
      Ref.~\onlinecite{Kulik}, with $M$-fold degenerate energy levels,
      where $M$ is the number of transverse channels (b); The 2D model
      of Ref.~\onlinecite{Kulik} with weakly irregular shape, then,
      channel mixing results in the lifted $M$-fold degeneracies (c);
      The continuous density of states (d). Compared to panel c, the
      broadening of the energy levels originates from the intrinsic
      inelastic electron-electron interactions, from the intrinsic
      electron-phonon coupling, or from the extrinsic coupling to the
      infinite $N_B$. The $x$-axis is the energy $E$. The $y$-axis on
      panels a-c is the degeneracy and it is the density of states on
      panel d.}
\label{fig:spectra}
}
\end{figure}
{\subsubsection{Semiclassics and periodic compression
    of the spectrum}\label{sec:periodic-compression}}

{In this subsection, we discuss the spectrum of highly
  transparent Josephson devices in the long-junction limit, starting
  from Ref.~\onlinecite{Kulik}. We start the discussion with the
  simplest 1D Andreev interferometer, consisting of a 1D tight-binding
  chain of length $L$ connected by the tunneling amplitudes $t_a$ and
  $t_b$ to the left and right grounded superconducting leads $S_a$ and
  $S_b$ having the phase variables $\varphi_a$ and $\varphi_b$. The
  tight-binding chain is additionally connected by the hopping
  amplitude $t_T$ to the top normal lead $N_T$ biased at the voltage
  $V_T$, see Fig.~\ref{fig:thedevices-bis}a. The fully grounded
  $S_a$-$N$-$S_b$ two-terminal Josephson junction is formed if $N_B$
  is disconnected from the remaining of the device, i.e. if $t_T=0$,
  see the 1D geometry in Fig.~\ref{fig:thedevices-bis}b and the 2D
  geometry in Fig.~\ref{fig:thedevices-bis}c.}

{The semi-classical calculation of
  Ref.~\onlinecite{Kulik} can be specialized to 1D, and, in the limit
  of small energies, its spectrum is shown in Fig.~\ref{fig:spectra}a,
  which illustrates the following expression of the energy levels
  \cite{Kulik}:
  \begin{equation}
    \label{eq:spectrum-Kulik-1D}
    E_n=\frac{\hbar
      v_F}{2L}\left[2\pi\left(n+\frac{1}{2}\right)\mp\chi \right] .
  \end{equation}
  The $2L$ factor in Eq.~(\ref{eq:spectrum-Kulik-1D}) reflects that,
  over one period, the 1D Andreev tube has to be traversed from $S_a$
  to $S_b$ and back from $S_b$ to $S_a$ by the spin-up electrons and
  the spin-down holes respectively. The periodicity of the spectrum
  upon changing the integer $n$ is generic of all types of
  Fabry-P\'erot interferometers. The half-period shift of the spectrum
  is the sum of the $\pi/2$-shifts that appear in Andreev reflection
  at each of the $S_a$-$N$ and $N$-$S_b$ interfaces, see for instance
  Eq.~(A11a) in Ref.~\onlinecite{BTK}, evaluated at zero-energy for a
  highly transparent interface. Using the notations of
  Ref.~\onlinecite{Kulik}, the Josephson energy is obtained as $\mp
  \chi$ for the electron- and hole-type ABS, with
  $\chi=\varphi_b-\varphi_a$ the difference between the
  superconducting phase variables of the right and left
  superconductors. This linear-in-phase Josephson energy contains
  contributions of all harmonics, which reflects the highly
  transparent interfaces between the central-$N$ and each of the
  superconducting lead.}

{Concerning the 2D or the three-dimensional (3D)
  geometries, Kulik obtains the same spectrum as in 1D, see
  Eq.~(\ref{eq:spectrum-Kulik-1D}).  Those energy levels now have
  degeneracy that is equal to the number of transverse channels, which
  is here denoted by the integer $M$. For instance, in 2D, the energy
  levels of Eq.~(\ref{eq:spectrum-Kulik-1D}) are now $M$-fold
  degenerate in the semiclassical limit, if the central normal metal
  consists of a cylinder of dimensions $N a_0\times M a_0$, where
  periodic boundary conditions are implemented along the $y$-axis,
  instead of the experimentally relevant open boundary conditions that
  are shown in Fig.~\ref{fig:thedevices-bis}c. A proposed
  interpretation is that the Andreev pairs transmitted from $S_a$ to
  $S_b$ or from $S_b$ to $S_a$ propagate along the 2D Andreev tubes
  that are parallel to the $x$-axis. Along the $y$-axis direction,
  they are semi-classically localized over the short Fermi wave-length
  $\lambda_F$. The resulting semiclassical Hilbert space is {\it
    fragmented} into those $M$ degenerate horizontal Andreev
  tubes. However, the open (instead of periodic) boundary conditions
  in the $y$-axis direction are expected to lift this semiclassical
  degeneracy because the tubes located on the bottom and top edges are
  localized on $\approx \lambda_F/2$ in a half-space instead of
  $\approx \lambda_F$ in the bulk. The Andreev tubes at the top and
  bottom edges are then expected to have spectrum that significantly
  differs from those in the bulk. However, from the point of view of
  discussing the qualitative physics, we replace the open boundary
  condition in the $y$-axis direction by the periodic ones, and make
  the approximation that the energy levels are $M$-fold
  degenerate. This approximation is within a relative error smaller
  than $\approx 1/M$ in the calculation of the current, as it is
  realistically the case in experiments with graphene where $M$ is
  large, such as the recent Penn State \cite{PSE} or Harvard
  \cite{Huang2022} group experiments. As a side remark, this type of
  semiclassical description in terms of Andreev tubes, see for
  instance Ref.~\onlinecite{Glazman-Falko}, successfully allows
  quantitative interpretation of the experimental magnetic oscillation
  pattern of bilayer graphene-based two-terminal Josephson junctions
  in many different geometries, see
  Ref.~\onlinecite{Danneau-nature-communications}.}

{Experimental implementation of Ref.~\onlinecite{Kulik}
  involves the ballistic 2D metal that is approximately realized with
  encapsulated graphene. The wave-guide geometry and the absence of
  channel mixing in Ref.~\onlinecite{Kulik} implies the experimental
  constraint of perfect control at the atomic scale over the geometry
  and over the transmission amplitudes along each of the interfaces,
  which, likely, is not possible within the present-time state-of-the
  art nanofabrication technology. It is then natural to conclude that,
  in the realistically nonideal situations, the approximate $M$-fold
  degeneracy of the 2D energy levels of Ref.~\onlinecite{Kulik} would
  significantly split into a collection of $M$ closely-spaced energy
  levels, see Fig.~\ref{fig:spectra}a for the spectrum of the long 1D
  $S_a$-$N$-$S_b$ Josephson junction, Fig.~\ref{fig:spectra}b in 2D
  without channel mixing, and Fig.~\ref{fig:spectra}c in 2D with
  channel mixing. The coarse-grained density of states defines a
  sequence of peaks that are plotted as a function of the energy, see
  Fig.~\ref{fig:spectra}d, or, in the Andreev interferometer, as a
  function of the bias voltage on the normal lead. The latter allows
  controlling the Fermi surface on the central-$N$, which is
  parameterized by the effective chemical potential $\mu_{Probe}$ in
  Fig.~\ref{fig:spectra}d according to the models that were proposed
  in the above section~\ref{sec:lowest-order}. Depending on whether
  electron-type of hole-type ABS are considered, each maximum is
  separated from the next by $2\chi$ (i.e. by the contribution of the
  Josephson effect to the energy) or by $\pi \hbar v_F/L-2\chi$
  (reflecting the elementary 1D Andreev tubes).}

{Those observations call for future investigations of a
  fully microscopic picture for the interplay between inelastic
  scattering and the 1D Andreev tubes in the 2D geometries. This
  emerging continuous spectrum is consistent with the assumptions of
  the Keldysh perturbation theory in the tunneling amplitudes, see the
  above section~\ref{sec:lowest-order}. The summation over the
  wave-vectors in the spectral representation was there technically
  replaced by a continuous integral, see the demonstration of
  Eqs.~(\ref{eq:J0-1})-(\ref{eq:J0-2}) in
  Appendix~\ref{app:methods}. This approximation mimics the
  enhancement of the line-width broadening produced by the
  electron-electron interactions or by the electron-phonon
  coupling. However, at present time, we do not fully theoretically
  understand to which extend the 1D Andreev tubes remain 1D in the
  presence of the stronger microscopic electron-phonon coupling that
  can be produced by increasing the temperature. Microscopic models of
  the electron-electron interactions would also be a significant step
  forward.}

{\subsubsection{Limiting cases of small and large interface transparencies}
\label{sec:final}}
{In this subsection, we summarize and compare the
  limits of small and large values of the interface transparencies,
  that are viewed as being representative of the perturbative and
  nonperturbative regimes of transport theory, see the above
  sections~\ref{sec:lowest-order} and~\ref{sec:intrinsic1}
  respectively. The conductance spectra in Fig.~\ref{fig:colorplot}
  were calculated in section~\ref{sec:lowest-order} in the tunnel
  limit for a continuous spectrum of the 2D tight-binding Hamiltonian
  of the central-$N$ coupled to $N_B$. In the absence of coupling to
  the infinite normal lead $N_B$, we relate the magnetic flux-$\Phi$
  sensitivity of the ABS spectrum to the circulating supercurrent
  transmitted from $S_a$ to $S_b$ or from $S_b$ to $S_a$, across the
  central-$N$. In perturbation theory in the tunneling amplitudes
  $t_a$ and $t_b$ with the superconductors $S_a$ and $S_b$, the
  circulating supercurrent can be expressed similarly to the above
  Eq.~(\ref{eq:I1-A1}) but now, roughly speaking, the Keldysh Green's
  function $\delta g^{+,-}$ is replaced by the advanced or by the
  retarded one. The circulating supercurrent is then proportional to
  the Andreev mode $g_{\alpha,\beta}^{1,1} g_{\beta,\alpha}^{2,2}$
  that oscillates in energy, with the same period $\sim \pi \hbar
  v_F/L$ as in the ABS spectrum with highly transparent interfaces,
  see the above subsection~\ref{sec:periodic-compression}. Each ABS
  carries supercurrent that is proportional to the derivative of its
  energy with respect to the phase difference.  We deduce from the
  periodic energy-oscillations of the circulating spectral
  supercurrent in the limit of small interface transparencies, that
  the ABS spectrum also shows the periodic compressions at both low
  and high values of the contact transparencies. In the tunnel limit,
  the periodic energy-compression of the spectrum has period $\sim
  \pi\hbar v_F/L$ and modulates the energy levels closely separated by
  $\sim 1/L^2$ if the central-$N$ has the geometry of a square. We
  conclude that the conductance plotted in the plane of the magnetic
  flux and bias voltage generically follows the ABS
  spectrum. {Not unexpectedly, the perturbative and the
    nonperturbative limits however yield {\it weak} and {\it strong}
    effects on the periodic magnetic flux- and bias voltage-dependence
    of the conductance spectra, in the sense of harmonic checkerboard
    pattern oscillations or sharp peaks in the conductance,
    respectively.}}

\begin{figure}[htb]
  \includegraphics[width=\columnwidth]{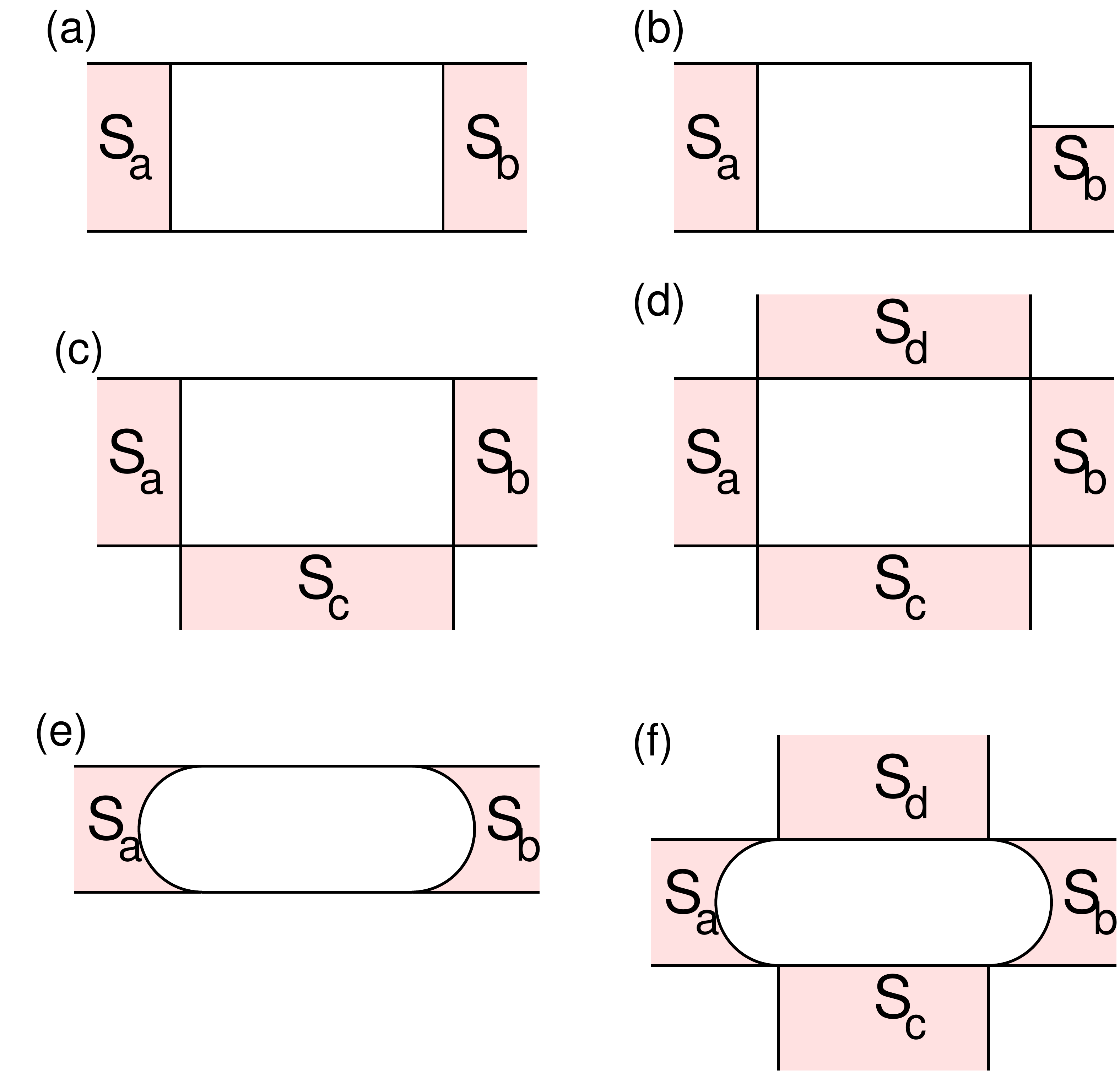}
  \caption{{Examples of multiterminal Josephson
      billiards generalizing the wave-guide configuration of
      Ref.~\onlinecite{Kulik} (panel a): Partially coated two- and
      three-terminal integrable billiards (b and c); Fully coated
      four-terminal integrable billiard (d); Partially coated
      two-terminal chaotic billiard (e); Fully coated four-terminal
      chaotic billiard (f).}
\label{fig:billiards}
}
\end{figure}
{
\subsubsection{Perspectives on multiterminal Josephson billiards}\label{sec:conjecture}}

{In Appendix~\ref{app:billiards}, we propose a
  conjecture on chaotic billiard becoming integrable once connected to
  the superconducting leads, within a set of specific assumptions that
  make the elementary 1D Andreev tube become the classical
  trajectories. This surprising statement calls for future
  investigations of the multiterminal Josephson billiards shown in
  Fig.~\ref{fig:billiards}.}

  {
\subsection{Conclusions}\label{sec:conclusions}}

{In this subsection, we summarize the paper and provide
  concluding remarks.} {We developed the theory
  according to the following framework: (i) We calculated the
  conductance $d I_B/dV_B$ of an Andreev interferometer from Keldysh
  perturbation theory in the tunneling amplitudes. (ii) We also
  evaluated the conductance to all orders in the tunneling amplitudes,
  and in the entire range of the device dimensions. (iii) We provided
  numerical results within physically-motivated approximations for
  dressing the perturbative expression of the conductance by the
  spectrum of bound states \cite{Kulik,Ishii,Bagwell}. Now, we
  summarize each of those items:}

{(i) We discussed {the microscopic theory of
    {a ballistic} $(S_a,S_b,N_B)$
    normal-superconducting Andreev interferometer}, where $S_a$ and
  $S_b$ are connected by a loop pierced by a magnetic flux $\Phi$, and
  $N_B$ is biased at the voltage $V_B$.}

{In our model, we assumed that the central-$N$ of large dimension
  produces a smooth energy dependence in its tunneling density of
  states. Then, we implemented the infinite planar geometry sustaining
  the finite local and nonlocal spectral densities of states of the
  central-$N$. In this geometry, we found that the conductance $d
  I_B/d V_B$ of the $(S_a,S_b,N_B)$ Andreev interferometer reveals a
  smooth oscillatory checkerboard pattern in the plane of the
  magnetic flux $\Phi$ and the bias voltage $V_B$. Microscopically,
  the effect was evaluated from the physical input of a nonequilibrium
  Fermi surface featuring a single or multiple steps in the
  nonequilibrium distribution function at energies that are sensitive
  {to} the bias voltage $V_B$. Those oscillations originated from the
  nonstandard microscopic process of the phase-AR that couples the
  current to both the superconducting phase variables and the
  nonequilibrium populations in the central-$N$.}

{(ii) The finite-size effects appear with a central-$N$ of smaller
  dimension and they produce sharp peaks in the tunneling density of
  states of the central-$N$ coupled to $N_B$. Then, we implemented a
  nonperturbative computational algorithm to uniquely separate the
  contributions of the one- and two-particle resonances to the
  differential conductance $d I_B/d V_B$ in the $(S_a,S_b,N_B)$
  device. We obtained compact formula for the differential conductance
  $d I_B/dV_B$, valid to all orders in tunneling and for all device
  geometries, ranging from quantum dots, to the short-junction limit,
  and to the long-junction limit. Again, the voltage $V_B$ could be
  used as a control parameter to perform the spectroscopy of those
  sharp resonances that related to the energy levels of a long
  Josephson junction \cite{Kulik,Ishii,Bagwell}.  We argued that the
  currents of the one- and two-particle resonances} {are} {not
  separately conserved in the regime of sequential tunneling, and that
  the sequential tunneling channel dominates for small or moderate
  interface transparencies and for devices that are much larger than
  the Fermi wave-length} {$\lambda_F$.}

{(iii) We provided physically-motivated approximations for the
  conductance of a $(S_a,S_b,N_B)$ Andreev interferometer, which are
  based on the perturbative and nonperturbative expressions of the
  current, see the above items (i) and (ii). We found that, at small
  line-width broadening, the conductance spectra are peaked at the
  energies of the ABS \cite{Kulik,Ishii,Bagwell}. Thus, measuring the
  differential conductance of the $(S_a,S_b,N_B)$} {or
  $(S_a,S_b,S_B)$} {Andreev interferometer as a function of the bias
  voltage $V_B$ allows} {for a} {spectroscopy of the spectrum of a
  long Josephson junction.} At higher values of the line-width
broadening, we recovered the smooth oscillatory checkerboard pattern
of the perturbative calculations.}

{We additionally proposed a semiclassical picture that
  establishes a connection between the
  experiments\cite{PSE,Huang2022}, perturbation theory in the
  tunneling amplitudes, the highly transparent interfaces of
  Ref~\onlinecite{Kulik}, and the chaotic or integrable Josephson
  billiards. In addition, we find the same periodic compression of the
  ABS spectrum in both limit of low and high contact
  transparencies. We conclude that the conductance spectra of the
  Andreev interferometer follows this energy-periodic modulation of
  the ABS spectrum.}

{Beyond (nonlocal) differential conductance or
  resistance measurements, a smoking gun of the phase-AR would be
  experimental evidence for bias voltage- and possibly also magnetic
  flux-sensitive electronic populations in the piece of graphene
  sustaining the 2D metal. Such populations can in principle be probed
  with superconducting tunneling spectroscopy \cite{Saclay}. In case
  the superconducting tunneling experiments would not be successful,
  future directions could involve the development of a microscopic
  framework that would be suitable to capture the relaxation of the
  electronic populations in the 2D metal, due to the electron-electron
  interaction or to the electron-phonon coupling. The present work
  will also have to be extended to the all-superconducting Andreev
  interferometers, involving the time-periodic dynamics of the
  superconducting phase variables as in Floquet theory. The phase-AR
  generally captures the coupling between the 2D metal density of
  states/populations, and a Josephson mode between the two
  superconducting interfaces on the central-$N$. We are now preparing a
  paper where we demonstrate that the phase-AR also captures a
  coupling between the normal density of states/populations, and a
  quartet mode in three- and four-terminal Josephson junctions, in
  connection with further theoretically discussing the Harvard group
  experiment \cite{Huang2022}. A fully open question is to provide a
  detailed understanding and to propose experiments to
  address the intermediate state formed at the interplay between the
  phase-AR coupling two normal-superconducting interfaces and the $4e$
  quartet modes coupling three of those. A kind of {\it quantum
    tomography} of the quartet and the phase-AR states could be
  proposed in the future. }

{Finally, we note Ref.~\onlinecite{vanWees} where the
  scattering approach is developed in a 1D configuration.}

\section*{Acknowledgements}
{R.M. wishes to thank Florence L\'evy-Bertrand and Klaus Hasselbach
  and for stimulating feedback during an informal blackboard seminar
  on this topic. R.M. wishes to thank to Romain Danneau for {useful
    discussions on the subject of this paper as well as on related
    issues.} Those discussions are within the collaboration between
  {the French} CNRS in Grenoble and {the German} KIT in Karlsruhe,
  which is funded by the International Research Project SUPRADEVMAT.}
{R.M. also acknowledges a fruitful discussion with
  Beno\^{\i}t Dou\c{c}ot.} M.K. and A.S.R. acknowledge support from
the Materials Research Science and Engineering Center supported by
{the US} National Science Foundation (DMR 2011839).

\appendix

\section{Details of the methods}
\label{app:methods}

This Appendix contains the necessary details of the methods on
which the paper relies. The structure of this Appendix is provided in
the main text, see section~\ref{sec:methods}.

\subsection{Advanced and retarded Green's functions in
 the normal region of the circuit}
\label{subsec:N}

In this subsection of Appendix~\ref{app:methods}, we first provide the
form of the {\it bare} Green's function $\hat{g}_{i,j}(\omega)$,
i.e. the Green's functions of the central region with vanishingly
small tunneling amplitudes to the superconducting or normal lead. In a
second step, we provide an approximation to the Green's function of
$N_{eff}$ consisting of the central-$N$ strongly coupled to $N_B$.

{The electronic} Green's functions of an infinite 2D
normal metal treated in the continuous limit {are given by}
\begin{eqnarray}
 \label{eq:J0-1}
 g^{A,1,1}(R,\omega)&=&-g^{R,1,1}(R)=\frac{i}{W} J_0(k_e R)\\
 g^{A,2,2}(R,\omega)&=&-g^{R,2,2}(R)=\frac{i}{W} J_0(k_h R)
 ,
 \label{eq:J0-2}
\end{eqnarray}
{where $J_0$ is a Bessel function,} $\omega$ is the
energy, $k_e$ and $k_h$ denote the electron and hole Fermi
wave-vectors $k_{e,h}=k_F\pm\frac{\delta \omega}{\hbar v_F}$, with $\delta
\omega$ the energy with respect to the equilibrium
{chemical potential}. The variable $R$ in
Eqs.~(\ref{eq:J0-1})-(\ref{eq:J0-2}) denotes the separation in real
space.

In the limit of large separation $R\equiv R_{\alpha,\beta}$ between
$\alpha$ and $\beta$, i.e. $R\gg 1/k_F$, we find the following
approximation to the {\it nonlocal} bare Green's functions:
\begin{eqnarray}
 \label{eq:A}
 g^{A,1,1}(R,\omega)&=&-g^{R,1,1}(R,\omega)\\&\simeq&\frac{i}{W\sqrt{k_e
   R}}\cos\left(k_e R
 -\frac{\pi}{4}\right)\nonumber\\  \label{eq:A-fin}
g^{A,2,2}(R,\omega)&=&-g^{R,2,2}(R,\omega)\\
 &\simeq&\frac{i}{W\sqrt{k_h
    R}}\cos\left(k_h R -\frac{\pi}{4}\right).
\nonumber
\end{eqnarray}

{Now, we carry out perturbation theory in the coupling
  to the superconducting leads, which holds for arbitrary dimensions
  of the central-$N$, which is attached both} to the superconducting
leads $S_a$ and $S_b$ at the left and right respectively, and to the
normal lead $N_B$. The {Nambu} hopping amplitudes are
denoted by $\hat{t}_{N,S_a}$, $\hat{t}_{N,S_b}$ and $\hat{t}_{N,N_b}$
respectively, where the diagonal matrix
$\hat{t}_{i,j}=(t_{i,j},\,-t_{i,j})$ has entries in Nambu and in the
set of the the tight-binding sites at the interfaces, see
Eq.~(\ref{eq:H-t}).

The Dyson equations for the fully dressed $\hat{G}_{N,N}(\omega)$ take
the following form:
\begin{eqnarray}
 \label{eq:Dyson-cal-C}
 \hat{G}_{N,N}(\omega)&=&\hat{g}_{N,N}(\omega) \\
\nonumber
 &+&
 \hat{g}_{N,N}(\omega)\hat{t}_{N,N_B} \hat{g}_{N_B,N_B}
 \hat{t}_{N_B,N} \hat{G}_{N,N}(\omega) \\&+& \sum_k
 \hat{g}_{N,N}(\omega)\hat{t}_{N,a_k} \hat{g}_{a_k,a_k}
 \hat{t}_{a_k,N} \hat{G}_{N,N}(\omega) .
 \nonumber
\end{eqnarray}
Perturbation theory in $t_{a_k,N}^2$ leads to
\begin{equation}
 \label{eq:GCC}
 \hat{G}_{N,N}(\omega)=\left(\left(\hat{G}^{(0)}_{N,N}\right)^{-1}(\omega)-\sum_k
 \hat{t}_{N,a_k} \hat{g}_{a_k,a_k}
 \hat{t}_{a_k,N}\right)^{-1}
 ,
\end{equation}
where $\hat{G}^{(0)}_{N,N}(\omega)$ is the component in the
central-$N$ coupled to $N_B$ of the fully dressed Green's function in
absence of coupling to the superconductors, i.e. with ${t}_{N,a_k}=0$:
\begin{equation}
 \hat{G}^{(0)}_{N,N}(\omega)=\left(g_{N,N}^{-1}(\omega)
 - \hat{t}_{N,N_B} \hat{g}_{N_B,N_B} \hat{t}_{N_B,N}\right)^{-1}
 .
\end{equation}
The Green's function $\hat{G}^{(0)}_{N,N}(\omega)$ features resonances
in the sense that the energy levels of the {\it bare} central-$N$ are
broadened by the coupling to the infinite normal metal $N_B$.

\subsection{Large-gap approximation}
\label{sec:large-gap}

In this subsection of Appendix~\ref{app:methods}, we introduce the
large-gap approximation that is {used throughout the
  paper. The local Green's functions} of a BCS superconductor takes
the following form:
\begin{eqnarray}
 \label{eq:g-supra}
 \hat{g}_0^A(\Delta,\omega)&=& \frac{1}{W
  \sqrt{\Delta^2-(\omega-i\eta_S)^2}}\times\\
&& \left(\begin{array}{cc} -(\omega-i\eta_S) & \Delta \exp(i\varphi) \\
  \Delta \exp(-i\varphi) & -(\omega-i\eta_S)
 \end{array}\right)
 ,
 \nonumber
\end{eqnarray}
where $W$, $\Delta$ and $\varphi$ are the band-width, the
superconducting gap and the superconducting phase variable respectively.
In Eq.~(\ref{eq:g-supra}), the energy $\omega$ of the advanced Green's
function acquires a small imaginary part $-i\eta_S$, where $\eta_S$ is
the Dynes parameter, which phenomenologically accounts for the
microscopic relaxation mechanisms such as electron-electron
interaction or electron-phonon coupling
\cite{Kaplan,Dynes,Pekola1,Pekola2}.

In what follows, we focus on energy or voltage scales that are much
smaller than the superconducting gap $\Delta$. Eq.~(\ref{eq:g-supra})
becomes independent on the energy $\omega$ in this {\it large-gap
  approximation}:
\begin{equation}
 \label{eq:g-supra-large-gap}
 \hat{g}_{S,S}= \frac{1}{W}
 \left(\begin{array}{cc} 0 & \exp(i\varphi) \\
  \exp(-i\varphi) & 0
 \end{array}\right)
 .
\end{equation}
The consistency \cite{Melin-Winkelmann-Danneau} between the
two-terminal Fraunhofer patterns calculated from the large-gap
approximation and known theoretical results \cite{Cuevas2007} further
supports the use of Eq.~(\ref{eq:g-supra-large-gap}) to capture the
qualitative behavior of 2D multiterminal devices such as in
Fig.~\ref{fig:thedevice}a.

\subsection{Notations for the populations}

\label{subsec:assumptions-popus}

In this subsection of Appendix~\ref{app:methods}, we fix the notations
regarding {the nonequilibrium distribution
  functions. Physically, we assume} that the electrons and holes in
the infinite $N_B$ can tunnel back and forth into the
central-$N$. Thus, in the regime of sequential tunneling, a collection
of steps $Z_p$ is produced at the energies $\mu_{N,p}$ in the
distribution function $\hat{n}_F(\omega)$. For instance,
$\hat{n}_F^{1,1}(\omega) = n_F^{(0)}(\omega-\mu_N)$ and
$\hat{n}_F^{2,2}(\omega) = n_F^{(0)}(\omega+\mu_N)$ for a single step
with $Z=1$ at the step-energy $\mu_N${, corresponding to strong
  electron-phonon coupling.} Multistep distribution functions with
discontinuities $\{Z_p\}$ at the energies $\{\mu_{N,p}\}$ are obtained
in the presence of two or more normal leads, see
Fig.~\ref{fig:thedevice}e.

\subsection{Nonequilibrium current and the Keldysh Green's function}
\label{sec:thecurrent}

In this subsection of Appendix~\ref{app:methods}, we present how the
currents are calculated {from the Keldysh Green's
  functions. We assume} that the superconducting leads $S_a$ and $S_b$
are connected to the central normal metal tight-binding sites
$(N_\alpha,\,N_\beta)$ by the single-channel contacts $S_a$-$N_\alpha$
and $N_\beta$-$S_b$, see Figs.~\ref{fig:thedevice}b
and~\ref{fig:thedevice}c. The spectral current transmitted into $S_a$
is the following \cite{CCNSJ1,CCNSJ2,Cuevas}:
\begin{equation}
 \label{eq:I1-model-A}
 I_a(\omega)= \mbox{Nambu-trace}
 \left\{\hat{\tau}_3 \left[ \hat{t}_{\alpha,a} 
  \hat{G}^{+,-}_{a,\alpha}(\omega) - \hat{t}_{a,\alpha}
  \hat{G}^{+,-}_{\alpha,a}(\omega)\right] \right\}
 ,
\end{equation}
where $\hat{G}^{+,-}(\omega)$ is the Keldysh Green's function, and
$\hat{t}$ is the self-energy provided by the hopping term at the
interfaces between the central-$N$ and the superconducting
leads. The physical current $I_a=\int d\omega I_a(\omega)$ is defined
as the integral over the energy $\omega$ of the spectral current
$I_a(\omega)$.

The {\it fully dressed} Keldysh Green's functions
$\hat{G}_{i,j}^{+,-}(\omega)$ are deduced as follows from the fully
dressed advanced and retarded Green's functions
$\hat{G}^A_{i,j}(\omega)$ and $\hat{G}^R_{i,j}(\omega)$ respectively,
and from the bare Keldysh Green's function $\hat{g}^{+,-}(\omega)$:
\begin{eqnarray}
 \label{eq:Gpm-1}
 &&
 \hat{G}_{a,\alpha}^{+,-}(\omega)=\\\nonumber&& \sum_k
 \left(\hat{I}+\hat{G}^R(\omega)\hat{t}\right)_{a,a_k}
 \hat{g}^{+,-}_{a_k,a_k}(\omega) \left(\hat{I}+\hat{t}
 \hat{G}^A(\omega)\right)_{a_k,\alpha}\\&+& \sum_{k,l}
 \left(\hat{I}+\hat{G}^R(\omega)\hat{t}\right)_{a,\alpha_k}
 \hat{g}^{+,-}_{\alpha_k,\alpha_l}(\omega) \left(\hat{I}+\hat{t}
 \hat{G}^A(\omega)\right)_{\alpha_l,\alpha}
 \nonumber
\end{eqnarray}
and
\begin{eqnarray}
 \label{eq:Gpm-3}
&& \hat{G}_{\alpha,a}^{+,-}(\omega)=\\\nonumber&&\sum_k
 \left(\hat{I}+\hat{G}^R(\omega)\hat{t}\right)_{\alpha,a_k}
 \hat{g}^{+,-}_{a_k,a_k}(\omega) \left(\hat{I}+\hat{t}
 \hat{G}^A(\omega)\right)_{a_k,a} \\&+& \sum_{k,l}
 \left(\hat{I}+\hat{G}^R(\omega)\hat{t}\right)_{\alpha,\alpha_k}
 \hat{g}^{+,-}_{\alpha_k,\alpha_l}(\omega) \left(\hat{I}+\hat{t}
 \hat{G}^A(\omega)\right)_{\alpha_l,a}
 \nonumber
 ,
\end{eqnarray}
where the summation over $k,\,l$ is such that the labels
$\alpha_k,\,\alpha_l$ take the values $\alpha$ or $\beta$, and the
label $a_k$ takes the values $a$ or $b$.

The fully dressed advanced and retarded Green's functions $\hat{G}^A(\omega)$ and
$\hat{G}^R(\omega)$ are the solution of the Dyson equations, which take the
following form in a compact notation:
\begin{equation}
 \label{eq:Dyson1}
 \hat{G}^{A,R}_{\alpha_k,\alpha_l}(\omega) = \hat{g}^{A,R}_{\alpha_k,\alpha_l}
 (\omega) + \sum_m \hat{g}^{A,R}_{\alpha_k,\alpha_m}(\omega)
\hat{t}_{\alpha_m,a_m} \hat{G}^{A,R}_{a_m,\alpha_l}(\omega)
,
\end{equation}
where $\hat{g}^{A,R}(\omega)$ are the bare advanced and retarded
Green's function. At {the} second order in the hopping
amplitudes {$\hat{t}$}, Eq.~(\ref{eq:Dyson1}) takes the form of a
closed set of equations for
$\{\hat{G}^{A,R}_{\alpha_k,\alpha_l}(\omega)\}$:
\begin{eqnarray}
 \label{eq:Dyson2}
 \hat{G}^{A,R}_{\alpha_k,\alpha_l}(\omega) &=&
 \hat{g}^{A,R}_{\alpha_k,\alpha_l} (\omega) \\&+& \sum_{m,n}
 \hat{g}^{A,R}_{\alpha_k,\alpha_m}(\omega) \hat{t}_{\alpha_m,a_m}
 \hat{g}^{A,R}_{a_m,a_n} \hat{t}_{a_n,\alpha_n}
 \hat{G}^{A,R}_{\alpha_n,\alpha_l}(\omega)
 .\nonumber
\end{eqnarray}

In order to distinguish the quasiequilibrium from the the
nonequilibrium currents, we add and subtract
\cite{Melin2009} the equilibrium Keldysh Green's
function
$\hat{G}^{+,-,eq}(\omega)=n_F^{eq}(\omega)\left\{\hat{G}^A(\omega) -
\hat{G}^R(\omega) \right\}$ to Eqs.~(\ref{eq:Gpm-1})-(\ref{eq:Gpm-3}):
\begin{eqnarray}
 \label{eq:decomposition-1}
 \hat{G}_{a,\alpha}^{+,-}(\omega)&=&\delta\hat{G}_{a,\alpha}^{+,-}(\omega)\\
 \nonumber
&+&
 n_F^{eq}(\omega)\left[\hat{G}^A_{a,\alpha}(\omega)-\hat{G}^R_{a,\alpha}(\omega)\right]\\
\label{eq:decomposition-2}
\hat{G}_{\alpha,a}^{+,-}(\omega)&=&\delta\hat{G}_{\alpha,a}^{+,-}(\omega)\\
\nonumber
&+&
n_F^{eq}(\omega)\left[\hat{G}^A_{\alpha,a}(\omega)-\hat{G}^R_{\alpha,a}(\omega)\right]
,
\end{eqnarray}
where
\begin{eqnarray}
 \label{eq:(1)}
&& \delta \hat{G}_{a,\alpha}^{+,-}(\omega)= \\\nonumber&&\sum_{k,l}
 \left(\hat{I}+\hat{G}^R(\omega)\hat{t}\right)_{a,\alpha_k}
 \delta \hat{g}^{+,-}_{\alpha_k,\alpha_l}(\omega) \left(\hat{I}+\hat{t}
 \hat{G}^A(\omega)\right)_{\alpha_l,\alpha}\\
 \label{eq:(1)-prime}
&& \delta\hat{G}_{\alpha,a}^{+,-}(\omega)=\\\nonumber&& \sum_{k,l}
 \left(\hat{I}+\hat{G}^R(\omega)\hat{t}\right)_{\alpha,\alpha_k}
 \delta\hat{g}^{+,-}_{\alpha_k,\alpha_l}(\omega)\left(\hat{I}+\hat{t}
 \hat{G}^A(\omega)\right)_{\alpha_l,a} ,
\end{eqnarray}
with $\delta\hat{g}^{+,-}_{\alpha_k,\alpha_l}(\omega)=
\hat{g}^{+,-}_{\alpha_k,\alpha_l}(\omega)-\hat{g}^{+,-,eq}_{\alpha_k,\alpha_l}(\omega)
$ and
\begin{eqnarray}
 \label{eq:aa1}
&& \hat{g}^{+,-}_{a_k,a_k}(\omega) = \hat{g}^{+,-,eq}_{a_k,a_k}(\omega) =
 n_F^{eq}(\omega) \left[ \hat{g}^A_{a_k,a_k} - \hat{g}^R_{a_k,a_k}
  \right]\\ &&\hat{g}^{+,-}_{\alpha_k,\alpha_l}(\omega) = \hat{n}_F(\omega)
 \left[ \hat{g}^A_{\alpha_k,\alpha_l}(\omega) - \hat{g}^R_{\alpha_k,\alpha_l}(\omega)
  \right] \mbox{and }\\&& \hat{g}^{+,-,eq}_{\alpha_k,\alpha_l}(\omega) =
 \hat{n}_F^{eq}(\omega) \left[ \hat{g}^A_{\alpha_k,\alpha_l}(\omega) -
  \hat{g}^R_{\alpha_k,\alpha_l}(\omega) \right]
 \label{eq:aa1-fin}
 .
\end{eqnarray}
Eq.~(\ref{eq:aa1}) results from the quasiparticle distribution
function in the reservoirs $S_a$ and $S_b$ [with
  $\hat{g}^{+,-}_{a_k,a_k}=\hat{g}^{+,-,eq}_{a_k,a_k}$]. The
nonequilibrium distribution function in the central-$N$
corresponds to $\hat{g}^{+,-}_{\alpha_k,\alpha_l}(\omega) \ne
\hat{g}^{+,-,eq}_{\alpha_k,\alpha_l}(\omega)$ in
Eq.~(\ref{eq:aa1-fin}).

We evaluate in the main text the {\it differential conductance}, i.e. the derivative of the
current $I_a$ with respect to $V_B$:
{
\begin{eqnarray}
  \label{eq:conductance}
 \frac{dI_a}{dV_B} = \frac{\partial I_a}{\partial V_B}+\sum_q \frac{\partial I_a}{\partial
   \mu_{N,q}}\frac{\partial \mu_{N,q}}{\partial V_B} .
\end{eqnarray}}
{It turns out that $\partial I_a/\partial V_b=0$ in
  our model of sequential tunneling, because the current $I_a$ depends
  on $V_B$ only via the effective chemical potentials $\mu_{N,q}$. We
  also note that} evaluating the partial derivative of the current
$I_a$ with respect to the energy $\mu_{N,p}$ involves differentiating
$\hat{g}^{+,-}(\omega)$ with respect to $\mu_{N,p}$ according to
$\partial \hat{g}^{+,-}(\omega)/\partial\mu_{N,p}$:
\begin{eqnarray}
  \nonumber
 \frac{\partial I_a}{\partial \mu_{N,p}}(\omega)&=& \mbox{Nambu-trace}
 \left\{\hat{\tau}_3 \left\{\hat{t}_{\alpha,a}
 \left[\frac{\partial}{\partial\mu_{N,p}}\hat{G}^{+,-}_{a,\alpha}\right](\omega)
 \right.\right.\\
 & -& \left.\left.
 \hat{t}_{a,\alpha} \left[\frac{\partial}{\partial\mu_{N,p}}
  \hat{G}^{+,-}_{\alpha,a}\right](\omega)\right\} \right\} .
\end{eqnarray}
where $\partial \hat{G}_{a,\alpha}^{+,-}(\omega)/\partial\mu_{N,p}$ and
$\partial \hat{G}_{\alpha,a}^{+,-}(\omega)/\partial\mu_{N,p}$ are deduced from
Eqs.~(\ref{eq:decomposition-1})-(\ref{eq:decomposition-2}),
Eqs.~(\ref{eq:(1)})-(\ref{eq:(1)-prime}) and
Eqs.~(\ref{eq:aa1})-(\ref{eq:aa1-fin}):
\begin{eqnarray}
  \nonumber
  \frac{\partial}{\partial\mu_{N,p}}\hat{G}_{a,\alpha}^{+,-}(\omega)&=&
 \sum_{k,l} \left(\hat{I}+\hat{G}^R(\omega)\hat{t}\right)_{a,\alpha_k}
 \left[\frac{\partial}{\partial
   \mu_{N,p}}\hat{g}^{+,-}_{\alpha_k,\alpha_l}\right](\omega)\times\\
 \label{eq:Gpm-3-bis-2}
&& \left(\hat{I}+\hat{t}
 \hat{G}^A(\omega)\right)_{\alpha_l,\alpha}\\
\nonumber
 \frac{\partial}{\partial
  \mu_{N,p}}\hat{G}_{\alpha,a}^{+,-}(\omega)&=& \sum_{k,l}
 \left(\hat{I}+\hat{G}^R(\omega)\hat{t}\right)_{\alpha,\alpha_k}
 \left[\frac{\partial}{\partial
   \mu_{N,p}}\hat{g}^{+,-}_{\alpha_k,\alpha_l}\right](\omega)\times\\
&& \left(\hat{I}+\hat{t} \hat{G}^A(\omega)\right)_{\alpha_l,a} .
 \label{eq:Gpm-3-bis-3}
\end{eqnarray}

We obtain the following for the bare Keldysh Green's functions
appearing in Eqs.~(\ref{eq:Gpm-3-bis-2})-(\ref{eq:Gpm-3-bis-3}):
 \begin{eqnarray}
  \label{eq:partial-gpm1}
&&\frac{\partial
    g^{+,-,1,1}(\{\mu_{N,q}\},\{Z_q\},R_{\alpha_k,\alpha_l},\omega)}{\partial\mu_{N,p}}\simeq\\ \nonumber
  &&2 i Z_p \frac{d n_F(\omega-\mu_{N,p})}{d\mu_{N,p}} J_0(k_e
  R_{\alpha_k,\alpha_l})\\ \label{eq:B} &&\frac{\partial
    g^{+,-,2,2}(\{\mu_{N,q}\},\{Z_q\},R_{\alpha_k,\alpha_l},\omega)}{\partial\mu_{N,p}}
  \simeq\\ \nonumber&&2 i Z _p \frac{d n_F(\omega+\mu_N)}{d\mu_{N,p}}
  J_0(k_h R_{\alpha_k,\alpha_l}) ,
\end{eqnarray}
where we assumed the steps $Z_q$ at the energies $\mu_{N,q}$ in the
central normal metal distribution function. We implement the
zero-temperature limit according to
\begin{eqnarray}
 \label{eq:dd-1}
\frac{d n_F(\omega-\mu_{N,p})}{d\mu_{N,p}}&=&\frac{d
  \theta(\mu_{N,p}-\omega)}{d\mu_{N,p}}\\&=&\delta(\omega-\mu_{N,p})
\nonumber\\\label{eq:dd-2}
\frac{d n_F(\omega+\mu_{N,p})}{d\mu_{N,p}}&=&\frac{d
  \theta(-\mu_{N,p}-\omega)}{d\mu_{N,p}}\\&=&-\delta(\omega+\mu_{N,p}).
\nonumber
\end{eqnarray}

\section{Expression of the current}
\label{app:thecurrent}

In this Appendix, we expand Eq.~(\ref{eq:(1)}) in terms of the central
normal metal-$N_{eff}$ fully dressed Green's functions:
\begin{eqnarray}
 \label{eq:DEBUT}
&& \mbox{$\alpha_k=\alpha$ and $\alpha_l=\alpha$ in
   Eq. }(\ref{eq:(1)})\\\nonumber
 &=& \hat{G}^R_{a,a}(\omega) \hat{t}_{a,\alpha} \delta
 \hat{g}^{+,-}_{\alpha,\alpha}(\omega) \left(\hat{I}+\hat{t}_{\alpha,a}
 \hat{G}^A_{a,\alpha}(\omega)\right)\\\nonumber
 &=& \left(\hat{g}^R_{a,a}
 +\hat{g}^R_{a,a} \hat{t}_{a,\alpha}
 \hat{G}^R_{\alpha,\alpha}(\omega)\hat{t}_{\alpha,a}
 \hat{g}^R_{a,a}\right)\hat{t}_{a,\alpha} \delta
 \hat{g}^{+,-}_{\alpha,\alpha}(\omega) \times\\ \nonumber
 &&\left(\hat{I} + \hat{t}_{\alpha,a}
 \hat{g}^A_{a,a} \hat{t}_{a,\alpha}
 \hat{G}^A_{\alpha,\alpha}(\omega)\right)\\ &=& \hat{g}^R_{a,a}
 \hat{t}_{a,\alpha} \delta \hat{g}^{+,-}_{\alpha,\alpha}(\omega) \mbox{
  [order-$0$ in powers of $G$]}\\\nonumber &&+ \hat{g}^R_{a,a}
 \hat{t}_{a,\alpha} \hat{G}^R_{\alpha,\alpha}(\omega) \hat{t}_{\alpha,a}
 \hat{g}^R_{a,a} \hat{t}_{a,\alpha} \times\\&&\delta
 \hat{g}^{+,-}_{\alpha,\alpha}(\omega) \mbox{[order-$1$ in powers of
   $G$]}\label{eq:O1-1}\\\nonumber &&+ \hat{g}^R_{a,a} \hat{t}_{a,\alpha}
 \delta \hat{g}^{+,-}_{\alpha,\alpha}(\omega) \hat{t}_{\alpha,a}
 \hat{g}^A_{a,a} \hat{t}_{a,\alpha} \times\\&&
 \hat{G}^A_{\alpha,\alpha}(\omega) \mbox{
  [order-$1$ in powers of $G$]}\label{eq:O1-2}\\\nonumber &&+ \hat{g}^R_{a,a}
 \hat{t}_{a,\alpha} \hat{G}^R_{\alpha,\alpha}(\omega) \hat{t}_{\alpha,a}
 \hat{g}^R_{a,a} \hat{t}_{a,\alpha} \delta
 \hat{g}^{+,-}_{\alpha,\alpha}(\omega) \hat{t}_{\alpha,a} \hat{g}^A_{a,a} \times\\
&& \hat{t}_{a,\alpha} \hat{G}^A_{\alpha,\alpha}(\omega) \mbox{[order-$2$ in
   powers of $G$]}\label{eq:O2-1}
\end{eqnarray}
\begin{eqnarray}
&& \mbox{$\alpha_k=\beta$ and $\alpha_l=\alpha$ in
    Eq. }(\ref{eq:(1)})\\\nonumber
  &=& \hat{G}^R_{a,b} \hat{t}_{b,\beta} \delta
  \hat{g}^{+,-}_{\beta,\alpha}(\omega)
  \left(\hat{I}+\hat{t}_{\alpha,a}
  \hat{G}^A_{a,\alpha}(\omega)\right)\\\nonumber
  &=& \hat{g}^R_{a,a}
  \hat{t}_{a,\alpha} \hat{G}^R_{\alpha,\beta}(\omega)
  \hat{t}_{\beta,b} \hat{g}^R_{b,b} \hat{t}_{b,\beta} \delta
  \hat{g}^{+,-}_{\beta,\alpha}(\omega) \times\\ \nonumber
  &&\left(\hat{I} + \hat{t}_{\alpha,a} \hat{g}^A_{a,a}
  \hat{t}_{a,\alpha} \hat{G}^A_{\alpha,\alpha}(\omega)\right)\\\nonumber &=&
  \hat{g}^R_{a,a} \hat{t}_{a,\alpha} \hat{G}^R_{\alpha,\beta}(\omega)
  \hat{t}_{\beta,b} \hat{g}^R_{b,b} \hat{t}_{b,\beta}
  \times\\&&\delta \hat{g}^{+,-}_{\beta,\alpha}(\omega)\mbox{
    [order-$1$ in powers of $G$]}\label{eq:O1-3}\\\nonumber &&+ \hat{g}^R_{a,a}
  \hat{t}_{a,\alpha} \hat{G}^R_{\alpha,\beta}(\omega)
  \hat{t}_{\beta,b} \hat{g}^R_{b,b} \hat{t}_{b,\beta} \delta
  \hat{g}^{+,-}_{\beta,\alpha}(\omega)\hat{t}_{\alpha,a}
  \hat{g}^A_{a,a} \times\\  && \hat{t}_{a,\alpha}
  \hat{G}^A_{\alpha,\alpha}(\omega) \mbox{[order-$2$ in powers of
      $G$]} \label{eq:O2-2}
\end{eqnarray}
\begin{eqnarray}
&& \mbox{$\alpha_k=\alpha$ and $\alpha_l=\beta$ in
    Eq. }(\ref{eq:(1)})\\\nonumber
  &=& \hat{G}^R_{a,a}(\omega) \hat{t}_{a,\alpha} \delta
 \hat{g}^{+,-}_{\alpha,\beta}(\omega) \hat{t}_{\beta,b} \hat{g}^A_{b,b}
 \hat{t}_{b,\beta} \hat{G}^{A}_{\beta,\alpha}(\omega)\\\nonumber&=&
 \left(\hat{g}^R_{a,a} +\hat{g}^R_{a,a}\hat{t}_{a,\alpha}
 \hat{G}^R_{\alpha,\alpha}(\omega) \hat{t}_{\alpha,a}\hat{g}^R_{a,a}\right)
 \hat{t}_{a,\alpha} \delta \hat{g}^{+,-}_{\alpha,\beta}(\omega)\times\\
 \nonumber
 &&
 \hat{t}_{\beta,b} \hat{g}^A_{b,b} \hat{t}_{b,\beta}
 \hat{G}^{A}_{\beta,\alpha}(\omega)\\ \nonumber &=& \hat{g}^R_{a,a} \hat{t}_{a,\alpha}
 \delta \hat{g}^{+,-}_{\alpha,\beta}(\omega) \hat{t}_{\beta,b}
 \hat{g}^A_{b,b} \times\\
 &&\hat{t}_{b,\beta}\hat{G}^A_{\beta,\alpha}(\omega) \mbox{
  [order-$1$ in powers of $G$]}\label{eq:O1-4}\\\nonumber &&+ \hat{g}^R_{a,a}
 \hat{t}_{a,\alpha} \hat{G}^R_{\alpha,\alpha}(\omega) \hat{t}_{\alpha,a}
 \hat{g}^R_{a,a} \hat{t}_{a,\alpha} \delta
 \hat{g}^{+,-}_{\alpha,\beta}(\omega)\hat{t}_{\beta,b} \hat{g}^A_{b,b}\times\\
 && 
 \hat{t}_{b,\beta} \hat{G}^A_{\beta,\alpha}(\omega)\mbox{[order-$2$ in
   powers of $G$]} \label{eq:O2-3}
\end{eqnarray}
\begin{eqnarray}
&& \mbox{$\alpha_k=\beta$ and $\alpha_l=\beta$ in
    Eq. }(\ref{eq:(1)})\\\nonumber&=& \hat{G}^R_{a,b}(\omega) \hat{t}_{b,\beta}
  \delta \hat{g}^{+,-}_{\beta,\beta}(\omega) \hat{t}_{\beta,b}
  \hat{G}^{A}_{b,\alpha}(\omega)\\ \nonumber &=& \hat{g}^R_{a,a}
  \hat{t}_{a,\alpha} \hat{G}^R_{\alpha,\beta}(\omega)
  \hat{t}_{\beta,b} \hat{g}^R_{b,b} \hat{t}_{b,\beta} \delta
  \hat{g}^{+,-}_{\beta,\beta}(\omega) \hat{t}_{\beta,b}
  \hat{g}^A_{b,b}\times\\  && \hat{t}_{b,\beta}
  \hat{G}^A_{\beta,\alpha}(\omega)\mbox{[order-$2$ in powers of
      $G$]} \label{eq:O2-4} ,
\end{eqnarray}
where the order-$0$, order-$1$ and order-$2$ terms correspond to zero,
one or two fully dressed Green's functions in the expression of the
current.

Eq.~(\ref{eq:(1)-prime}) is expressed as follows:
\begin{eqnarray}
&& \mbox{$\alpha_k=\alpha$ and $\alpha_l=\alpha$ in
    Eq. }(\ref{eq:(1)-prime})\\\nonumber&=&
  \left(\hat{I}+\hat{G}^R_{\alpha,a}(\omega)\hat{t}_{a,\alpha}\right)
  \delta \hat{g}^{+,-}_{\alpha,\alpha}(\omega) \hat{t}_{\alpha,a}
  \hat{G}^A_{a,a}(\omega)\\ \nonumber&=&
  \left(\hat{I}+\hat{G}^R_{\alpha,\alpha}(\omega)\hat{t}_{\alpha,a}\hat{g}^R_{a,a}\hat{t}_{a,\alpha}\right)
  \delta \hat{g}^{+,-}_{\alpha,\alpha}(\omega) \times\\\nonumber
  &&\hat{t}_{\alpha,a} \left(\hat{g}^A_{a,a} +
  \hat{g}^A_{a,a}\hat{t}_{a,\alpha}\hat{G}^A_{\alpha,\alpha}(\omega)\hat{t}_{\alpha,a}
  \hat{g}^A_{a,a}\right)\\ &=& \delta
  \hat{g}^{+,-}_{\alpha,\alpha}(\omega) \hat{t}_{\alpha,a}
  \hat{g}^A_{a,a} \mbox{[order-$0$ in powers of $G$]}\\ \nonumber&&+
  \hat{G}^R_{\alpha,\alpha}(\omega) \hat{t}_{\alpha,a} \hat{g}^R_{a,a}
  \hat{t}_{a,\alpha} \delta \hat{g}^{+,-}_{\alpha,\alpha}(\omega)
  \times\\&&\hat{t}_{\alpha,a} \hat{g}^A_{a,a} \mbox{
    [order-$1$ in powers of $G$]}\label{eq:O1-1-bis}\\ \nonumber&&+
  \delta \hat{g}^{+,-}_{\alpha,\alpha}(\omega) \hat{t}_{\alpha,a}
  \hat{g}^A_{a,a} \hat{t}_{a,\alpha}
  \hat{G}^A_{\alpha,\alpha}(\omega)\times\\&&
  \hat{t}_{\alpha,a} \hat{g}^A_{a,a} \mbox{[order-$1$ in powers of
      $G$]}\label{eq:O1-2-bis}\\ \nonumber &&+
  \hat{G}^R_{\alpha,\alpha}(\omega)\hat{t}_{\alpha,a}\hat{g}^R_{a,a}\hat{t}_{a,\alpha}
  \delta \hat{g}^{+,-}_{\alpha,\alpha}(\omega)\hat{t}_{\alpha,a} \hat{g}^A_{a,a}
  \times\\&&
  \hat{t}_{a,\alpha} \hat{G}^A_{\alpha,\alpha}(\omega)
  \hat{t}_{\alpha,a} \hat{g}^A_{a,a} \mbox{[order-$2$ in powers of
      $G$]} \label{eq:O2-prime-1}
\end{eqnarray}
\begin{eqnarray}
&& \mbox{$\alpha_k=\beta$ and $\alpha_l=\alpha$ in
  Eq. }(\ref{eq:(1)-prime})\\\nonumber&=& \hat{G}^R_{\alpha,b}(\omega)\hat{t}_{b,\beta}
 \delta \hat{g}^{+,-}_{\beta,\alpha}(\omega) \hat{t}_{\alpha,a}
 \hat{G}^A_{a,a}(\omega)\\ \nonumber&=& \hat{G}^R_{\alpha,\beta}(\omega) \hat{t}_{\beta,b}
 \hat{g}^R_{b,b} \hat{t}_{b,\beta} \delta
 \hat{g}^{+,-}_{\beta,\alpha}(\omega)\times\\\nonumber
 && \hat{t}_{\alpha,a} \left(
 \hat{g}^A_{a,a} + \hat{g}^A_{a,a} \hat{t}_{a,\alpha}
 \hat{G}^A_{\alpha,\alpha}(\omega) \hat{t}_{\alpha,a} \hat{g}^A_{a,a}
 \right)\\ \nonumber &=& \hat{G}^R_{\alpha,\beta}(\omega) \hat{t}_{\beta,b}
 \hat{g}^R_{b,b} \hat{t}_{b,\beta} \delta
 \hat{g}^{+,-}_{\beta,\alpha}(\omega) \times\\&&\hat{t}_{\alpha,a}
 \hat{g}^A_{a,a}\mbox{[order-$1$ in powers of $G$]}\label{eq:O1-3-bis}\\
 \nonumber &&+
 \hat{G}^R_{\alpha,\beta}(\omega) \hat{t}_{\beta,b} \hat{g}^R_{b,b}
 \hat{t}_{b,\beta} \delta \hat{g}^{+,-}_{\beta,\alpha}(\omega) \hat{t}_{\alpha,a} \hat{g}^A_{a,a}
 \times\\&& \hat{t}_{a,\alpha}
 \hat{G}^A_{\alpha,\alpha}(\omega) \hat{t}_{\alpha,a} \hat{g}^A_{a,a}\mbox{
  [order-$2$ in powers of $G$]} \label{eq:O2-prime-2}
\end{eqnarray}
\begin{eqnarray}
&& \mbox{$\alpha_k=\alpha$ and $\alpha_l=\beta$ in
  Eq. }(\ref{eq:(1)-prime})\\\nonumber &=&
 \left(\hat{I}+\hat{G}^R_{\alpha,a}(\omega)\hat{t}_{a,\alpha}\right) \delta
 \hat{g}^{+,-}_{\alpha,\beta}(\omega) \hat{t}_{\beta,b} \hat{G}^A_{b,a}(\omega)\\
 \nonumber &=&
 \left(\hat{I}+\hat{G}^R_{\alpha,\alpha}(\omega) \hat{t}_{\alpha,a}
 \hat{g}^R_{a,a} \hat{t}_{a,\alpha}\right) \delta
 \hat{g}^{+,-}_{\alpha,\beta}(\omega)\times\\\nonumber && \hat{t}_{\beta,b} \hat{g}^A_{b,b}
 \hat{t}_{b,\beta} \hat{G}^A_{\beta,\alpha}(\omega) \hat{t}_{\alpha,a}
 \hat{g}^A_{a,a}\\\nonumber &=& \delta \hat{g}^{+,-}_{\alpha,\beta}(\omega)
 \hat{t}_{\beta,b} \hat{g}^A_{b,b} \hat{t}_{b,\beta}
 \hat{G}^A_{\beta,\alpha}(\omega) \times\\&&\hat{t}_{\alpha,a} \hat{g}^A_{a,a}\mbox{
  [order-$1$ in powers of $G$]}\label{eq:O1-4-bis}\\ &&+\hat{G}^R_{\alpha,\alpha}(\omega)
 \hat{t}_{\alpha,a} \hat{g}^R_{a,a} \hat{t}_{a,\alpha} \delta
 \hat{g}^{+,-}_{\alpha,\beta}(\omega)\hat{t}_{\beta,b} \hat{g}^A_{b,b}
 \times \nonumber\\&& 
 \hat{t}_{b,\beta} \hat{G}^A_{\beta,\alpha}(\omega) \hat{t}_{\alpha,a}
 \hat{g}^A_{a,a}\mbox{[order-$2$ in powers of
   $G$]}\label{eq:O2-prime-3}
\end{eqnarray}
\begin{eqnarray}
&&\mbox{$\alpha_k=\beta$ and $\alpha_l=\beta$ in
    Eq. }(\ref{eq:(1)-prime})\\\nonumber&=&
  \hat{G}^R_{\alpha,b}(\omega)\hat{t}_{b,\beta} \delta
  \hat{g}^{+,-}_{\beta,\beta}(\omega) \hat{t}_{\beta,b}
  \hat{G}^A_{b,a}(\omega)\\ \nonumber&=& \hat{G}^R_{\alpha,\beta}(\omega)
  \hat{t}_{\beta,b} \hat{g}^R_{b,b} \hat{t}_{b,\beta} \delta
  \hat{g}^{+,-}_{\beta,\beta}(\omega)\hat{t}_{\beta,b} \hat{g}^A_{b,b}
  \times\\&& \hat{t}_{b,\beta} \hat{G}^A_{\beta,\alpha}(\omega)
  \hat{t}_{\alpha,a} \hat{g}^A_{a,a}\mbox{[order-$2$ in powers of
      $G$]}. \label{eq:O2-prime-4}
\end{eqnarray}

\section{Zero-dimensional limit}

\label{sec:0D}

{In this Appendix, we present} {a demonstration of
  Eq.~(\ref{eq:final}) that is physically discussed in
  section~\ref{sec:num-dot}.  We consider the simple 0D limit} {with
  vanishingly small Dynes parameter}{, see} the device in
Fig.~\ref{fig:thedevice}f consisting of a single 0D tight-binding site
connected by the hopping amplitude $t_{x,a}$, $t_{x,b}$ and
$t_{x,N_B}$ to the superconducting leads $S_a$, $S_b$ and to the
normal-metallic $N_B$ respectively. We assume low bias voltage $V_B$
and treat the superconductors $S_a$ and $S_b$ in the large-gap
approximation. In a standard notation, we parameterize the interface
transparencies by $\Gamma_a=t_{x,a}^2/W$, $\Gamma_b=t_{x,b}^2/W$ and
$\gamma_B=t_{x,N_B}^2/W$, where $W$ is the band-width, taken identical
in all of the leads. The two superconducting leads $S_a$ and $S_b$ are
assumed to be grounded, and the third normal lead $N_B$ is biased at
the voltage $V_B$.

{{\it ABS energies:} The Dyson
  Eqs.~(\ref{eq:Dyson1})-(\ref{eq:Dyson2}) take the following form for
  the device in Fig.~\ref{fig:thedevice}f:
\begin{equation}
 \label{eq:Gxx}
 \hat{G}_{x,x}(\omega)=\left[ \left(\hat{g}_{x,x}\right)^{-1}(\omega)
  -\hat{\Gamma}_{N_B,N_B}
  -\hat{\Gamma}_{S,S} \right]^{-1}
 .
\end{equation}
We implemented the large-gap approximation and used the following notations:
\begin{eqnarray}
 \hat{g}_{x,x}(\omega)&=&\left(\begin{array}{cc} \omega-\epsilon_0 & 0\\
  0 & \omega+\epsilon_0 \end{array}\right)^{-1}\\
 \hat{\Gamma}_{N_B,N_B}&=&\left(\begin{array}{cc} i \gamma_B & 0\\
  0 & i \gamma_B \end{array}\right)\\
 \hat{\Gamma}_{S,S}&=&\left(\begin{array}{cc}0 &-\Gamma_a e^{i\varphi_a}-\Gamma_b e^{i\varphi_b}\\
  -\Gamma_a e^{-i\varphi_a}-\Gamma_b e^{-i\varphi_b}& 0 \end{array}\right)
 \label{eq:derniere-eq1}
 ,
\end{eqnarray}
where $\hat{g}_{x,x}(\omega)$ and $\epsilon_0$ are the bare Green's
function on the single tight-binding site $x$ and the on-site energy.
The resolvent given by Eq.~(\ref{eq:Gxx}) is related to a
non-Hermitian {effective} Hamiltonian:
$\hat{G}_{x,x}(\omega)=\left(\omega-\hat{\cal H}_{eff}\right)^{-1}$,
with
\begin{equation}
 \label{eq:H-eff}
 \hat{\cal H}_{eff}=\left( \begin{array}{cc}\epsilon_0 +i\gamma_B
  & -\Gamma_a e^{i\varphi_a}-\Gamma_b e^{i\varphi_b} \\ -\Gamma_a
  e^{-i\varphi_a}-\Gamma_b e^{-i\varphi_b}&
  -\epsilon_0 +i\gamma_B\end{array} \right) .
\end{equation}
The ABS energies are at $\omega=\Omega_\pm$, with
\begin{equation}
 \label{eq:ABS-ener}
 \Omega_\pm = i \gamma_B \pm \sqrt{\epsilon_0^2 + \left|\Gamma_a
   e^{i\varphi_a}+\Gamma_b e^{i\varphi_b}\right|^2}
 .
\end{equation}}

{The coupling to the normal lead produces
  nonvanishingly small line-width broadening captured by the
  complex-valued energies of the non-Hermitian Hamiltonian, see
  Eq.~(\ref{eq:H-eff}).}

{{\it Andreev reflection current $I_B$:} The current
  transmitted from the normal lead $N_B$ to the 0D single
  tight-binding site $x$ connected to the superconducting leads $S_a$
  and $S_b$ involves an incoming spin-$\sigma$ electron and an
  outgoing hole in the spin-$(-\sigma)$ band. The resulting Andreev
  reflection current scales like $\gamma_B^2$ if $\gamma_B$ is
  small, because each of the spin-$\sigma$ electron and
  spin-($-\sigma$) hole have to cross the interfaces.}

{{\it Small coupling to the normal lead $N_B$:} Let us
  first demonstrate that the current $I_B=0$ is vanishingly small at
  the lowest-order-$\gamma_B$. We practically implement
  perturbation theory in $\gamma_B$ assumed to be much smaller
  than $\Gamma_a$ and $\Gamma_b$, i.e. $\gamma_B\ll \Gamma_a$ and
  $\gamma_B\ll\Gamma_b$:
  {
\begin{eqnarray}
 \hat{t}_{N_B,x} \hat{G}^{+,-}_{x,N_B}(\omega)&=&
 \hat{t}_{N_B,x} \hat{G}^{+,-,(0)}_{x,x}(\omega) \hat{t}_{x,N_B} \hat{g}^A_{N_B,N_B}\\
 &+& 
 \hat{t}_{N_B,x} \hat{G}^{R,(0)}_{x,x}(\omega) \hat{t}_{x,N_B} \hat{g}^{+,-}_{N_B,N_B}(\omega)
\nonumber
 ,
\end{eqnarray}}
with
\begin{equation}
 \hat{G}^{+,-,(0)}_{x,x}(\omega) = n_F(\omega) \left[ G^{A,(0)}_{x,x}(\omega)
  - G^{R,(0)}_{x,x}(\omega) \right]
\end{equation}
and
\begin{eqnarray}
 \hat{g}^{+,-,1,1}_{N_B,N_B}(\omega) &=& n_F(\omega-eV_B) \left[ g^A_{N_B,N_B} -
  g^R_{N_B,N_B}\right]\\ \hat{g}^{+,-,2,2}_{N_B,N_B}(\omega) &=& n_F(\omega+eV_B)
 \left[ g^A_{N_B,N_B} - g^R_{N_B,N_B}\right] ,
\end{eqnarray}
and where $G^{R,(0)}_{x,x}(\omega)$ is the fully dressed Green's
function in the absence of coupling to the normal lead, i.e. with
$\gamma_B=0$:
\begin{eqnarray}
 \label{eq:G(0)-1}
 G^{R,(0),1,1}_{x,x}(\omega)&=&\\&&\nonumber
 \frac{\omega+\epsilon_0}{\omega^2-\epsilon_0^2 - |\Gamma_a
   \exp(i\varphi_a)+\Gamma_b\exp(i\varphi_b)|^2}\\ \label{eq:G(0)-2}
 G^{R,(0),2,2}_{x,x}(\omega) &=&\\&&\nonumber
 \frac{\omega-\epsilon_0}{\omega^2-\epsilon_0^2 - |\Gamma_a
   \exp(i\varphi_a)+\Gamma_b\exp(i\varphi_b)|^2} .
\end{eqnarray}
We deduce
\begin{widetext}
\begin{eqnarray}
 \frac{d}{d(eV_B)} \left(\hat{t}_{N_B,x}
 \hat{G}^{+,-}_{x,N_B}(\omega)\right)^{1,1} &\simeq&
 \delta(\omega-eV_B)\left[t_{N_B,x}^{1,1} G^{R,(0),1,1}_{x,x}(\omega)
  t_{x,N_B}^{1,1} \left( g^{A,1,1}_{N_B,N_B} -
  g^{R,1,1}_{N_B,N_B}\right)\right]\\ \frac{d}{d(eV_B)}
 \left(\hat{t}_{N_B,x} \hat{G}^{+,-}_{x,N_B}(\omega)\right)^{2,2} &\simeq&-
 \delta(\omega+eV_B)\left[t_{N_B,x}^{2,2} G^{R,(0),2,2}_{x,x}(\omega)
  t_{x,N_B}^{2,2} \left( g^{A,2,2}_{N_B,N_B} -
  g^{R,2,2}_{N_B,N_B}\right)\right] .
\end{eqnarray}
Integrating over the energy $\omega$ leads to
\begin{equation}
 \int d\omega \left\{\frac{d}{d(eV_B)} \left(\hat{t}_{N_B,x}
 \hat{G}^{+,-}_{x,N_B}(\omega)\right)^{1,1} -\frac{d}{d(eV_B)}
 \left(\hat{t}_{N_B,x} \hat{G}^{+,-}_{x,N_B}(\omega)\right)^{2,2} \right\} \simeq
 \frac{2it_{x,N_B}^2}{W}\left[ G^{R,(0),1,1}_{x,x}(\omega=eV_B) +
  G^{R,(0),2,2}_{x,x}(\omega=-eV_B)\right] .
\end{equation}
\end{widetext}
It is deduced from Eqs.~(\ref{eq:G(0)-1})-(\ref{eq:G(0)-2}) that
$G^{R,(0),1,1}_{x,x}(\omega=eV_B) +
G^{R,(0),2,2}_{x,x}(\omega=-eV_B) = 0$, which implies
$dI_B/dV_B=0$ at the lowest-order-$\gamma_B$ because the
Andreev processes are of order $\gamma_B^2$, as it was
anticipated in the above discussion.
}

{{\it Strong coupling to the normal lead $N_B$:} Now, we evaluate an
  approximation to $dI_B/dV_B$ beyond the order $\gamma_B$, starting
  from the opposite {Andreev} limit of weak coupling
  to the superconducting leads, i.e. we implement the following
  assumptions about $\Gamma_a$, $\Gamma_b$ and $\gamma_B$:
  $\Gamma_a\ll \gamma_B$ and $\Gamma_b \ll \gamma_B$.}

{Similarly to Eq.~(\ref{eq:I1-model-A}), the current
  is expressed as
\begin{equation}
 \label{eq:I-NB}
 I_B(\omega)=\mbox{Nambu-trace}\left\{\hat{\tau}_3
 \left[\hat{t}_{N_B,x}\hat{G}^{+,-}_{x,N_B}(\omega)-
  \hat{t}_{x,N_B} \hat{G}^{+,-}_{N_B,x}(\omega)\right]\right\}
 ,
\end{equation}
where $\hat{t}_{N_B,x}\hat{G}^{+,-}_{x,N_B}(\omega)$ and
$\hat{t}_{x,N_B} \hat{G}^{+,-}_{N_B,x}(\omega)$ are given by
\begin{eqnarray}
&& \hat{t}_{N_B,x}\label{eq:M1}
  \hat{G}^{+,-}_{x,N_B}(\omega)=\\
\nonumber
  &&\hat{t}_{N_B,x}\hat{G}^R_{x,x}(\omega)\hat{t}_{x,N_B}
 \hat{g}^{+,-}_{N_B,N_B} \hat{t}_{N_B,x} \hat{G}^A_{x,x}(\omega)
 \hat{t}_{x,N_B} \hat{g}^A_{N_B,N_B}\\   \nonumber
&+&
 \hat{t}_{N_B,x}\hat{G}^R_{x,x}(\omega)\hat{t}_{x,N_B}
 \hat{g}^{+,-}_{N_B,N_B}(\omega)\\ \label{eq:N2}
&& \hat{t}_{x,N_B} \hat{G}^{+,-}_{N_B,x}(\omega)=\\
\nonumber
&& \hat{t}_{x,N_B} \hat{g}^R_{N_B,N_B} \hat{t}_{N_B,x}
 \hat{G}^R_{x,x}(\omega) \hat{t}_{x,N_B} \hat{g}^{+,-}_{N_B,N_B}(\omega)
 \hat{t}_{N_B,x} \hat{G}^A_{x,x}(\omega)\\&+&\hat{t}_{x,N_B}
 \hat{g}^{+,-}_{N_B,N_B}(\omega) \hat{t}_{N_B,x} \hat{G}^A_{x,x}(\omega)
 .\nonumber
\end{eqnarray}
The fully dressed advanced
Green's function of the 0D single tight-binding site
$\hat{G}_{x,x}(\omega)$ takes the following form at the orders
$\Gamma_a$ and $\Gamma_b$:
\begin{eqnarray}
 \label{eq:Gxx-debut}
&& G_{x,x}^{1,1}(\omega)\simeq \left(\omega-\epsilon-i\gamma_B\right)^{-1}
 +{\cal O}\left(\Gamma^2\right)\\
 && G_{x,x}^{1,2}(\omega)\simeq \\\nonumber
 &-&\left(\omega-\epsilon-i\gamma_B\right)^{-1}
 \left(\Gamma_a e^{i\varphi_a}+\Gamma_b e^{i\varphi_b}\right)
 \left(\omega+\epsilon-i\gamma_B\right)^{-1}\\&+&{\cal O}\left(\Gamma^3\right)
 \nonumber\\
&& G_{x,x}^{2,1}(\omega)\simeq\\\nonumber &-&\left(\omega+\epsilon-i\gamma_B\right)^{-1}
 \left(\Gamma_a e^{-i\varphi_a}+\Gamma_b e^{-i\varphi_b}\right)
 \left(\omega-\epsilon-i\gamma_B\right)^{-1}\\\nonumber
 &+&{\cal O}\left(\Gamma^3\right)\\
&& G_{x,x}^{2,2}(\omega)\simeq \left(\omega+\epsilon-i\gamma_B\right)^{-1}
 +{\cal O}\left(\Gamma^2\right)
 .
 \label{eq:Gxx-fin}
\end{eqnarray}
{Inserting
  Eqs.~(\ref{eq:Gxx-debut})-(\ref{eq:Gxx-fin}) into
  Eqs.~(\ref{eq:M1})-(\ref{eq:N2}) leads to Eq.~(\ref{eq:final}) for
  the final expression of the differential conductance
  $dI_B/dV_B$. The above subsection~\ref{sec:num-dot} presents a
  physical discussion of this Eq.~(\ref{eq:final}).}

\section{Adiabatic limit}
\label{app:adiab}
{In this Appendix, we present {calculations for} the current in the
  adiabatic limit of low bias voltage $V_B$, for an
  all-superconducting Andreev interferometer {biased
    at $(0,\,0,\,V)$. Specifically,} we consider a 0D quantum dot
  connected to three superconducting leads with the phase variable
  $\varphi_1=-\Phi/2$, $\varphi_2=\Phi/2$ and
  $\varphi_3=\varphi_B$. We assume that the energy/voltage scales are
  small compared to the superconducting gap. Then, the $2\times 2$
  Bogoliubov-de Gennes Hamiltonian takes the following form
\begin{equation}
  {\cal H}=\left(\begin{array}{cc} 0 & \sum_{k=1}^3 \Gamma_k
    \exp\left(i\varphi_k\right)\\ \sum_{k=1}^3 \Gamma_k
    \exp\left(-i\varphi_k\right) & 0 \end{array}\right) ,
\end{equation}
where we use the standard notation $\Gamma_k=t_k^2/W$, with $t_k$ the
hopping amplitude between the quantum dot and the superconducting lead
$S_k$, and $W$ the band-width.}

{The ABS energies are given by
  {
    \begin{eqnarray}
      \label{eq:E1}
      && E_\pm\left(\frac{\Phi}{\Phi_0},\varphi_B\right)\\ &=&\pm
      \left|\Gamma_a \exp\left(-\frac{2i\pi\Phi}{\Phi_0}\right) +
      \Gamma_b \exp\left(\frac{2i\pi\Phi}{\Phi_0}\right)+ \Gamma_B
      \exp(i\varphi_B)\right|.  \nonumber
\end{eqnarray}}
We use the notation $\varphi_B=\varphi_B^{(0)} + \frac{2 e V_B t}{\hbar}$,
with $V_B$ a (small) voltage on $S_B$ and average Eq.~(\ref{eq:E1})
over $\varphi_B$ in the adiabatic limit:
{
\begin{eqnarray}
&& \langle E_\pm\rangle\left(\frac{\Phi}{\Phi_0}\right)= \int \frac{d
    \varphi_B}{2\pi} E_\pm\left(\frac{\Phi}{\Phi_0},\varphi_B\right)=
  \pm \int \frac{d \varphi_B}{2\pi} \times\\ \nonumber &&
  \left|\Gamma_a \exp\left(-\frac{2i\pi\Phi}{\Phi_0}\right) + \Gamma_b
  \exp\left(\frac{2i\pi\Phi}{\Phi_0}\right) + \Gamma_B
  \exp(i\varphi_B)\right| .
\end{eqnarray}}
Then, we obtain two ladders of Floquet states at the energies
{$\langle E_\pm\rangle\left(\frac{\Phi}{\Phi_0}\right)+2nV_B$,}
with $n$ an integer, which is consistent with our previous paper on
the spectroscopy of the Floquet-Andreev ladders using finite frequency
noise, see Ref.~\onlinecite{Melin2019}.}

{The supercurrent is obtained from differentiating
  Eq.~(\ref{eq:E1}) with respect to $\varphi_B$:
  {
\begin{eqnarray}
  && I_\pm\left(\frac{\Phi}{\Phi_0},\varphi_B\right)=\mp\frac{2e}{\hbar}
  \times\\ \nonumber &&\frac{\partial \left|\Gamma_a
    \exp\left(-\frac{2i\pi\Phi}{\Phi_0}\right) + \Gamma_b
    \exp\left(\frac{2i\pi\Phi}{\Phi_0}\right) +
    \Gamma_B \exp(i\varphi_B)\right|} {\partial \varphi_B} .
\end{eqnarray}}
}

{Considering symmetric contacts with
  $\Gamma_a=\Gamma_b\equiv\Gamma$, we obtain
  {
\begin{eqnarray}
  \label{eq:E2}
  &&  E_\pm\left(\frac{\Phi}{\Phi_0},\varphi_B\right)=\\
\nonumber
  &&\pm \sqrt{
    4 \Gamma^2 \cos^2\left(\frac{\pi\Phi}{\Phi_0}\right) + \Gamma_B^2+ 4 \Gamma \Gamma_B \cos\left(\frac{\pi\Phi}{\Phi_0}\right) \cos\varphi_B}
\end{eqnarray}}
  and
  {
\begin{eqnarray}
  \label{eq:Ipm}
&&  \langle I_\pm\rangle\left(\frac{\Phi}{\Phi_0}\right)=\pm \frac{2e}{\hbar}\int_0^{2\pi} \frac{d \varphi_B}{2\pi}
\times\\
\nonumber
&&  \frac{2 \Gamma
    \Gamma_B \cos\left(\frac{\pi\Phi}{\Phi_0}\right) \sin\varphi_B}{\sqrt{4 \Gamma^2
      \cos^2\left(\frac{\pi\Phi}{\Phi_0}\right) + \Gamma_B^2 + 4 \Gamma \Gamma_B \cos\left(\frac{\pi\Phi}{\Phi_0}\right)
      \cos\varphi_B}} .
\end{eqnarray}}
Due to the antisymmetry in $\varphi_B\rightarrow - \varphi_B$, we
conclude that the supercurrent {$\langle I_\pm
  \rangle\left(\frac{\Phi}{\Phi_0}\right)$} in Eq.~(\ref{eq:Ipm}) {is vanishingly
  small} in the adiabatic limit for symmetric contacts.}

{In the general case of asymmetric contacts, we find
  {
\begin{eqnarray}
  \label{eq:Ipm2}
  && \langle I_\pm\rangle\left(\frac{\Phi}{\Phi_0}\right)=
  -\frac{2}{\hbar} \int_0^{2\pi} \frac{d \varphi_B}{2\pi}
  \times\\ \nonumber && \frac{\partial \left|\Gamma_a
    \exp\left(-\frac{2i\pi\Phi}{\Phi_0}\right) + \Gamma_b
    \exp\left(\frac{2i\Phi}{\Phi_0}\right) + \Gamma_B
    \exp(i\varphi_B)\right|} {\partial \varphi_B}\\ &=&
  -\frac{2}{\hbar} \frac{1}{2\pi}\left[\left|\Gamma_a
    \exp\left(-\frac{2i\pi\Phi}{\Phi_0}\right) + \Gamma_b
    \exp\left(\frac{2i\pi\Phi}{\Phi_0}\right)\right.\right.\\\nonumber&&
    +\left.\left.  \Gamma_B
    \exp(i\varphi_B)\right|\right]_{\varphi_B=0}^{\varphi_B=2\pi}\\ &=&0
  .
\end{eqnarray}}
Thus, {$\langle
  I_\pm\rangle\left(\frac{\Phi}{\Phi_0}\right)$} vanishes for
arbitrary values of the asymmetry in the contact transparencies and in
the adiabatic limit.}

{The interpretation is in terms of energy conservation: A
  DC-supercurrent of Cooper pair does not flow between the grounded
  loop and $S_B$ biased at the voltage $V_B$. Instead, we have
  AC-Josephson oscillations of the current} {as soon
  as the superconducting phases are running.}

{\section{Conjecture on multiterminal Josephson
    billiards}\label{app:billiards}}

{In this Appendix, we discuss an arbitrary shape of the
  central-$N$ connected to an arbitrary number of superconducting
  leads, thus forming a {\it multiterminal Josephson billiard}, see
  Fig.~\ref{fig:billiards}. Related models were recently developed in
  Ref.~\onlinecite{Pankratova2020} to interpret their experiments on
  the critical current contours of four-terminal Josephson junctions
  as a function of the magnetic field. In the absence of coupling to
  the superconductors, billiards can be {\it integrable} or {\it
    chaotic}, such as the square and rectangular billiards, or the
  stadium and the Sinai billiards, respectively.  The classical
  trajectories of integrable billiards form closed orbits, which
  produce as many conserved quantities as the number of energy levels,
  thus resulting in independent energy levels that are free to cross
  each other once plotted as a function of physical
  parameters. Conversely, the classical trajectories of the chaotic
  billiards do not form closed orbits in general, thus with a number
  of the conserved quantities that is smaller than the number of
  energy levels. This yields repulsion between the energy levels in
  the corresponding blocks of the Hamiltonian.}

{Here, we suggest that the one-dimensionality of the
  Andreev tube classical trajectories can qualitatively change the
  spectrum of the ABS, in the sense of the emergence of energy level
  crossings in the billiard connected to the superconducting leads,
  with respect to avoided crossings in the billiard disconnected from
  the superconducting leads. Thus, we conjecture that, under specific
  conditions, a chaotic billiard can become integrable once coupled to
  the superconducting leads, for the reasons that are now discussed.}

{Now, we specify the three assumptions of our
  conjecture. (i) The limit of a large superconducting gap yields
  absence of nonlocality in the superconductors. The billiard coupled
  to the infinite leads is described by the large-gap Nambu
  Hamiltonian that is solely sustained in space by the boundary and by
  the interior of the billiard. (ii) We additionally assume perfectly
  transparent interfaces between the central billiard and each of the
  superconducting leads.  (iii) We also assume that the entire
  boundary of the billiard is connected to superconducting leads,
  which we equivalently call as {\it full coating} by an arbitrary
  number of superconducting leads. Examples of partially coated
  billiards are shown in Figs.~\ref{fig:billiards}a-c and
  Fig.~\ref{fig:billiards}e, and fully coated billiards are shown in
  Figs.~\ref{fig:billiards}d and~\ref{fig:billiards}f. An example of
  fully coated chaotic billiard is shown in
  Fig.~\ref{fig:billiards}f.}

{We conjecture that the above assumptions (i)-(iii)
  imply that a chaotic billiard (in the absence of coupling to the
  superconductors) can transmute into an integrable multiterminal
  Josephson billiard (once coupled to an arbitrary number of
  superconducting leads). In addition, in the multiterminal Josephson
  device, the Josephson couplings solely operate in pairs, and the
  Josephson current and the ABS spectra do not involve terms that are
  sensitive to the three-body quartet phases or to the higher-order
  multi-body combinations of the superconducting phase variables.}

{Finally, we provide remarks that support those
  statements. The geometric constraint resulting from the condition of
  perfect momentum conservation in the quartet channel makes disappear
  the sensitivity on the three-terminal quartet phase combinations of
  the $\varphi_1+\varphi_2-2\varphi_3$-type. Those three-body
  combinations are replaced by the higher-order two-terminal harmonics
  of the $p(\varphi_1-\varphi_2)$-type with $p$ an integer. The
  conjecture is motivated by the observation that, within the above
  (i)-(iii) assumptions, the infinite-gap Josephson Hamiltonian is
  block-diagonal on the basis of the Andreev tubes, even after taking
  Andreev reflection in the quartet channel into
  account. Ref.~\onlinecite{Kulik} gives all signs that the resulting
  1D channel of an Andreev tube coupled to two superconducting leads
  at its extremities is integrable, and its spectrum with equally
  spaced energy levels reveals more symmetries at low energy than in a
  generic integrable Hamiltonian, see Eq.~(\ref{eq:spectrum-Kulik-1D})
  and Fig.~\ref{fig:spectra}a. We conclude that each energy level of
  the classical billiard is associated with an independent set of
  quantum numbers, which is specified by the location of the tube and
  by the discretized wave-vector of the corresponding 1D modes. This
  possibly can make the multiterminal Josephson billiard integrable,
  even if its shape makes it chaotic in the absence of coupling to the
  superconducting leads.}

\end{document}